\documentclass[12pt]{article}

\let\OLDitemize\itemize
\renewcommand\itemize{\OLDitemize\addtolength{\itemsep}{7pt}}

\usepackage[outline]{contour}
\usepackage{lmodern}
\usepackage[customcolors]{hf-tikz}
\usetikzlibrary{shapes.geometric, positioning, calc, shadows, through}
\usepackage{pgfplots}
\pgfplotsset{compat=newest}
\usepackage{multirow}
\usepackage{xcolor,colortbl}
\usepackage[]{natbib}
\usepackage{caption}
\usepackage{subcaption}
\usepackage{lscape}
\usepackage{placeins} %flotbarrier
\usepackage{graphicx} % Allows including images
\usepackage{booktabs} % Allows the use of \toprule, \midrule and \bottomrule in tables
\usepackage{array}
\usepackage[top=2.54cm, left=2.54cm, right=2.54cm, bottom=2.54cm]{geometry}
\usepackage[flushleft]{threeparttable}
\usepackage{centernot}
\usepackage{amsmath,amsthm,amssymb,amsfonts}
\usepackage[export]{adjustbox} %adjust table size
\usepackage{varwidth}
\usepackage{mathtools}
\usepackage[capposition=top]{floatrow}
\usepackage[onehalfspacing]{setspace}
\usepackage{epigraph} %opening quote
\usepackage{tcolorbox} %border around figures
\usepackage{float} %force figure position with [H]
\usepackage{longtable} %make tables and figures spanning pages
\usepackage{fontawesome} %fancy link

\usepackage{hyperref} %links

\makeatletter
\def\thanks#1{\protected@xdef\@thanks{\@thanks
		\protect\footnotetext{#1}}}
\makeatother

\usepackage[T1]{fontenc}
\usepackage[sc]{mathpazo} % font

\definecolor{LightCyan}{rgb}{0.88,1,1}
\definecolor{dark-red}{rgb}{0.4,0.15,0.15}
\definecolor{mygreen}{RGB}{34,120,34}
\definecolor{mygreen2}{RGB}{0,100,51} %dark green, default
\definecolor{mygreen3}{RGB}{210,255,220} %lighter
\definecolor{myblue}{RGB}{0,80,178} %citations
\definecolor{mybluedark}{RGB}{0,20,90} %citations
\definecolor{mypurple}{RGB}{102,0,102} %deep purple
\definecolor{myorange}{RGB}{255,128,0} %orange
\colorlet{lightgreen}{mygreen3!20} %light background green
\colorlet{darkblue}{myblue!80!black}

\hypersetup{
	colorlinks=true,
	linkcolor=black,
	filecolor=mybluedark,      
	urlcolor=mybluedark,
	citecolor=mybluedark,
	pdfpagemode=FullScreen,
}
\newcommand{\sss}[1]{\medskip\noindent \textbf{#1.~}}

\newcommand{\E}{\mathbb{E}}

\newtheorem{theorem}{Theorem}
\newtheorem{definition}{Definition}

\newtheorem{assumption}{Assumption}
\newtheorem*{assumption*}{Assumption}

\newtheorem{prediction}{Prediction}
\newtheorem*{prediction*}{Prediction}
\newtheorem{corollary}{Corollary}
\newtheorem*{corollary*}{Corollary}
\newtheorem{proposition}{Proposition}
\DeclareMathOperator*{\argmax}{arg\,max}
\DeclareMathOperator*{\argmin}{arg\,min}
\graphicspath{{/Users/raphaelraux/Dropbox/AI Project/Analysis/priors}{/Users/raphaelraux/Dropbox/PHD/G3/Early stage ideas/MU/Main/Analysis/}{/Users/raphaelraux/Dropbox/PHD/G3/Early stage ideas/MU/Paper/Pictures/}{/Users/raphaelraux/Dropbox/AI Project/Analysis/}{/Users/raphaelraux/Dropbox/AI Project/Analysis/human_perf}{/Users/raphaelraux/Dropbox/AI Project/Analysis}{/Users/raphaelraux/Dropbox/AI Project/Images}{/Users/raphaelraux/Dropbox/AI Project/Images/GIFs}{/Users/raphaelraux/Dropbox/AI Project/Standardized tests/TIMSS/Qualtrics questions/4th grade/4th grade 2003}}

\title{Human Learning about AI} 

\author{Bnaya Dreyfuss \\ \small{Harvard University} \and Rapha\"el Raux\thanks{\protect\linespread{1}\protect\selectfont
		\noindent Corresponding author:   raphraux@mit.edu.   We thank Katie Coffman, Benjamin Enke, Jesse Shapiro, and David Yang for guidance throughout this project. We thank Nikhil Agarwal, David Almog, Isaiah Andrews,  Leonardo D'Amico, Jean-François Bonnefon, Yannai Gonczarowski, Alex Imas, David Laibson, Daniel Martin, Dev Patel,  Matthew Rabin, Tobias Salz, Josh Schwartzstein, as well as conference and seminar participants at ASSA 2026, Bar-Ilan University, Columbia, EEA 2025, EC'25, ECBE, Harvard, Hebrew University, FSU, MIT, Monash, NBER Digital \& AI Meeting, TSE, and UIUC for helpful comments, and Kaitlyn Ernst for excellent research assistance. Finally, we are particularly thankful to Emily Oster and Alex Tryon for access to chatbot conversation data. All experiments were approved by Harvard IRB (ID 23-0588), and pre-registered at AEA RCT Registry (IDs 0012253; 0012622; 0014039; 0014086).
		Math tasks were obtained from TIMSS: $\copyright$ 2024 International Association for the Evaluation of Educational Achievement (IEA); TIMSS \& PIRLS International Study Center, Lynch School of Education, Boston College.
		This project was financially supported by the Chae Family Fund and the IQSS.} \\ \small{MIT Sloan} }

\date{\small First version: June 8, 2024 \\ This version: \today} % Date, can be changed to a custom date

\begin{document}
	\setstretch{1.5}
	%no page number for abstract
	
	\maketitle
	\thispagestyle{empty}
	%------------------------------------------------%
	%                     ABSTRACT                   %
	%------------------------------------------------%
	
	\vspace{-.7cm}
	\begin{abstract}\singlespacing\noindent
		\noindent
		
		\noindent
		We study \emph{Human Projection} (HP): people's tendency to evaluate AI using the same frameworks they use for humans---treating features such as task difficulty and the reasonableness of mistakes as diagnostic of overall ability. We formalize HP and its consequences for equilibrium adoption, testing its predictions experimentally.
		First, people project human difficulty onto AI, overestimating performance on human-easy tasks, underestimating it on human-hard ones, and over-updating after easy failures and hard successes---leading to systematic misspecification when AI performance is jagged rather than human-ordered. Second, HP interprets observed performance through a single ability index, inducing all-or-nothing adoption even when AI outperforms humans on only  some tasks; experimentally stripping AI of human-like cues weakens cross-task generalization and reduces over-adoption. Finally, a field experiment with a parenting-advice chatbot shows that less humanly reasonable mistakes cause larger drops in trust and future engagement. Anthropomorphic AI design can amplify HP, misaligning beliefs and distorting adoption.

	\end{abstract}
	\clearpage

	%------------------------------------------------%
	%                  INTRODUCTION                  %
	%------------------------------------------------%
	\pagenumbering{arabic}
	
	\vspace{-.5cm}
	\epigraph{The main lesson of thirty-five years of AI research is that the hard problems are easy and the easy problems are hard.}{Steven Pinker, \emph{The Language Instinct}}
	\nocite{pinker2007language}
	
	\vspace{-.5cm}
	\section{Introduction}
	\label{sec:intro}

	Recent AI systems are remarkably versatile, but their performance is often \emph{jagged}: they can excel at some tasks while failing at  closely related ones \citep{bengio2026international}, making evaluation challenging. Misjudging where AI is reliable has important economic consequences: highly trained consultants can incur net productivity losses when they trust AI on tasks outside its competence \citep{dell2023navigating}, and incorrect clinical recommendations can steer radiologists toward erroneous diagnoses and suboptimal follow-up decisions \citep{bernstein2023can,yu2024heterogeneity}.

	Yet these jagged systems are increasingly deployed in \emph{anthropomorphic} form: LLMs power user-facing chatbot  ``assistants'' that converse in the first person, use human turn-taking cues, and increasingly incorporate human-like features such as voice and personalization \citep{hurst2024gpt,anthropic2025system}.
	In this paper, we ask:  when jagged AI is wrapped in a human interface, do users evaluate it by human yardsticks?  What are the implications for performance beliefs and adoption decisions?

	%%% IN THIS PAPER, 1-2-3 structure (intuition, model, data)
	Our main hypothesis is that people evaluate AI using the same assessment frameworks  they use for humans---a phenomenon we call \emph{Human Projection} (HP). 
	They treat features diagnostic of \emph{human} competence, such as task difficulty or whether a mistake appears reasonable, as informative for AI. They use these features to infer an overall notion of AI ``ability,'' which generalizes across tasks and shapes adoption decisions.
	While this simplifying structure is very useful when assessing humans, it becomes misspecified when ``ability,''  ``difficulty,'' and ``reasonableness'' differ between humans and machines.

	We develop a formal framework for HP comprising two components: (i) treating features diagnostic of human performance, such as task difficulty, as informative for AI (\emph{difficulty projection});\footnote{Task difficulty is the main feature we study; Section~\ref{sec:parentdata} extends the logic to settings without ground-truth difficulty, using the reasonableness of failures as another human-diagnostic feature.} (ii) imposing a human-like generalization structure in which a single latent ability indexes performance across tasks (\emph{ability model  projection}). We use this framework to derive testable predictions for belief formation. We then embed HP in a Berk-Nash adoption model \citep{esponda2016berk} where delegation choices shape which performance signals are observed. 
	We show that HP induces \emph{all-or-nothing adoption} in equilibrium: perceiving a single-indexed ability precludes comparative advantage, so partial adoption cannot be sustained even when AI is only superior on some tasks.
	As a result, the same projection mechanism can generate \emph{both} over- and under-adoption relative to the task-by-task optimum.

	We test these predictions in a series of pre-registered experiments. In a lab study on performance beliefs, we find evidence consistent with Human Projection: priors fall steeply with human difficulty, and updating is stronger after observing easy failures and hard successes. Because difficulty is only weakly informative about AI performance in our setting, this yields systematic miscalibration---overestimating AI on human-easy tasks and underestimating it on human-hard tasks.
	In a repeated-delegation experiment mirroring the adoption model, 
	we find that removing human-like cues and presenting AI as a ``black box'' reduces the prevalence of HP:  all-or-nothing adoption falls sharply, consistent with weaker projection of a single-index ability,  which limits generalizations across tasks.
	In a framed field experiment with a chatbot providing parenting advice, we find that engagement drops sharply when observed AI mistakes are less humanly ``reasonable''---i.e., less similar to a helpful reply. This pattern is consistent with users projecting human-relevant notions of error reasonableness onto AI outputs, which can lead them to disengage from an otherwise helpful tool.

	% ----- HP FRAMEWORK ----- %
	We begin in Section~\ref{sec:framework} by formalizing Human Projection. A decision-maker observes an agent's binary performance in a domain of tasks and makes forecasts based on observed performance. Human Projection has two components. \emph{Difficulty projection} means treating human task difficulty as informative for AI performance. \emph{Ability model projection} means interpreting outcomes through a human ``ability model'': success is increasing in a unidimensional latent ability and decreasing in task difficulty, and signals satisfy a diagnosticity property (MLRP) so that hard successes and easy failures are especially informative about latent ability.  
	HP is a structured way of mapping task features into beliefs and extrapolating from limited feedback, and need not be a ``bias'' \emph{per se}---it is a sensible default when applied to humans. It can become misspecified, however, when transported to AI: its performance 
	need not admit the assumed single-index ability representation, and human difficulty may only be weakly informative for AI. Unlike distortions that compress the \emph{correct} signal mapping \citep{augenblick2025overinference}, predictions under HP rely on the \emph{wrong} mapping: people apply a human-diagnostic signal structure where it does not belong. 
	The framework yields two testable predictions. First, prior beliefs about success should decline with human difficulty, including for AI. Second, belief updating should follow the human-diagnostic ranking of signals: hard-task successes and easy-task failures should move beliefs more than easy successes and hard failures.
	
	We empirically test these predictions in a pre-registered experiment in Section \ref{sec:beliefs}. We focus on the math domain and construct a new benchmark of 414 multiple-choice math items drawn from the Trends in International Mathematics and Science Study (TIMSS).  
	We obtain item-level human performance data from an incentivized online test averaging over 41 answers per item, and measure the \emph{human difficulty} of each item as the share of test-takers answering it incorrectly. 
	We then evaluate ChatGPT~3.5 on the same items and find that it is jagged: item-level human difficulty is only weakly informative about AI success (slope $-0.07$, s.e.\ $0.08$, $R^2=0.002$).\footnote{This weak alignment is not specific to ChatGPT 3.5: Section~\ref{sec:math} documents similarly flat human--AI difficulty slopes across frontier models and prompt variants, consistent with recent evidence on AI's jagged capability patterns, even for frontier models \citep{li2025can,bengio2026international}.}

	We then elicit incentivized beliefs about the success rates of either humans  or AI (framed as ChatGPT).\footnote{Before stage-specific comprehension exclusions, the sample includes $n=240$ respondents predicting human performance and $n=957$ predicting AI performance.} We present two main findings, confirming our predictions. First, participants' prior beliefs about AI performance on an item strongly decrease with the item's human difficulty (OLS slope $-0.31$, s.e. $0.01$, $R^2=0.12$; about half the human slope), consistent with difficulty projection. %This gradient increases significantly with participants' own math performance---consistent with higher-performing participants perceiving question difficulty more accurately---though the relationship is attenuated in the AI arm, suggesting they also rely less on human difficulty as a cue for AI.
	Second, belief updating is significantly stronger after observing easy failures (vs. hard failures) and hard successes (vs. easy successes), and the size of this effect increases with the participant's difficulty projection.\footnote{The link between a participant's difficulty projection, measured from their priors, and the strength of their updating is consistent with reliance on a human-diagnostic mapping rather than compression of the AI-correct one: under a compression account as in \cite{augenblick2025overinference}, updating asymmetries should track the true (AI) signal structure, not the participant's prior gradient over human difficulty.} 
	Therefore, because AI performance is empirically jagged (uncorrelated with  difficulty), participants' reliance on human difficulty for belief formation yields miscalibration.

	In Section~\ref{sec:adoption} we study AI adoption when signals about performance are endogenous to delegation decisions. We model a decision-maker observing feedback only on tasks they choose to delegate to AI, and interpreting it through the HP model. 
	Projecting a single-index ability forces a binary conclusion: 
	as task-level feedback is aggregated into a single ability comparison, the decision-maker infers that AI  either has higher or lower ability than humans in the domain, leading to \emph{all-or-nothing} adoption in Berk-Nash equilibrium \citep{esponda2016berk}.  In particular, partial adoption can never be sustained even when it is optimal under task-by-task evaluation.  Thus, HP can generate distortions toward under-adoption (no adoption despite AI being superior on some tasks) or over-adoption (full adoption despite AI being inferior on some tasks); we provide precise conditions under which each arises. %In the appendix, we also trace out the dynamic implications for adoption by allowing technology to progress over time.
	Because human-like cues are known to induce anthropomorphic reasoning about machines \citep{chugunova2022we}, removing them should weaken Human Projection. The model predicts this would shift choices away from all-or-nothing adoption toward selectively adopting  AI in tasks where it outperforms humans.
	
	%for all-nothing: (baseline: $\Delta=-0.13$, $p=0.01$; replication: $\Delta=-0.20$, $p=0.02$),
	We experimentally test this prediction in Section~\ref{sec:adoption}. Using the same universe of math questions as before, we construct two question pools: easy and hard. Human performance is better in the easy pool, and AI performance is better in the hard one, making partial adoption (delegating easy questions to humans, and hard ones to AI) the task-by-task optimum. The experiment has two main stages, where all decisions are incentivized. In the learning stage, participants repeatedly delegate randomly selected questions from each pool to either a human or to AI, observing the realized performance of the chosen agent.  Then, in the adoption stage,  participants make a one-shot adoption decision separately for each pool. We randomize whether the AI agent is framed anthropomorphically (a named assistant with a human-like icon) or as a non-anthropomorphic ``black box'' (anthropomorphic: $n=149$; black box: $n=157$).
	The black box framing significantly reduces all-or-nothing adoption ($\Delta=-0.13$, $p=0.01$). In the baseline sample, however, the treatment-induced shift toward partial adoption is concentrated in the wrong choice---delegating the human-easy pool to AI. We show that this pattern is consistent with misperceptions of the pools' \emph{human} difficulty, and that the treatment raises the share of participants who select the optimal adoption choice, conditional on their (mis-)perception of human difficulty.
	To test this interpretation directly, we replicate the experiment on a new sample, making the human difficulty ordering salient. We find the same reduction in all-or-nothing adoption ($\Delta=-0.20$, $p=0.02$) but this time delegation shifts toward the human-hard tasks (the optimal choice).\footnote{Sample sizes in the replication experiment  are $n=57$ for anthropomorphic and $n=59$ for black box. The two experiments share the same design and interface, with the only difference being that the replication experiment explicitly states that the human-easy pool is composed of grade-school problems, and the human-hard pool of high-school problems.} 
	Evidence suggests weaker ability model projection is the mechanism behind the reduction in all-or-nothing adoption. The black box framing reduces the coupling of AI beliefs across task pools---namely, easy-hard belief co-movement is more than halved---suggesting that participants are more likely to treat AI performance as pool-specific rather than driven by a single ability index.

	In Section~\ref{sec:parentdata} we extend the HP logic to a field setting, studying engagement with a parenting-advice chatbot. 
	In this context, notions of ground-truth correctness and task difficulty are harder to define. We instead study a different feature diagnostic of human competence: the \emph{reasonableness of failures}---how closely an unhelpful response matches what a helpful answer to a related question would have looked like. This plays a parallel role in our framework to human task difficulty in settings without ground-truth correctness, allowing us to extend the same projection logic to deployed AI systems.
	We study whether users project this human-relevant feature onto AI when deciding whether to engage further. For instance, responding to ``What is the best car seat brand?'' with advice on car seat installation is a more reasonable failure than recommending brands of baby food, and may therefore depress engagement less. The inferential logic is similar: users may view an unreasonable mistake as more diagnostic of low ability, and generalize from it to the chatbot's overall competence.
	In a framed field experiment ($n=654$) using matched failure pairs that are equally unhelpful but differ only in reasonableness (both rated by a separate sample of potential users), less reasonable failures generate substantially larger drops (almost twice as large) in beliefs and trust and reduce subsequent engagement, both in a revealed-preference link choice (8.9 pp, $p=0.021$) and in actual chatbot use over the following weeks (2.6 pp, $p=0.030$). Consistent with HP, users treat the human-assessed reasonableness of (empirically rare) mistakes as diagnostic of  AI competence, disengaging from an otherwise helpful tool.

	%%comp stat
	\sss{HP distortions as models improve}
	A natural question is whether the distortions we document attenuate as models improve. Our framework predicts that distortions diminish only to the extent that errors become more human-ordered as capability grows---that is, if failures concentrate increasingly on human-hard items. Higher mean accuracy alone is not enough. Jaggedness, however, appears to persist beyond any specific generation of  AI systems: the 2026 International AI Safety Report notes that ``capabilities remain jagged: leading systems may excel at some difficult tasks while failing at other, simpler ones'' \citep{bengio2026international}. In Section \ref{sec:misalignment-general} we show that human-AI difficulty alignment remains weak for several more recent and capable models. 
	As an illustration, in Online Appendix~\ref{appendix:distortion}\footnote{The Online Appendix (OA) contains theoretical details and proofs, additional empirical results and robustness checks, as well as details on benchmark construction and experimental procedures.} we apply the elicited belief rule (from ChatGPT 3.5) to these models' realized performance and show that---holding beliefs fixed---the gap between belief-predicted and actual difficulty slopes does not narrow with capability. Beliefs themselves may of course evolve with user experience or across model generations, and studying this development is an important direction for future work.

	\section{Related Literature}
	Our paper builds on a growing literature studying \emph{mental models} as simplified---and often misspecified---representations that shape inference and choice.\footnote{Examples of misspecification channels include biased recall \citep{bordalo2023memory,enke2024associative,bohren2025mental}, selective attention \citep{graeber2023inattentive,bastianello2026biases}, improper coarsening \citep{graeber2025coarse}, or cognitive noise \citep{khaw2021cognitive,enke2023cognitive,andre2024mental}.}
	Among them, \emph{projection} heuristics---agents imposing their own attributes onto other people---shape  consumption choices \citep{loewenstein2003projection}, social learning \citep{gagnon2024quality} and strategic behavior \citep{madarasz2023projective}.
	We extend this line of work by showing people project human inferential structure onto AI, with consequences for beliefs and adoption. 
	Closest to our study is \citet{vafa2024large}, who document that people generalize from observed LLM performance in structured, sparse ways: seeing a model succeed at math raises beliefs about physics, but not literature. We view HP as a microfoundation for this pattern: ability model projection is the within-domain mechanism that moves beliefs across tasks. Observed sparsity  is consistent with HP extended through a perceived correlation structure of domain-specific abilities,  where people treat abilities for math and physics as relatively more correlated than for math and literature.
	More recently, \cite{he2025human} document that people also project human \emph{preferences}---their own and the human average---onto AI.
	
	Whether wrong mental models can persist in the face of (sometimes disconfirming) evidence is central to understanding their economic consequences.
	Theoretical work has shown they can, by shaping data acquisition and interpretation \citep{gagnon2023channeled, fudenberg2021limit,fudenberg2024selective,gans2026model}. Berk-Nash equilibrium (BN-E) \citep{esponda2016berk} formalizes equilibrium behavior when agents optimize within a fixed, potentially misspecified class of models.\footnote{This captures situations where agents have limited concern for their own misspecification \citep{lanzani2025dynamic} or seek to learn about coarse latent structures from observations \citep{samuelson2025robust}. These theoretical situations are consistent with evidence that people favor simpler models over richer but more complex ones  \citep{kendall2024complexity}.} 
	While the theoretical foundations are well developed, experimental tests of misspecified-model dynamics remain limited \citep[with notable exceptions, e.g.,][]{hanna2014learning,esponda2024mental}.
	We add to this work by applying BN-E to study AI adoption, showing HP misspecification induces all-or-nothing adoption equilibria and then providing experimental evidence for these patterns in an environment with repeated endogenous feedback.

	HP causes over- and under-inference from performance signals. This connects our paper to a body of work that documents empirical patterns of distorted inference from signals in various settings,\footnote{Including experimental belief updating \citep{griffin1992weighing,enke2023cognitive,ba2024over}, risky choice \citep{khaw2021cognitive},  financial forecasting \citep{fan2024inference}, stereotyping \citep{bai2025learning}, and survey expectations \citep{bordalo2026people}.}
	with under-inference being the empirically dominant direction \citep{benjamin2019errors}. \citet{augenblick2025overinference} show these patterns arise from misperceptions of signal \emph{diagnosticity}: people compress their estimate of signal strength toward an average, generating over-inference from weak signals and under-inference from strong ones. We contribute a distinct channel: rather than distorting the scale of the correct mapping, HP substitutes the \emph{wrong mapping}, evaluating AI signals through a human-diagnostic inference structure. This is a structured form of ``system neglect''  \citep{massey2005detecting}: agents not only fail to adjust to the AI signal-generating process, they import the human one. Human task difficulty is thus perceived to affect signal strength, which  systematically  distorts inference when difficulty is weakly informative of AI performance.

	Finally, our work speaks to the literature on AI adoption, where recent evidence shows heterogeneous and context-dependent effects on economic outcomes, with sometimes null or even negative effects.\footnote{Productivity gains vary across settings: writing \citep{noy2023experimental}, customer support \citep{brynjolfsson2025generative}, clinical decision-making \citep{agarwal2023combining}. Null or negative effects can appear in consulting \citep{dell2023navigating}, microenterprise \citep{otis2024uneven}, and white-collar work \citep{humlum2025large}. These variable effects are often related to user characteristics \citep{caplin2024abc}, motivating user-specific calibration \citep{dreyfuss2025calibrated}.}
	A key factor is that users often misjudge which tasks jagged AI systems are reliable on \citep{bernstein2023can,yu2024heterogeneity,chevrier2024algorithm}. We provide a theoretical account of why jaggedness systematically misleads users and how it generates sub-optimal adoption.
	Our unified mechanism helps reconcile the coexistence of algorithm aversion \citep{dietvorst2015algorithm,dietvorst2020people} and appreciation \citep{logg2019algorithm}: under HP, the same projection process can generate over-adoption when AI is humanly impressive (succeeding on hard tasks) and under-adoption when it appears humanly incompetent (failing easy ones).
	Our experimental manipulation of anthropomorphic cues speaks directly to the literature on machine anthropomorphism \citep{chugunova2022we,schimmelpfennig2025humanlike}, providing causal evidence that human-like design shapes belief formation and adoption.

	%\textbf{Mixed economic effects of AI adoption:}
	%Productivity gains are heterogeneous \citep{noy2023experimental,otis2024uneven,brynjolfsson2025generative,cui2025effects,fang2025generative}
	%Adoption can have weak effects \citep{humlum2025large,dillon2025shifting}, or even worsen outcomes \citep{becker2025measuring,wiles2025generative,krakowski2026human}

	\section{A Framework of Human Projection}
	\label{sec:framework}
	
	In this section we develop the Human Projection (HP) framework: a structure in which inferences about human performance within a domain of tasks are projected onto---potentially non-human---agents.
	HP decomposes into two distinct, complementary components: ability model projection, in which the DM summarizes outcomes with a unidimensional latent ability, and difficulty projection, where they project the human task difficulty onto the agent.

	\subsection{Framework}
	\sss{Setup} A \emph{decision-maker} (DM) assesses the performance of an \emph{agent} $i$ drawn from a superpopulation composed of two \emph{groups} $g$: humans ($g=H$) and AI models ($g=A$).\footnote{We use ``agent'' to refer to any entity whose performance is being evaluated, distinct from the autonomous AI agents from computer science terminology.} The DM knows each agent's group membership. The domain of assessment is a set of $n$ tasks denoted $\mathcal{T}=\{t_1,\dots,t_n\}$. Performance is binary and stochastic: agent $i$ either succeeds or fails on task $t_j$, where $y_{ij}=1$ denotes success on task $j$. Let $q_{ij}=\Pr(y_{ij}=1)$ denote the true success probability of agent $i$
	on task $t_j$.

	The DM does not observe $\mathbf q_i=(q_{i1},\dots,q_{in})$.
	Instead, they infer performance through a mental model in which success
	probabilities are determined by known task difficulty and unknown agent ability. Crucially, the model is tailored for evaluating humans, but may be misspecified for AI.
	
	\sss{Mental model}
	Each agent $i$ in group $g$ has a unidimensional \emph{type} 
	$\theta_i\in\Theta\subseteq\mathbb{R}$, unobserved by the DM, that represents 
	latent \emph{ability} within the domain; $\theta_i$ is distributed according 
	to group-specific CDF $F^g$. Each task $t\in \mathcal T$ has a 
	group-specific \emph{difficulty} represented by the function $\delta^g: 
	\mathcal T \to \Delta\subseteq \mathbb R$, which the DM treats as known. 
	The DM believes the success probability of agent $i$ in group $g$  at 
	task $t_j$ is determined by a group-specific \emph{success rate function}  
	$p^g:\Theta\times\Delta\to(0,1)$, which satisfies the following structure.
	\begin{assumption}[Ability Model Projection]\label{assn:human-model} 
		The success-rate function $p^g(\cdot,\cdot)$ satisfies: 
		\begin{enumerate} 
			\item \textbf{Ability:} $p^g$ is strictly increasing in $\theta$.
			\item \textbf{Difficulty:} $p^g$ is strictly decreasing in $\delta$.
			\item \textbf{MLRP:} For $\theta'>\theta$ and $\delta'>\delta$: 
			$\frac{p^g(\theta',\delta')}{p^g(\theta,\delta')}\geq 
			\frac{p^g(\theta',\delta)}{p^g(\theta,\delta)}$ and 
			$\frac{1-p^g(\theta,\delta')}{1-p^g(\theta',\delta')}\leq 
			\frac{1-p^g(\theta,\delta)}{1-p^g(\theta',\delta)}$. 
		\end{enumerate} 
	\end{assumption}

	The first two conditions reduce dimensionality by placing agents and tasks each on a unidimensional scale: agents are ordered by ability, tasks by difficulty. MLRP adds structure to how these interact: successes on harder tasks are more
	diagnostic of high ability, while failures on easier tasks are more diagnostic
	of low ability.
	Together, these conditions constitute the first component of Human Projection: an \emph{Ability Model} used to assess performance within a domain, which closely matches models in Item Response Theory \citep{lord2008statistical}, the standard framework for estimating student proficiency from standardized test performance. Most modern tests, including TIMSS (the source of our math tasks in Section \ref{sec:math}), estimate proficiency using models satisfying Assumption \ref{assn:human-model}.\footnote{Commonly used logistic and ogive functions in IRT, mapping ability and difficulty to success probability, satisfy all parts of Assumption \ref{assn:human-model}. See the TIMSS Technical Report: \url{https://timssandpirls.bc.edu/timss2019/methods/pdf/T19_MP_Ch11-scaling-methodology.pdf}.}
	
	Relying on this model greatly simplifies learning: rather than updating on success rates task by task, the DM updates on a single  variable, ability, which determines all success rates. This simplification is natural within a domain of similar tasks---e.g., mathematics---where a single latent ability plausibly underlies performance across subdomains like arithmetic, algebra, and calculus.
	Both the mappings ($p^A$, $p^H$) and priors ($F^A$, $F^H$) can be group-specific and  can allow for different gradients of success with respect to ability or difficulty, as well as differences in the average level of performance and its dispersion.
	
	The second component of Human Projection is \emph{difficulty projection}, which imposes the same difficulty ordering of tasks across groups:
	\begin{assumption}[Difficulty Projection] \label{assn:proj}
		AI and human difficulty have the same ordering within the domain: for any $t,t'\in \mathcal T$
		$\delta^H(t)> \delta^H(t')$ if and only if $\delta^A(t)> \delta^A(t')$.
	\end{assumption}
	
	In Appendix \ref{app:theory} we highlight one potential microfoundation for HP---consistent with empirical evidence---via a cognitive imprecision model \citep{woodford2020modeling}. When the DM receives a noisy signal of AI difficulty and combines it with a prior anchored on human difficulty, perceived AI difficulty is shrunk toward human difficulty.
	
	\sss{Model misspecification} Assumptions~\ref{assn:human-model} and~\ref{assn:proj} together constitute 
	\emph{Human Projection} (HP). The Ability Model compresses performance onto a single latent  dimension, shaping both priors and belief updating; difficulty projection anchors 
	perceived AI task hardness on human difficulty, governing baseline success-rate gradients and signal informativeness. While the model need not hold exactly, it is widely used and well-supported empirically when applied to humans: within a domain, the Ability Model provides a good approximation of human performance, and human difficulty rankings are broadly consistent across individuals.

	Applied to AI, however, each component may be misspecified.
	First, the Ability Model imposes two orderings that need not hold: one AI agent's success rate vector may not dominate another's (violating ability ordering) and task difficulty need not be consistently ordered across AI agents (violating difficulty ordering).\footnote{Formally, Assumption \ref{assn:human-model} is violated whenever $q_{ij}>q_{i'j}$ but $q_{ij'}<q_{i'j'}$ for some $t_j,t_{j'}\in \mathcal T$ and $i,i'\in A$.} Second, even when AI performance does satisfy the Ability Model, the difficulty ranking may still diverge from humans: human-easy tasks need not be AI-easy, and vice versa---we provide evidence for this in Section \ref{sec:math}.\footnote{AI is thus a natural application, but the projection logic is more general: people may project human-relevant features onto any agent whose competence profile differs from the types of humans they are familiar with. Basic grammar mistakes by a non-native speaker, for instance, may be read as evidence of poor education when ``basic'' is judged from a native speaker's perspective.} Yet HP is a natural default: learning about an agent's performance without imposing any structure is difficult, and, as we discuss below (Section \ref{sec:adoption-exp}), AI agents often exhibit human-like cues that further invite its use.
	
	\sss{Results} HP yields two main results: one about priors, the other about learning. First, priors over success rates decrease with human task difficulty:
	\begin{proposition}\label{prop:gradient}  Define the prior expected success rate of agent $i\in g$ on task $t$ as $\pi^g(t)=\mathbb{E}_{F^g}[p^g(\theta_i,\delta^g(t))]$, where the expectation is over $\theta_i$. For any  $t_j, t_{j'}\in\mathcal{T}$ with $\delta^H(t_j)>\delta^H(t_{j'})$: $\pi^g(t_j)<\pi^g(t_{j'}).$  \end{proposition}
	All proofs are in the Appendix. Difficulty projection carries human difficulty over to AI, so the DM predicts that tasks that are harder for humans will be harder for AI as well, and that the success rates of \emph{both} groups decrease with human difficulty.
	The second prediction concerns learning.  The DM observes data
	\[
	D_i=((s_\ell,y_{i\ell}))_{\ell=1}^m,
	\]
	where each $s_\ell\in\mathcal T$ is an observed task and
	$y_{i\ell}\in\{0,1\}$ indicates success or failure. The DM updates their prior $F^g$ over $\theta_i$ via
	Bayes' rule through the lens of the Ability Model, yielding posterior
	$F^g(\cdot\mid D_i)$. When the data consist of a single observation, we write $(s,y)$ instead of $((s,y))$.
	
	\begin{proposition}\label{prop:cross}  Define the posterior expected success rate of agent $i$ in group $g$ on task $t$, given observed performance $D_i$, as $\pi^g(t\mid D_i)=\mathbb{E}_{F^g(\cdot\mid D_i)}[p^g(\theta_i,\delta^g(t))]$.  Take any two tasks $t_j, t_{j'}\in\mathcal{T}$ with $\delta^H(t_j)>\delta^H(t_{j'})$.  For any task $t\in\mathcal{T}$: \begin{enumerate} \item $\pi^g(t\mid (t_{j'},1)) \leq \pi^g(t\mid (t_j,1))$, \item $\pi^g(t\mid (t_{j'},0)) \leq \pi^g(t\mid (t_j,0))$. \end{enumerate} \end{proposition}
	
	Proposition \ref{prop:cross} captures two asymmetric updating patterns: a success on a harder task raises posterior success rates more than a success on an easier task, while a failure on an easier task lowers them more than a failure on a harder task. This follows from MLRP: because a failure on an easy task is more diagnostic of low ability than a failure on a hard task, the DM's posterior on ability drops further, dragging down expected success rates across the domain; symmetrically, a success on a harder task is more diagnostic of high ability, raising posteriors more. Difficulty projection ensures both patterns apply equally to AI by carrying the human difficulty ordering over to AI tasks.

	\sss{Experimental predictions}
	Propositions~\ref{prop:gradient} and~\ref{prop:cross} yield two experimental predictions, tested in Section~\ref{sec:beliefs}. The experiment operationalizes human task difficulty using a benchmark sample (Section~\ref{sec:math}), and the design maps onto the predictions directly: prior beliefs across items test Prediction~\ref{pred:gradient}, while an updating task---where participants revise beliefs after observing performance on a separate signal item---tests Prediction~\ref{pred:cross}. Both predictions apply to beliefs about human and AI agents alike.
	
	\begin{prediction}\label{pred:gradient}
		Expected success rates decrease with human difficulty.
	\end{prediction}
	
	\begin{prediction}\label{pred:cross}
		For any two tasks with unequal human difficulty, and any prediction task $t$:
		\begin{enumerate}
			\vspace{-.2cm}\item[(i)] After a \textbf{success}, posterior beliefs increase more if the success occurred on the harder task.
			\vspace{-.3cm}\item[(ii)] After a \textbf{failure}, posterior beliefs decrease more if the failure occurred on the easier task.
		\end{enumerate}
	\end{prediction}

	%====================================================%
	%      HUMAN vs GENAI DIFFICULTY IN MATHEMATICS       %
	%====================================================%
	
	\section{Human and Generative AI Difficulty in Mathematics}
	\label{sec:math}
	
	Human Projection generates belief misspecification precisely when \emph{actual} generative AI (GenAI) performance departs from the human ability model that people use to form expectations. In this section, we document such departures in the domain of mathematics. We first introduce a benchmark of standardized tasks and an empirical measure of \emph{human} difficulty. We then present our main case study, ChatGPT 3.5, where AI performance is  uncorrelated with human difficulty in our domain, contrasting with the anthropomorphic expectations we document in our experiments (Sections~\ref{sec:beliefs} and~\ref{sec:adoption}). Finally, we provide evidence that human-AI difficulty alignment on our benchmark remains weak for more recent models, echoing findings that jaggedness appears as a stable feature of GenAI models, unresolved by capability improvements alone \citep{bengio2026international}. Appendix~\ref{appendix:math} provides details on benchmark construction and performance data.
	
	\subsection{Benchmark of Tasks and Human Difficulty}
	\label{sec:beliefs-math-tasks}
	
	We collect and manually re-transcribe released items from the \textit{Trends in International Mathematics and Science Study} (TIMSS). These international standardized tests assess student proficiency at the 4\textsuperscript{th}-grade, 8\textsuperscript{th}-grade, and high-school level. We focus on multiple-choice items, which have either 4 or 5 possible answers (A--E), yielding a binary and objectively graded outcome.
	We obtain a final dataset of 414 items (29\% from 4\textsuperscript{th} grade, 58\% from 8\textsuperscript{th} grade, and 13\% from high-school) spanning a range of topics.
	
	This benchmark is well-suited for comparing human difficulty to AI performance. First, items are designed to measure mathematical proficiency using test-scaling methodologies consistent with the Ability Model. Second, question format is stable across items, and tasks span a broad range of topics and human difficulty. Third, a significant portion of items are not accessible online and were obtained through direct request to the IEA, reducing the risk of benchmark contamination. 
	
	\subsection{Main Case Study: ChatGPT and the Human Difficulty Gradient}
	\label{sec:beliefs-math-diff}
	
	\sss{Human performance data and difficulty measure}
	We collect incentivized human performance data on our benchmark using adult participants from Prolific. We pool data  from two samples collected under  similar test-like conditions (October 2023 and December 2023). Participants are instructed to approach the questions as they would a real math test and to remain time-conscious.
	
	We obtain human performance for each item: the average item is answered 41.5 times (median = 34), allowing us to measure ``human difficulty'' with low sampling noise at the item level. Following the standardized testing literature \citep{bachman1990fundamental}, we define human task difficulty by the share incorrect: $\text{Difficulty}_t \;=\; 100\times(1-\text{ShareCorrect}_t)$,
	reported in percentage points. Hereafter, ``difficulty'' refers to this empirical measure of \emph{human} difficulty. The average item difficulty is 30.3 and the median is 24; the most difficult item has difficulty 92, and the easiest has difficulty 0 (7 items).
	
	\begin{figure}[t]
		\centering
		\caption{ChatGPT Performance by Human Difficulty Deciles}
		\label{fig:perf-gpt}
		\includegraphics[width=.95\textwidth]{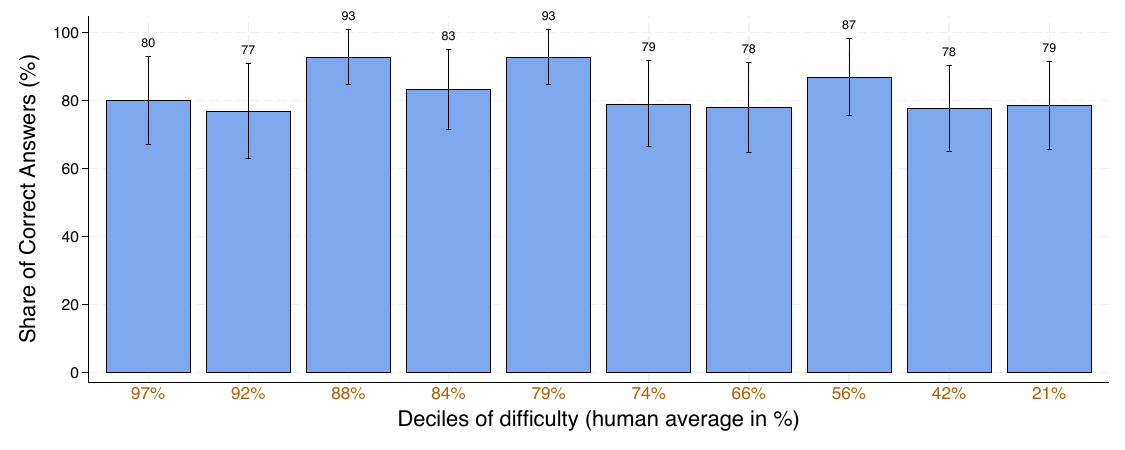}
		\vspace{-0.5cm}\floatfoot{\textit{Notes:} The figure plots ChatGPT 3.5's share of correct answers within each decile of human difficulty. Each decile contains between 40 and 42 items. Numbers under each bar report the average human success rate in the corresponding decile. Human difficulty is computed from an average of 41.5 answers per item.}
	\end{figure}
	
	\sss{ChatGPT performance}
	The AI performance data used in our experiments come from ChatGPT 3.5 (August 2023 version), evaluated on all items under zero-shot text prompting through the web chat interface. We classify an answer as correct if the model selects the correct answer key, and as incorrect otherwise. Figure~\ref{fig:perf-gpt} reports ChatGPT accuracy by deciles of human difficulty. ChatGPT accuracy is approximately flat across the human difficulty gradient: errors occur throughout the distribution at roughly the same rate rather than concentrating on human-hard items.
	Indeed, human difficulty has no predictive power for ChatGPT performance: Regressing an indicator for a correct answer (scaled by 100) on the 0--100 human-difficulty index yields an estimated slope of $-0.07$ (s.e.\ $0.08$, $p=0.38$, $R^2=0.002$). This finding is robust to a more recent API-based  replication (May 2025), which yields an even flatter slope (slope $-0.03$, s.e.\ $0.09$, $p=0.72$, $R^2<0.001$), with slightly lower average accuracy; Appendix Figure~\ref{fig:perf-gpt-api} reports the corresponding decile-accuracy plot.

	\subsection{Broader Evidence on Human--AI Difficulty Misalignment}
	\label{sec:misalignment-general}
	
	To assess whether the difficulty mismatch we document generalizes beyond ChatGPT 3.5, we evaluate six more recent and advanced models on the same TIMSS item set, quantifying human-likeness via item-level difficulty alignment and answer-choice overlap (Appendix Table~\ref{tab:humanlike}). The central pattern is stable across model families and prompt variants: high mean accuracy does not imply a human-like difficulty ordering. Correlations between human item easiness and model correctness are small,\footnote{Across the six models and three prompting variants (standard, no-chain-of-thought, and image) we tested, the highest Spearman correlation is 0.255 (median = 0.133).} and answer-choice overlap is substantially less consistent on mistakes---conditional on error, models do not  concentrate on the same distractors humans choose. Human-likeness of performance is also sensitive to presentation: suppressing chain-of-thought or switching to image prompts significantly shifts both accuracy and the human-model difficulty mapping, suggesting alignment is also sensitive to context.

	These patterns are corroborated by a growing literature: LLMs can solve complex questions yet fail simple ones \citep{williams2024easy}, trivial perturbations sharply affect model correctness \citep{berglund2023reversal,wang2025reversal}, and human-machine difficulty rankings are only weakly correlated across domains \citep{li2025can}, including agentic benchmarks \citep{mialon2023gaia,xie2024osworld}.
	
	Human difficulty is thus an imperfect guide to GenAI performance: even strong models can exhibit weak, context-sensitive alignment with human difficulty, and treating  human difficulty as informative about AI ability---as HP predicts---would entail systematic belief misspecification. We study this next.

	%------------------------------------------------%
	%                     BELIEFS EXP                %
	%------------------------------------------------%
	
	\section{Documenting Human Projection: Beliefs in Performance}
	\label{sec:beliefs}
	
	We design an experiment to elicit beliefs in performance over the domain of mathematical tasks presented in Section \ref{sec:math}. We test whether beliefs are consistent with the predictions of our Human Projection framework (Section \ref{sec:framework}), i.e., whether people see \emph{human} task difficulty as a relevant feature in predicting AI performance.

	%--------------------  EXPERIMENT ------------------%
	
	\subsection{Experimental Design}
	\label{sec:beliefs-exp}

	\sss{Overview and randomization}
	Participants predict an agent's performance on math items sampled from our task domain.
	Participants are randomly assigned to one of two arms varying the identity of the agent: in \emph{Human}, beliefs concern a randomly selected participant from our initial human benchmark; in \emph{AI}, beliefs concern ChatGPT 3.5.
	To minimize demand effects, we never reference ``difficulty'' in the  instructions, and each treatment describes only the relevant agent (AI is not mentioned in \emph{Human}, and human test-takers are not mentioned in \emph{AI}).
	The experiment has two stages: (i) elicitation of prior beliefs (i.e., before observing signals) across a set of items spanning the difficulty distribution, and (ii) a belief-updating task where participants update beliefs about performance on one prediction item after observing performance on a separate signal item.
	Within each agent treatment, we additionally randomize the updating signal arm between participants: Easy Fail, Hard Fail, Easy Success, or Hard Success.
	Overall, each participant reports 11 prior beliefs (10 in Stage~1 and 1 in Stage~2) and 1 posterior belief (Stage~2).

	\sss{Introducing the agent}
	We begin by providing brief background information about the agent to increase realism and reduce noise in elicited beliefs.
	In \emph{AI}, we introduce large language models and illustrate ChatGPT responding to a small set of generic prompts unrelated to the math items.
	In \emph{Human}, we describe the human benchmark sample and testing conditions from the initial data collection, and provide basic aggregate demographics of the test-taker pool.
	
	\sss{Stage 1: prior beliefs}
	Each participant is shown 10 items on the same page and provides two incentivized responses for each item: (i) their own answer, and (ii) a probabilistic belief about whether the agent answered correctly.
	Beliefs are elicited with the same wording across treatments except for the agent label: ``What do you think is the \% chance that [a random participant / ChatGPT] answered correctly?''
	Item draws are stratified to expose each participant to a broad range of difficulties: 2 items are sampled from the high-school (HS) pool (excluding the subset reserved for prediction items in the updating stage), and 8 items are sampled from the combined 4th- and 8th-grade pool.
	Eliciting participants' own answers before beliefs provides them with a  signal of task difficulty and generates additional performance data that we use to refine the human-difficulty index.
	
	\sss{Stage 2: belief updating}
	The updating stage elicits one prior and one posterior belief for a single  prediction item with high human difficulty.
	Participants first draw one \emph{prediction} item from a set of 10 high-school items reserved for this stage, answer it, and report their prior belief that the agent answered it correctly.
	Second, participants observe the agent's performance on a separate \emph{signal} item (distinct from the prediction item) drawn from one of four pools corresponding to the randomized arm (easy vs.\ hard $\times$ success vs.\ failure). ``Easy'' and ``hard'' signal pools are defined using the human-difficulty index (lowest and highest deciles), and success/failure is implemented by drawing from pools where the agent's realized answer is  correct or incorrect, respectively.
	Participants see the signal item, the agent's answer, and whether that answer is correct.
	Third, the prediction item is shown again and participants report their posterior belief: ``Given what you saw, what do you think is the \% chance that [the same participant / ChatGPT] answered this question correctly?''
	We define belief movement as posterior minus prior for the prediction item, and test Prediction~\ref{pred:cross} by comparing movement across signal-difficulty arms within successes and within failures.
	%\footnote{\% APPX: In the \emph{AI} treatment we also display a screenshot of the prompt and ChatGPT output used to generate the signal, to increase credibility of errors that may be surprising ex ante.}
	
	\sss{Incentives and logistics}
	Correct answers to math items are rewarded at \$0.05 each. We keep those stakes low to be comparable to the initial benchmark and to limit incentives to use outside help. 
	%given prior evidence on the small effect of stake size on belief updating errors  \citep{enke2023cognitive}.
	Belief elicitation is incentivized using a binarized scoring rule \citep{hossain2013binarized}, described in intuitive terms; participants can earn up to \$0.10 per prior belief report (10 reports) and up to \$0.30 for the updating belief report.
	%\footnote{\% APPX: We explain the scoring rule in intuitive terms on-screen and provide full details behind a clickable information button; see Appendix~\ref{appendix:beliefs} for screenshots.}
	The study was implemented in Qualtrics and run on Prolific in December 2023, following an initial human benchmark collection (which involved question-solving only) in October 2023.
	%full sample is 244 (human) and 971 (AI), but we exclude failing the easy) attention checks
	A total of 240 participants in \emph{Human} and 957 in \emph{AI} passed basic attention checks. After excluding participants who failed  stage-specific comprehension questions, the analysis samples are 231 (\emph{Human}) and 940 (\emph{AI}) for the priors stage, and 188 (\emph{Human}) and 784 (\emph{AI}) for the updating stage.\footnote{We include separate comprehension questions for each stage which can yield different sample sizes across stages and conditions.} As stated in the IRB submission and instructions, the experiment does not rely on deception: performance signals are generated using real human or model answers.

	%The initial test was designed to be completed in around 30 minutes with a base pay of \$4 with a maximum potential bonus of $30 \times 0.05 =$ \$1.5. The experiment was designed to last around 15 minutes, with a base pay of \$2 and a maximum potential bonus of $10\times( 0.05 +0.1)+0.3$ = \$1.8.  

	%---------------- BELIEF RESULTS ----------------%
	\subsection{Results}
	\label{sec:beliefs-results}
	
	% ---- MAIN FIGURE PRIORS
	\begin{figure}[t]
		\centering
		\caption{Prior Beliefs in Performance and Implied Difficulty Projection}
		\label{fig:bin-belief}
		\begin{subfigure}[t]{0.49\textwidth}
			\centering
			\caption{Binned Beliefs in Performance}
			\includegraphics[width=\textwidth]{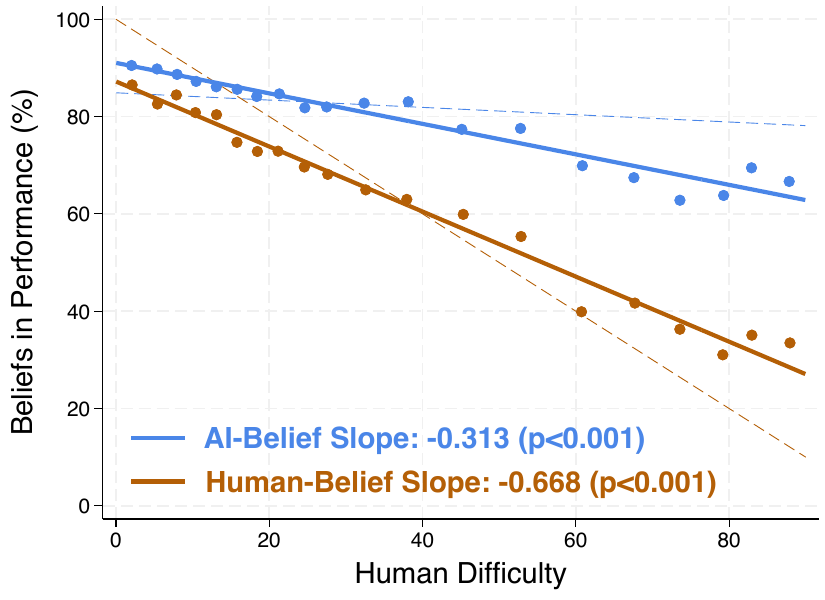}
		\end{subfigure}
		\begin{subfigure}[t]{0.49\textwidth}
			\caption{Histogram of Belief Slopes}
			\centering
			\resizebox{1\textwidth}{!}{
				\includegraphics[width=\textwidth]{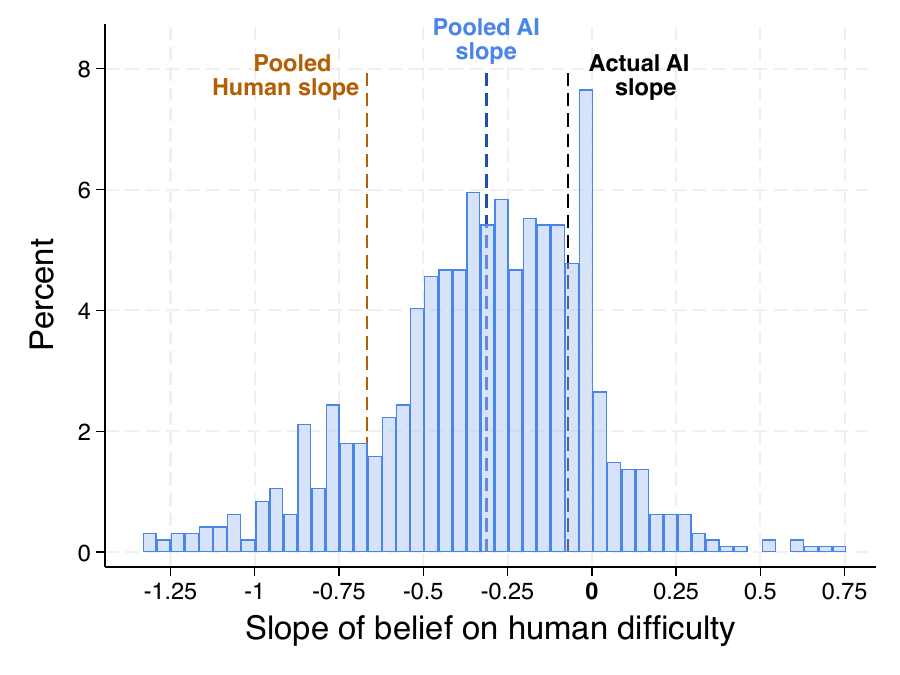}}
		\end{subfigure}
		\vspace{-0.4cm}
		\floatfoot{\textit{Notes:} Panel (a) presents binned scatter plots and regression lines (solid) of prior beliefs about human (blue) and ChatGPT (orange) performance. The sample excludes participants who failed the comprehension question related to performance prediction. $n=2541$ for humans, and $n=10340$ for ChatGPT. Dashed lines represent actual human and AI performance. We report coefficients from the basic OLS regression with standard errors clustered at the participant level. Panel (b) plots the distribution of participant-specific slopes from regressing ChatGPT-belief on human difficulty (11 elicitation items per participant); vertical lines denote the true ChatGPT performance slope, the pooled ChatGPT-belief slope, and the pooled human-belief slope.}
	\end{figure}

	\sss{Result 1: difficulty gradients in prior beliefs}
	Consistent with Prediction~\ref{pred:gradient}, prior beliefs decline sharply with human difficulty for both humans and AI, as shown in Figure~\ref{fig:bin-belief} and Table~\ref{tab:belief-perf}.  
	The AI gradient is about half as steep as the human gradient: the pooled OLS slopes in Figure~\ref{fig:bin-belief} imply a difficulty gradient ratio $\hat\beta^{AI}/\hat\beta^{H} \approx (-0.313)/(-0.668)=0.47$, and this ratio is similar when adding covariates and estimating both slopes in a pooled interaction specification.
	Beliefs are also higher on average for AI (mean $80\%$) than for humans (mean $64\%$).
	Overall, consistent with difficulty projection, participants treat human difficulty as highly informative for AI success rates: difficulty alone explains about $12\%$ of the variation in AI priors (Table~\ref{tab:belief-perf}).\footnote{This pattern is consistent with subjects applying a generic heuristic that hard-looking problems are hard for any solver. We view this as substantively equivalent to difficulty projection in our framework, whose central claim is precisely that intuitions regarding human difficulty are misapplied to AI rather than that subjects engage in AI-specific reasoning. The updating results below, where the strength of HP-consistent updating scales with each subject's own difficulty projection, further support this interpretation.}
	
	% --------- MAIN TABLE PRIORS	
	\begin{table}[t]
		\caption{Correlations of Beliefs in Performance with Item Difficulty}
		\label{tab:belief-perf}
		\centering
		\resizebox{\textwidth}{!}{
			\begin{threeparttable}
				\begin{tabular}{lcccccc} \toprule
& \multicolumn{3}{c}{\textbf{Beliefs in Human Perf.}} & \multicolumn{3}{c}{\textbf{Beliefs in AI Perf.}} \\
\cmidrule(lr){2-4} \cmidrule(lr){5-7}
& (1) & (2) & (3) & (4) & (5) & (6) \\
\hline
Task difficulty     &      -0.668***&      -0.668***&      -0.667***&      -0.313***&      -0.313***&      -0.317***\\
                    &     (0.020)   &     (0.021)   &     (0.022)   &     (0.010)   &     (0.010)   &     (0.011)   \\
\midrule
Controls            &               &  \checkmark   &  \checkmark   &               &  \checkmark   &  \checkmark   \\
Subject FE          &               &               &  \checkmark   &               &               &  \checkmark   \\
R-squared           &       0.391   &       0.393   &       0.604   &       0.116   &       0.125   &       0.442   \\
Observations        &        2541   &        2541   &        2541   &       10340   &       10340   &       10340   \\
\bottomrule \end{tabular}

				\smallskip\footnotesize
				\begin{tablenotes}
					\item \textit{Notes:} The table reports OLS coefficients for task difficulty. Controls include demographics of age, gender, college degree, math education, and prior AI familiarity (for \textit{AI}). Standard errors clustered at the participant level in parentheses. * $p<0.1$, ** $p<0.05$, *** $p<0.01$ 
				\end{tablenotes}
		\end{threeparttable}}
	\end{table}	
	
	In the context of our ChatGPT 3.5 case study (the model whose performance participants were asked to predict), this induces systematic misspecification: relative to realized performance, participants overestimate AI success on easier items and underestimate it on harder items (solid vs.\ dashed lines in Figure~\ref{fig:bin-belief}, panel~A), and elicited beliefs have very limited predictive power for correctness across items (bivariate $R^2\simeq 0.002$). Consistent with this aggregate pattern, Figure~\ref{fig:bin-belief}, panel~B shows that the distribution of participant-specific AI belief slopes is centered below the true AI slope, so a majority of participants impose an overly steep difficulty gradient.

	% ------ MAIN UPDATING FIGURE
	\begin{figure}[t]
		\centering
		\caption{Belief Updating after Exogenous Signals of Performance}
		\label{fig:belief-updating}
		\includegraphics[width=\textwidth]{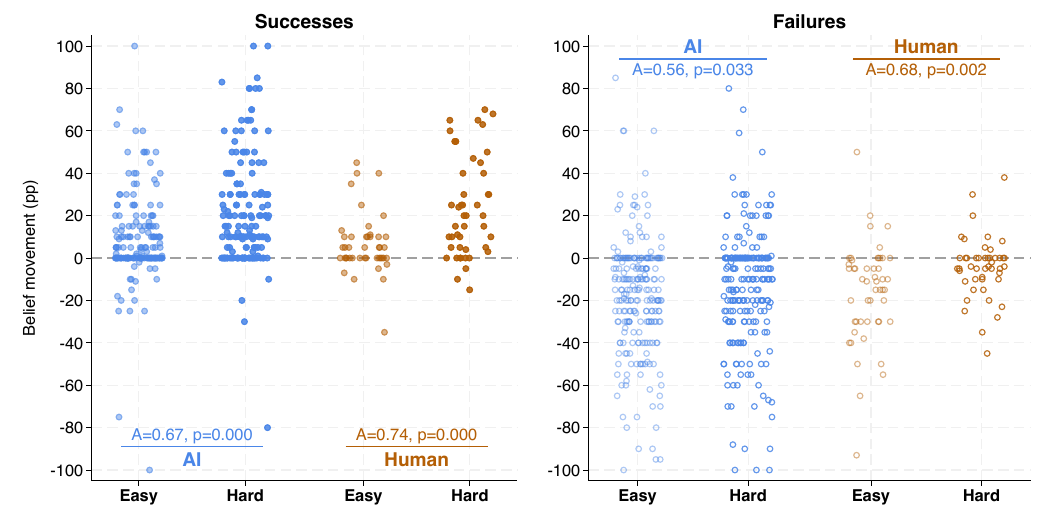}
		\vspace{-0.7cm}
		\floatfoot{\textit{Notes:} The figure shows jittered dot plots of belief movement ($\Delta$; posterior minus prior, in pp). Sample sizes from left to right are $n=167$, $n=161$, $n=44$, $n=47$, $n=230$, $n=226$, $n=50$, $n=47$. Reported $A = \Pr(\Delta_H > \Delta_E) + 0.5\Pr(\Delta_H = \Delta_E)$ is the common-language effect size from a Mann--Whitney $U$-test ($A = 0.5$ under the null; $A > 0.5$ indicates the hard-question group stochastically dominates the easy-question group); $p$-values are from two-sided Wilcoxon rank-sum tests.}
	\end{figure}
	
	\sss{Result 2: difficulty affects belief updating}
	Consistent with Prediction~\ref{pred:cross}, the human difficulty of the signal task strongly affects how participants update about both agents (human and AI).
	Figure~\ref{fig:belief-updating} plots the distribution of belief movements across arms. In both \emph{Human} and \emph{AI}, after a \emph{success}, posterior beliefs increase more when the success occurs on a harder task; after a \emph{failure}, posterior beliefs decrease less when the failure occurs on a harder task.
	Table~\ref{tab:belief-update} shows the same pattern using a continuous  difficulty measure (rather than a binary easy/hard classification) and remains qualitatively unchanged when adding task fixed effects and controls.
	Interestingly, for both \emph{Human} and \emph{AI}, the difficulty effect is larger after successes than after failures.%\footnote{One mechanical contributor is that the realized separation in difficulty between the ``easy'' and ``hard'' signal pools is smaller for failures than for successes: using the refined difficulty index, the hard--easy gap is $77.0-9.1=67.9$ for failures versus $87.3-7.1=80.2$ for successes. This discrepancy arises because pools were constructed using  performance data from our initial test, which we then refined using the experiment sample (see Section~\ref{sec:beliefs-exp}).} 

	\begin{table}[t]
		\caption{Belief Updating - Effect of Signal Difficulty on Beliefs}
		\label{tab:belief-update}
		\centering
		\resizebox{.85\textwidth}{!}{
			\begin{threeparttable}
				\setlength{\tabcolsep}{12pt}
\renewcommand{\arraystretch}{0.9}
\begin{tabular}{lcccc} \toprule
& \multicolumn{4}{c}{\textbf{Belief change after performance signal}} \\
\cmidrule(lr){2-5}
& \multicolumn{2}{c}{\textbf{Success}} & \multicolumn{2}{c}{\textbf{Failure}} \\
\cmidrule(lr){2-3} \cmidrule(lr){4-5}
& \textbf{Human} & \textbf{AI} & \textbf{Human} & \textbf{AI} \\
& (1) & (2) & (3) & (4) \\
\midrule
\multicolumn{5}{l}{\textit{Panel A. Mean effect (OLS)}} \\ [.3em]

Signal task difficulty&       0.268***&       0.154***&       0.128** &       0.066*  \\
                    &     (0.061)   &     (0.033)   &     (0.056)   &     (0.036)   \\

\midrule
\multicolumn{5}{l}{\textit{Panel B. Median effect (median regression)}} \\ [.3em]

Signal task difficulty&       0.202***&       0.126***&       0.108*  &       0.103***\\
                    &     (0.062)   &     (0.029)   &     (0.055)   &     (0.036)   \\

\midrule
Controls          & \checkmark & \checkmark  & \checkmark  & \checkmark \\
Prediction task FE & \checkmark & \checkmark & \checkmark & \checkmark  \\
R-squared (OLS)       & 0.254 & 0.092 & 0.197 & 0.032 \\
Observations      & 91  & 328  & 97  & 456  \\
\bottomrule \end{tabular}

				\smallskip\footnotesize
				\begin{tablenotes}
					\item \textit{Notes:} The table reports estimates of the effect of task difficulty on belief movement (posterior minus prior), estimated separately by outcome and by agent. Panel A reports OLS estimates and Panel B reports median regression estimates. All specifications include prediction-task fixed effects and participant controls (age, gender, college degree, math education, and prior AI familiarity). R-squared is from Panel A. Robust standard errors in parentheses. * $p<0.1$, ** $p<0.05$, *** $p<0.01$
				\end{tablenotes}
		\end{threeparttable}}
	\end{table}

	\sss{Additional results}
	We first separate two sources of the human--AI gap in priors: higher 
	baseline beliefs about AI ability (level), and weaker difficulty 
	sensitivity (gradient), consistent with partial difficulty projection.\footnote{Because we elicit probabilistic beliefs rather than the subjective difficulty-performance mappings, differences in average beliefs can reflect differences in perceived ability even assuming the same mappings.} As a diagnostic, we compare overall beliefs about AI to beliefs about humans  restricted to participants from the top decile of average beliefs about human performance. Beliefs in this restricted sample closely  resemble beliefs in \emph{AI}, in both level and slope 
	(Appendix Figure~\ref{fig:bin-humanlike}), suggesting that the difference in gradient between the two arms is at least partly driven by a higher perceived AI ability and not just differences in  difficulty gradients.
	
	Turning to heterogeneity, we ask which participants exhibit stronger difficulty projection onto AI. Participants who perform better at solving the math items show steeper belief-difficulty gradients in both arms, consistent with sharper perception of item difficulty (Appendix Figure~\ref{fig:hetero_slope}). The positive relationship between own performance and perceived difficulty gradient is significantly attenuated in \emph{AI}, consistent with lower difficulty projection among high-performing participants.
	The difficulty gradient is uncorrelated with self-reported AI familiarity, as well as with every demographic characteristic we examine: math background, education, income, gender, and age.
	
	Finally, our framework implies that participants whose priors exhibit stronger difficulty projection should also exhibit greater sensitivity of updating to signal difficulty. Indeed, we find that the hard-easy updating gap is systematically larger among high-projection participants (i.e., those with above-median difficulty gradient) than among low-projection ones; the pattern holds for both successes and failures, again with weaker effects for the latter (Appendix Figure~\ref{fig:belief-updating-proj}). 
	Rather than generic misperceptions of signal precision  as in \cite{augenblick2025overinference}, these findings reflect the use of a coherent mental model---consistent with HP---that relies on  human-relevant features, which leads to potential misperceptions of signal informativeness.

	\sss{Takeaways}
	The results support Human Projection in two distinct senses. First, priors about AI 
	inherit a human-difficulty gradient 
	(Prediction~\ref{pred:gradient}), consistent with difficulty projection. 
	Second, the same performance signal is interpreted differently depending 
	on task difficulty (Prediction~\ref{pred:cross}), consistent with 
	participants combining difficulty projection with ability model 
	projection---treating outcomes as informative about a latent, cross-task 
	ability. 
	This anthropomorphic structure yields 
	systematically misspecified beliefs in our ChatGPT case study, and does so predictably in any context where AI's performance is significantly more jagged than what people expect. We illustrate this point in Appendix \ref{appendix:distortion}, quantifying the misspecification that would result from applying our participants' beliefs  to more recent models.

	Having documented that beliefs exhibit HP, we now ask how these beliefs shape adoption decisions when feedback is endogenous to delegation choices.

	%------------------------------------------------%
	%                     DYNA--MICS                   %
	%------------------------------------------------%
	
	\section{AI Adoption in Equilibrium under Human Projection}
	\label{sec:adoption}
	
	So far, we have investigated beliefs about performance. How do these translate to adoption decisions? Adoption and beliefs are usually jointly determined: adoption shapes the information available, which in turn shapes adoption. To study this action-belief endogeneity in a principled way, we embed our HP framework in a Berk-Nash equilibrium (BN-E) adoption model where a DM evaluates AI performance through HP, and decides which tasks are delegated to AI. We show that HP can generate both over- and under-adoption compared to an optimal benchmark. 
	We then test whether experimentally reducing HP at the population level can reduce the likelihood of over-adoption. 
	
	\subsection{Framework for Equilibrium AI Adoption}
	\label{sec:adoption-model}
	
	% =========================
	\sss{Setup}
	A DM manages a  domain $\mathcal{T}$ of $n$ tasks, varying in difficulty and indexed by $j\in\{1,\dots,n\}$. The DM delegates each task to either a human or an AI;  $x=(x_1,\dots,x_n)$ denotes the \emph{delegation profile} where $x_j\in\{A,H\}$. The human has \emph{known} ability $\theta^H$, which induces success rates $q_j^H=p^H(\theta^H,\delta^H(t_j))$. Thus the human data-generating process is known to the DM. By contrast, the AI's success rates $\mathbf{q}^A\in(0,1)^n$ are unknown.
	Observed \emph{outcomes} are summarized by $y=(y_1,\dots,y_n)\in\{0,1\}^n$, where $y_j=1$ denotes success on task $j$. Profit is additively separable: $\Pi(y)=\sum_{j=1}^n y_j\cdot R_j$, where $R_j>0$ is the reward for success on task $j$.\footnote{We abstract away from adoption costs for simplicity. Including them should only shift adoption thresholds without qualitatively changing the  results.}

	% =========================
	\sss{Human Projection}
	The DM's subjective model follows HP as defined in Section~\ref{sec:framework}. For tractability, we focus on the case of \emph{full projection}: the DM evaluates AI using exactly the same success-rate function and difficulty mapping as for humans, so that $p^A=p^H=p$ and $\delta^A(t_j)=\delta^H(t_j)=\delta_j$. Without loss of generality,  tasks are indexed by difficulty so that $\delta_1<\delta_2<\dots<\delta_n$. The DM holds a prior $F^A$ over AI ability $\theta^A$, with full support on $\Theta$. Importantly, the true AI success-rate vector $\mathbf{q}^A$ need not correspond to any ability level in the DM's subjective (mental) model.
	
	% =========================

	\sss{Equilibrium}
	A delegation profile $x$ induces an \emph{objective} distribution $Q(\cdot\mid x)$ over outcomes $y\in\{0,1\}^n$, where a success on task $j$ ($y_j=1$) occurs  with probability $q_j^{x_j}$, independently across tasks. Under the DM's mental model, if task $j$ is assigned to the AI of believed ability $\theta$, success occurs with probability $p(\theta,\delta_j)$; if assigned to the human, success occurs with  probability $q_j^H$. Each ability level $\theta$ thus induces a \emph{subjective} distribution over outcomes $Q_\theta(\cdot\mid x)$. Given a delegation profile $x$ and a believed AI ability level $\theta$, the discrepancy between the objective and subjective  distributions is captured by the Kullback-Leibler (KL) divergence: \[
	K(x,\theta)=\E_{Q(\cdot\mid x)}\!\left[\ln\frac{Q(y\mid x)}{Q_\theta(y\mid x)}\right]
	= \sum_{j:x_j=A}\!\left[q_j^A\ln\frac{q_j^A}{p(\theta,\delta_j)}+(1-q_j^A)\ln\frac{1-q_j^A}{1-p(\theta,\delta_j)}\right].
	\]
	Following \cite{esponda2016berk}, a pure Berk-Nash equilibrium is a pair $(x^*,\theta^*)$ such that belief $\theta^*$ ``best explains'' (in the KL sense) the data generated by the delegation vector $x^*$, and $x^*$ is optimal given belief $\theta^*$.
	\begin{definition} A pure Berk-Nash equilibrium (BN-E) is a pair $(x^*,\theta^*)$ such that:
		\begin{enumerate}
			\item $x^*\in\argmax_x\,\mathbb{E}_{Q_{\theta^*}(\cdot \mid x)}[\Pi(y)]$: delegation maximizes subjective expected profit given $\theta^*$.
			\item $\theta^*\in\argmin_\theta\, K(x^*,\theta)$: perceived ability minimizes KL divergence given delegation $x^*$. \end{enumerate} \end{definition}
	As a benchmark, the task-by-task optimal allocation assigns task $j$ to the AI whenever $q_j^A>q_j^H$.  This may yield no adoption, partial adoption, or full adoption depending on how $\mathbf{q}^A$ compares to $\mathbf{q}^H$. In particular, partial adoption is optimal whenever AI is better than humans on some but not all tasks---consistent with the evidence in Section~\ref{sec:math}. The theorem below characterizes which delegation profiles can arise in a BN-E, and how BN-E adoption deviates from this benchmark.
	For this result, we impose two regularity conditions. 
	First, ties in delegation are broken in favor of the human, ensuring a unique delegation choice whenever the DM is indifferent. Second, $p(\theta,\delta)$ and $1-p(\theta,\delta)$ are log-concave in $\theta$ for each $\delta$, with one of them being strictly log-concave; this ensures the KL divergence typically has a unique minimizer $\theta^*$.\footnote{This condition is satisfied by standard IRT specifications, e.g., logistic and normal-ogive functions.}
	% =========================
	\begin{theorem}[All-or-Nothing]\label{thm:BN-E}
		Let $\mathcal{H}$ be the hyperplane passing through $\mathbf{q}^H$ defined by
		\[
		\mathcal{H}=\left\{\mathbf{q}\in[0,1]^n:\sum_{j=1}^n w_j\,(q_j-q_j^H)=0\right\}, \qquad w_j=\frac{\partial p(\theta^H,\delta_j)/\partial\theta}{q_j^H(1-q_j^H)}.
		\]
		Under full HP:
		\vspace{-.3cm}
		\begin{enumerate}
			\item \textbf{Partial adoption:} No BN-E features partial adoption, even when it is optimal.
			\vspace{-.3cm}
			\item \textbf{Full adoption:} A BN-E with full adoption exists if and only if $\mathbf{q}^A$ lies strictly above $\mathcal{H}$.
			\vspace{-.3cm}
			\item \textbf{No adoption:} A BN-E with no adoption always exists; when $\mathbf{q}^A$ lies below $\mathcal{H}$, every BN-E features no adoption.
		\end{enumerate}
	\end{theorem}

	% =========================
	\sss{Intuition}
	The all-or-nothing result arises because the DM infers a single latent ability parameter from observed AI performance and then uses it to evaluate AI on every task through the success function $p(\cdot,\delta_j)$. Under partial adoption, the KL-minimizing belief $\theta^*$  lies either above or below the human benchmark $\theta^H$. Under HP, this comparison determines the ranking of AI relative to the human on \emph{all} tasks: if $\theta^*>\theta^H$, the DM concludes that AI  outperforms the human everywhere and switches to full adoption; if $\theta^*\leq\theta^H$, the DM concludes that AI (weakly) underperforms and switches to no adoption. 
	Partial adoption therefore cannot be sustained: the DM is pushed toward the coarse conclusion that AI is either globally better or globally  worse, even when the true technology $\mathbf{q}^A$ is better on some tasks and worse on others (which implies partial adoption is  task-by-task optimal).
	
\begin{figure}[t]
		\centering
		\caption{KL Minimization and Adoption Equilibrium Regions Under Human Projection}
		\label{fig:adoption-regions}
		
		\begin{minipage}[t]{0.47\textwidth}
			\centering
			\resizebox{\linewidth}{!}{
				\includegraphics[]{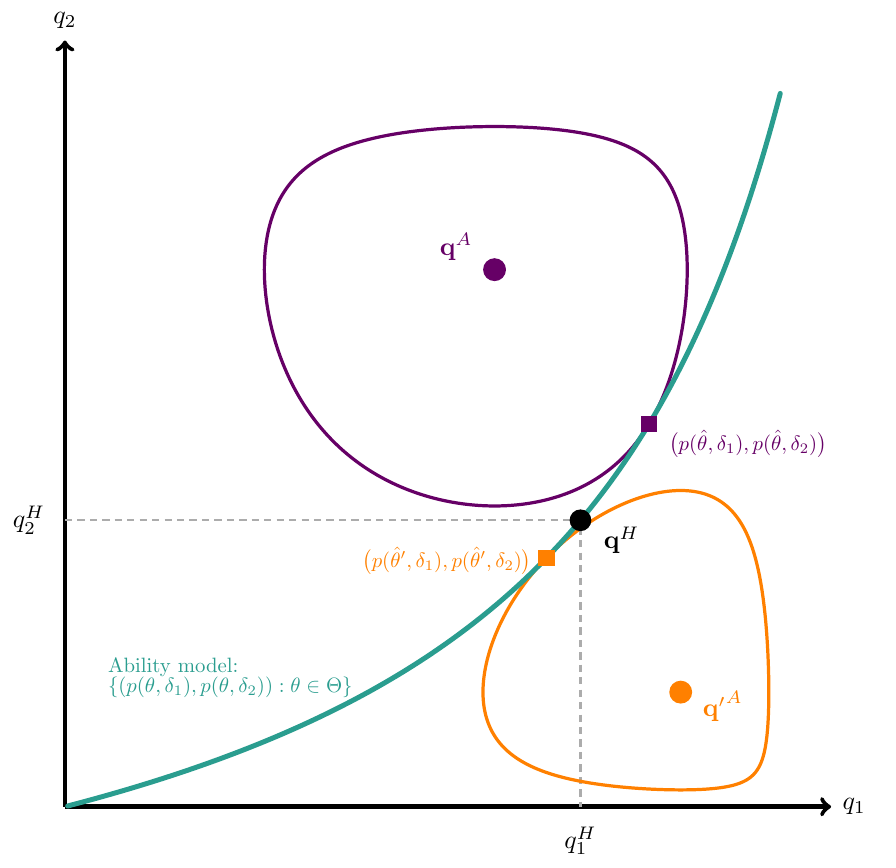}
			}
			
			\vspace{0.1cm}
			{\small \textbf{(a) KL minimization under HP}}
		\end{minipage}
		\hfill
		\begin{minipage}[t]{0.47\textwidth}
			\centering
			\resizebox{\linewidth}{!}{
				\includegraphics[]{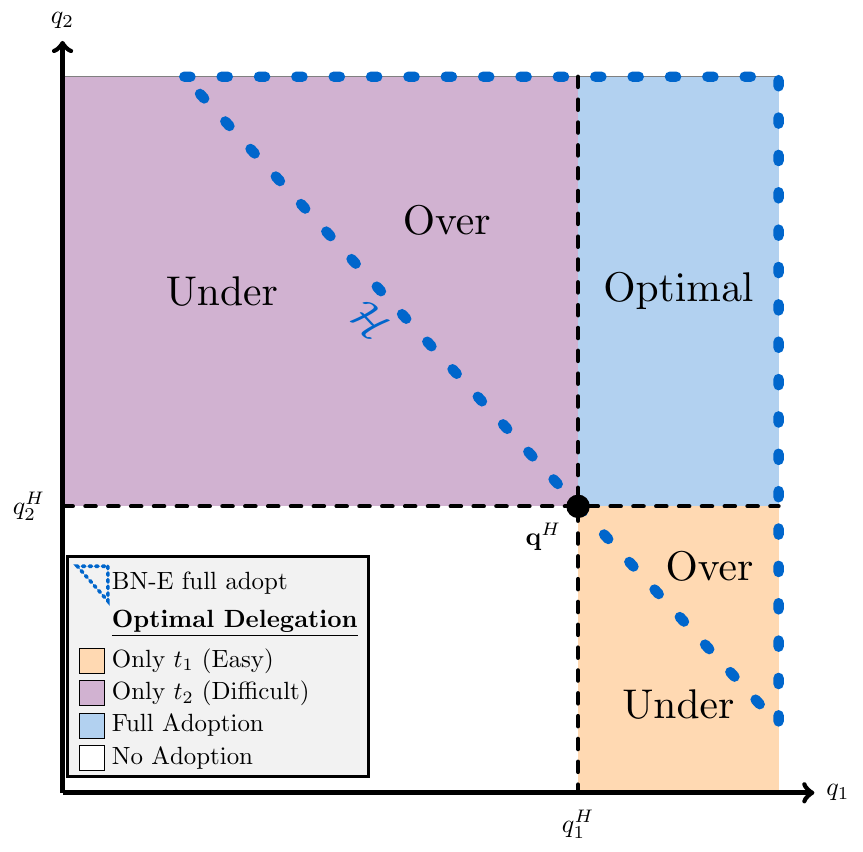}
			}
			
			\vspace{0.1cm}
			{\small \textbf{(b) Adoption regions}}
		\end{minipage}
		
		\vspace{-0.15cm}
		\floatfoot{\textit{Notes:}
			Panel (a) illustrates KL minimization under Human Projection. The green curve plots HP-consistent success-rate pairs, $(p(\theta,\delta_1),p(\theta,\delta_2))$, in the $(q_1,q_2)$ plane. Circles denote true AI success-rate pairs, and squares denote the projected success rate  implied by the KL-minimizing $\hat \theta$. Contours are level sets of the corresponding KL objective. $q^A$ (purple) projects to a belief that induces full adoption; $q'^A$ (orange) projects to a belief that induces no adoption. Panel (b) traces out the equilibrium vs.\ optimal adoption across the success-rate plane.  The dashed blue triangle marks the set of pairs for which full adoption is sustained as a Berk-Nash equilibrium; outside this region, only no adoption is sustained. The colored regions show the task-by-task optimal  delegation decision relative to the human benchmark $\mathbf q^H$. Comparing the equilibrium outcome with the optimal delegation regions identifies over-adoption and under-adoption.}
	\end{figure}

	The hyperplane $\mathcal{H}$ in Theorem~\ref{thm:BN-E} partitions the technology space $[0,1]^n$ into two regions: above it, a BN-E with full adoption exists; below it, every BN-E involves no adoption.\footnote{No adoption generates no data on AI performance and therefore a constant KL divergence. Hence, no adoption together with a sufficiently low $\theta$ (which trivially minimizes the KL divergence) always constitutes a BN-E, even above $\mathcal{H}$ where full adoption is also sustainable. However, any refinement that selects an equilibrium involving exposure to data instead selects full adoption there. For example, in that region, full adoption is the unique \emph{uniform BN-E} \citep{fudenberg2021limit}.} Figure~\ref{fig:adoption-regions} illustrates this in the two-task case: Panel (a) shows how true AI success-rate pairs are mapped to HP-consistent beliefs by KL minimization. Panel (b) traces this logic across the success-rate plane: $\mathcal{H}$ is a downward-sloping line through $\mathbf{q}^H$, and HP generates both over-adoption (full adoption despite AI being
	inferior on some tasks), and under-adoption (no adoption despite AI being superior on some tasks), relative to the task-by-task optimum.
	
	As $n$ grows, the result also becomes more realistic in practice: tracking task-specific AI success rates becomes increasingly demanding, making it natural to heuristically rely on an index that summarizes performance across tasks in the domain.
	
	Finally, in Appendix~\ref{app:adoption-model-dyn} we consider how HP distorts adoption along a path of continuous AI improvement. Along this path, the benchmark task-by-task optimal policy evolves gradually from no adoption, to partial adoption, and eventually to full adoption. Under HP, this intermediate phase breaks down. Initial adoption is delayed relative to the optimal benchmark, but once AI improves enough to cross $\mathcal{H}$, the equilibrium jumps directly to full adoption.
	As a result, HP generates under-adoption early in the technology path and over-adoption later on, with full adoption sustained even while humans still outperform AI on some tasks.

	\sss{Experimental prediction}
	Theorem~\ref{thm:BN-E} provides a benchmark under \emph{full} HP, which identifies a qualitative force at play: HP generates a tendency toward all-or-nothing adoption, which should weaken as projection weakens. If the DM instead perceives multiple AI-relevant subdomains (rather than a single-index ability), or correctly relies on AI-specific task difficulty, then partial adoption can in principle be sustained in equilibrium. 
	
	This suggests a natural empirical question: can visible features of an AI system affect the extent to which people engage in Human Projection, and thereby the prevalence of all-or-nothing adoption?
	We focus on one particularly salient and widespread  design feature: anthropomorphic cues. Many consumer-facing AI systems are presented with human-like cues (e.g., dialogue interfaces, names, voices, etc). A natural hypothesis is that these cues increase reliance on HP when reasoning about AI performance, consistent with evidence from experimental psychology \citep{chugunova2022we}. 
	\begin{assumption}\label{ass:anthro}
		Anthropomorphic presentation of AI increases the likelihood that decision-makers engage in Human Projection.
	\end{assumption}
	
	\noindent
	In our setting, we expect a non-anthropomorphic (``black box'') presentation  to weaken HP through either ability model projection, difficulty projection, or both.
	Combined with Theorem~\ref{thm:BN-E}, Assumption~\ref{ass:anthro} yields our main experimental prediction:
	\begin{prediction}\label{pred:adopt}
		A non-anthropomorphic presentation of AI decreases the share of all-or-nothing adoption decisions by reducing Human Projection.
	\end{prediction}

	% --------------- ADOPTION EXP ----------------- %
	
	\subsection{Adoption Experiment}
	\label{sec:adoption-exp}
	
	% --------------- DESIGN ----------------- %
	\subsubsection{Design}
	\label{sec:adoption-exp-design}
	Our experiment uses the benchmark of tasks described in Section~\ref{sec:math}, in a context where AI outperforms humans in only some tasks (the human-harder ones).
	Participants are introduced to the tasks and AI, then go through a delegation phase where they acquire costly and endogenous signals about human and AI performance, and make a one-shot high-stakes adoption decision at the end, testing Prediction~\ref{pred:adopt}.
	Our experimental variation shifts the prevalence of Human Projection at the treatment level during instructions, by manipulating anthropomorphic cues in AI framing.
	
	\sss{Tasks}
	We introduce the tasks as belonging to two pools, ``blue'' and ``green'', and familiarize participants with them during instructions: we show 6 examples (held fixed across treatments) from each pool, with human success rates and AI performance (a success) for one of these problems. 
	Problems in the blue pool are human-easy (mostly 4\textsuperscript{th}- and 8\textsuperscript{th}-grade; 78\% human success), while problems in the green pool are human-hard (mostly high-school; 23\% human success).\footnote{The blue pool is relatively human-easy while the green one is relatively human-hard. In the baseline version of the experiment, we never explicitly refer to the difficulty of each pool, and find evidence of significant misperceptions among participants. We run a replication with an identical design where we reduce misperceptions  by explicitly describing blue tasks as ``mostly of 8th-grade level'' and green tasks as ``mostly of high-school level.'' See Section \ref{sec:adoption-exp-results} for details.}
	AI success is held constant at 66\% across both pools, so humans outperform the AI on blue tasks and the AI outperforms humans on green tasks. 
	This design specifically tests for the treatment effect on the share of Full Adoption and makes Only Hard the optimal adoption choice given our incentive scheme.

	\sss{Treatment variation}
	Our experimental variation shifts the prevalence of Human Projection at the treatment level by varying the human-likeness of AI's framing.
	In the \emph{Anthropomorphic} condition, we use a standard consumer-facing AI framing: we present the AI as ``Morgan,'' a neutral AI based on ``currently available LLM technology,'' describe its behavior using active, agent-like language, and depict it with a human-like logo. This framing resembles interfaces used by leading LLM companies (e.g., Anthropic's \emph{Claude} or OpenAI's \emph{ChatGPT}).
	In the \emph{Black Box} condition, we present the AI as a ``black box'': we use the passive voice when describing its behavior and omit any reference to a human being.

	We instruct participants to use information from the delegation phase to inform their final adoption decision. Using language inspired by \citet{esponda2024mental}, we induce participants to be more agnostic about the black box's performance: ``100 green problems were put into the black box. A certain number (which you won't be told) came out as successes, and the rest as failures'' (\emph{Black Box}), compared to: ``Morgan attempted 100 green problems. It successfully solved a certain number (which you won't be told), and failed the remaining ones'' (\emph{Anthropomorphic}).

	\sss{Delegation phase and final adoption}
	The delegation phase consists of 60 problems (30 from each pool) shown in random order. On each problem, the participant chooses whether to delegate to a random human or to the AI, and then observes whether their chosen agent succeeds or fails on the next screen.\footnote{Performance corresponds to ChatGPT 3.5's benchmark in Section~\ref{sec:math}. For humans, we draw outcomes from a Bernoulli distribution with the empirical success rate for that pool.} Delegation is incentivized via a small bonus for each success observed. 
	After delegating all 60 problems, participants make a one-shot adoption decision with larger incentives: they choose whether to adopt a human or the AI to solve 10 tasks from each pool. This yields four adoption patterns: No Adoption, Only Easy, Only Hard, and Full Adoption.
	
	We elicit probabilistic beliefs in human and AI performance on each pool at three points: before the first delegation (initial beliefs), after 30 delegations (interim), and after the final adoption decision (final), using the following language: ``What do you think is the success rate of [Morgan / the black box] on the [blue / green] problems?''.

	\sss{Main sample}
	We programmed the experiment in Qualtrics and recruited participants on Prolific in August 2024. The pre-registration specifies the design, sample sizes, and main hypotheses. We collected two samples. The baseline sample includes 149 participants in \emph{Anthropomorphic} and 157 in \emph{Black Box}. 
	
	\sss{Replication sample} As discussed below, the baseline sample delivers the main pre-registered reduction in all-or-nothing adoption, but it also exhibits unanticipated reallocation of choices toward non-optimal partial adoption, consistent with misperceptions of the relative \emph{human} difficulty of the two pools, for which we provide evidence below. We therefore collected an additional replication sample (57 and 59 participants, respectively) with an identical design and interface except that human difficulty is made more salient: blue problems are truthfully labeled ``grade school,'' and green problems ``high-school.'' We report results from both samples in the main text (Section~\ref{sec:adoption-exp-results}) and present evidence on these misperceptions in Appendix~\ref{app:misperceptions}.
	%2.5 bonus average, only among those passing checks
	
	\sss{Incentives}
	On average, participants earned \$4.2 for a median completion time of around 15 minutes. The base fee was \$1.7 with a potential bonus of \$4.4 (\$0.03 per delegation-phase success observed, \$0.05 per accurate belief report, and up to \$2 in the adoption phase).
	
	% --------------- RESULTS ----------------- %
	\subsubsection{Results}
	\label{sec:adoption-exp-results}

\begin{figure}[t]
		\centering
		\caption{Treatment Effect on Adoption Decisions}
		\label{fig:adoption-main}
		\begin{subfigure}{.49\textwidth}
			\caption{Baseline Sample}
			\centering
			\includegraphics[width=\textwidth]{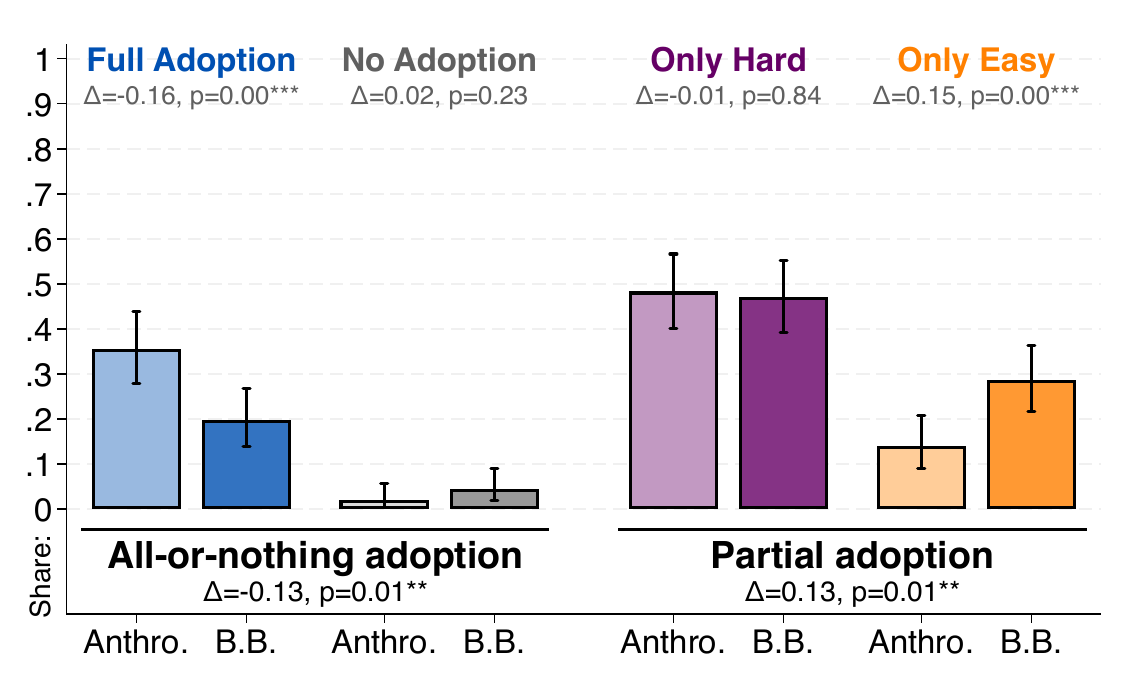}
		\end{subfigure}
		\begin{subfigure}{.49\textwidth}
			\caption{Replication Sample}
			\centering
			\includegraphics[width=\textwidth]{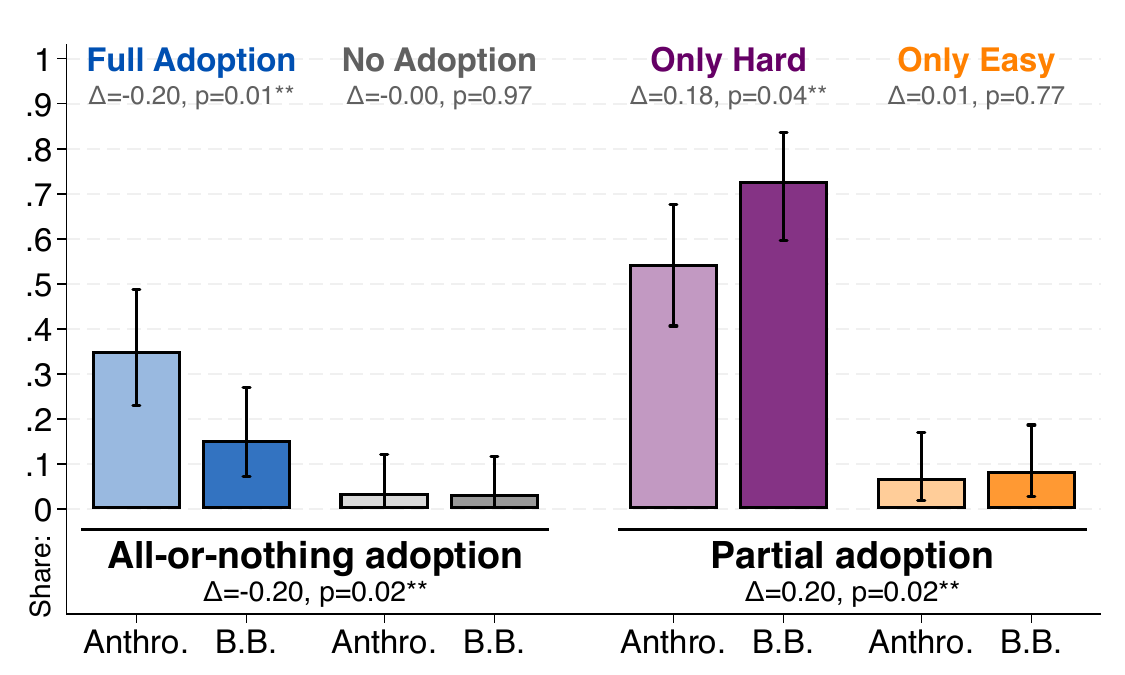}
		\end{subfigure}
		\vspace{-.2cm}
		\floatfoot{\textit{Notes:} The figure reports shares (0--1) of participants choosing each final adoption choice, across treatments \emph{Anthropomorphic} and \emph{Black Box}. Sample sizes are respectively $n=149$ and $n=157$ (baseline) and $n=57$ and $n=59$ (replication). All-or-nothing adoption corresponds to either Full or No adoption, and Partial adoption to either Only Hard or Only Easy. p-values from two-sided tests of proportions across treatments are reported. }
	\end{figure}

	\sss{Main effect: all-or-nothing adoption}
	Figure \ref{fig:adoption-main} reports the shares of participants selecting each final adoption choice, among the baseline (left) and replication samples (right). Consistent with Prediction~\ref{pred:adopt} and our pre-registration, we find that all-or-nothing adoption is significantly lower under \emph{Black Box} than under \emph{Anthropomorphic} in both samples (baseline: $\Delta=-0.13$, $p=0.01$; replication: $\Delta=-0.20$, $p=0.02$), driven by a reduction in Full Adoption (baseline: $\Delta=-0.16$, $p<0.01$; replication: $\Delta=-0.20$, $p=0.01$).
	This shift is offset by an increase in partial adoption. In the baseline sample, the increase is concentrated in Only Easy rather than Only Hard (the task-by-task optimal choice given realized AI performance across pools); this unanticipated pattern motivated the replication and the analyses below.

	\sss{Perceptions of human difficulty and optimal adoption}
	We argue that the increase in Only Easy is driven by misperceptions of \emph{human} task difficulty. We classify participants as misperceiving difficulty if they report initial beliefs---elicited pre-delegation, so before receiving any endogenous feedback---about \emph{human} performance that are weakly higher on the human-hard pool than the human-easy pool. In the baseline sample, a large share (56\%) of participants misperceive difficulty in this sense. Under such a misperception, delegating the human-easy pool to AI is optimal. The top panel of Table~\ref{tab:belief-opt-misp} shows that \emph{Black Box} induces partial adoption consistent with participants' (mis-)perceptions:  we find no effect on normatively optimal adoption (Only Hard, column 2) in the baseline sample, consistent with the large share of misperceptions, but large increases in belief-consistent partial adoption, defined as delegating the \emph{perceived} human-hard pool to  AI (columns 3--5). In other words, \emph{Black Box} leads participants to delegate the pool they believe is human-hard to  AI, and the pool that they believe is human-easy  to humans.
	
	To further test this interpretation, we collected a replication sample where the human difficulty of pools is made more salient within an otherwise identical design. The rate of misperception in this sample falls sharply by 39 pp ($p<0.001$; Table~\ref{tab:misperceptions}). In this sample, we find a similar decrease in all-or-nothing adoption, now matched by strong increases in both optimal and belief-consistent partial adoption (right panel of Figure \ref{fig:adoption-main} and bottom panel of Table \ref{tab:belief-opt-misp}). Appendix~\ref{app:misperceptions} provides further  analyses of the proxy and belief-conditional effects on adoption. We show that the treatment does not affect misperception in either wave, and that misperception  predicts \emph{which} type of partial adoption participants select but not \emph{whether} they partially adopt. 
	While initial beliefs are our preferred proxy throughout (as they are uncontaminated by endogenous performance signals),  our results are robust to using interim or final beliefs as proxies (columns 4--5 in Table \ref{tab:belief-opt-misp} and Appendix \ref{app:misperceptions}).
	% HP and lead users to 
	%holding accurate perceptions of human task difficulty are distinct conditions: the former is achieved by , but the latter is not. 
	%Welfare-relevant adoption improvements will likely require both — and since difficulty perceptions appear to vary systematically with user performance (see Section \ref{sec:beliefs-results}), the second condition may not be uniformly met in practice.

	\begin{table}[t]
		\caption{Treatment Effect on Adoption Decisions}
		\label{tab:belief-opt-misp}
		\centering
		\resizebox{\textwidth}{!}{
			\begin{threeparttable}
				{\setlength{\tabcolsep}{3pt}
\renewcommand{\arraystretch}{.9}
\begin{tabular}{l@{\hspace{17pt}} l@{\hspace{0pt}} c@{\hspace{8pt}} c @{\hspace{13pt}}cccc}
\toprule
 & & \multirow[c]{3}{2.6cm}{\centering\emph{All-or-nothing adoption}} & \multirow[c]{3}{1.5cm}{\centering\emph{Optimal adoption}} & \multicolumn{3}{c}{\emph{Belief-consistent partial adoption}} \\
\cmidrule(lr){5-7}
 & & & & \hspace{10pt}Initial\hspace{10pt} & \hspace{10pt}Interim\hspace{10pt} & Final\hspace{10pt} \\ [.2em]
 \multicolumn{1}{l}{Sample} & Statistic  & (1) & (2) & (3) & (4) & (5) \\
\midrule
\multirow{4}{2cm}{\centering\textbf{Baseline} \\ \textbf{sample}} & Black Box & -14.6*** & -2.3 & \phantom{-}17.5*** & \phantom{-}19.9*** & \phantom{-}16.1*** \\
 & & (5.2) & (5.6) & (5.6) & (5.6) & (5.7) \\ [.5em]
 & R-squared & 0.048 & 0.065 & 0.040 & 0.069 & 0.050 \\
 & Observations   & 306  & 306  & 306  & 306  & 306 \\
\midrule
\multirow{4}{2cm}{\centering\textbf{Replication} \\ \textbf{sample}} & Black Box & -19.8**\phantom{*} & \phantom{-}19.8**\phantom{*} & \phantom{-}16.6*\phantom{**} & \phantom{-}24.2*** & \phantom{-}26.8*** \\
 & & (8.4) & (9.0) & (9.4) & (9.2) & (9.0) \\ [.5em]
 & R-squared & 0.067 & 0.060 & 0.043 & 0.088 & 0.077 \\
 & Observations   & 116  & 116  & 116  & 116  & 116 \\
\bottomrule
\end{tabular}}

				\smallskip\footnotesize
				\begin{tablenotes}
					\item \textit{Notes:} The table reports OLS estimates of the effect of the Black Box treatment on adoption decisions, estimated separately by outcome and sample. Column (1) reports effects on all-or-nothing adoption (full or no adoption). Column (2) reports effects on normatively optimal partial adoption (Only Hard). Columns (3)--(5) report effects on belief-consistent partial adoption, defined as Only Hard for participants whose beliefs in human performance are higher on the human-easy pool, and Only Easy otherwise. Beliefs are elicited at the initial, interim, and final phases. All specifications include participant controls (age, gender, education, race, familiarity with AI). Robust standard errors in parentheses. * $p<0.1$, ** $p<0.05$, *** $p<0.01$.
				\end{tablenotes}
		\end{threeparttable}}
	\end{table}

	\begin{table}[t]
		\caption{Adoption Experiment -- Projection Mechanism}
		\label{tab:adoption_mechanism}
		\centering
		\resizebox{\textwidth}{!}{
			\begin{threeparttable}
				{\setlength{\tabcolsep}{4pt}
\renewcommand{\arraystretch}{0.85}
\begin{tabular}{l @{\hspace{35pt}} l @{\hspace{8pt}} c c c @{\hspace{2pt}} c}
\toprule
 &  & \multicolumn{2}{c}{\textbf{Treatment}} &  &  \\
\cmidrule(lr){3-4} 
Measure & Beliefs & \textit{Anthrop.} & \textit{Black Box} & \textbf{Difference} & \textit{p} \\
\midrule 
\multicolumn{1}{l}{\textbf{Ability-Model Projection}}  & & & & &  \\
\addlinespace[1pt]
\multirow{3}{*}{\shortstack[l]{\textit{Cross-task belief}\\\textit{correlation:} \(\rho(p^{AI}_{easy},p^{AI}_{hard}) \) }} & Initial &  0.69 &  0.31 & \textbf{-0.38} & <0.001 \\
 & Interim &  0.38 &  0.18 & \textbf{-0.21} & 0.021 \\
 & Final &  0.41 &  0.18 & \textbf{-0.23} & 0.011 \\
\midrule
\multicolumn{1}{l}{\textbf{Difficulty Projection}} & & & & & \\
\addlinespace[1pt]
\multirow{3}{*}{\shortstack[l]{\textit{Human-AI difficulty}\\\textit{correlation:} \(\rho(\Delta^{AI},\Delta^H) \)}} & Initial &  0.53 &  0.48 & \textbf{-0.05} & 0.483 \\
 & Interim &  0.15 &  0.22 & \textbf{ 0.08} & 0.409 \\
 & Final &  0.08 & -0.03 & \textbf{-0.11} & 0.245 \\
\midrule
\textit{Observations} &  & 206 & 216 &  &  \\
\bottomrule
\end{tabular}}

				\smallskip\footnotesize
				\begin{tablenotes}[flushleft]
					\footnotesize
					\item \textit{Notes:} The table reports belief-based projection proxies at three elicitation phases: Initial (pre-delegation), Interim (after 30 problems), and Final (after final adoption). \textit{Cross-task belief correlation} is the Spearman correlation between AI beliefs across the easy and hard task pools. \textit{Human--AI difficulty correlation} is the Spearman correlation between $\Delta^A$ and $\Delta^H$ across participants, where $\Delta^i$ is the difference in beliefs about agent $i$ on the easy pool minus the hard pool. Higher values are consistent with more projection for both measures. $p$-values are from Fisher $z$-tests for the difference in correlations between treatments.
				\end{tablenotes}
		\end{threeparttable}}
	\end{table}
	
	\sss{Mechanism: weakening of Human Projection}
	Table~\ref{tab:adoption_mechanism} reports belief-based proxies for ability model projection (the extent to which participants summarize AI performance with a single latent ability index) and difficulty projection (the extent to which they map human difficulty differences onto AI). The \emph{Black Box} framing consistently reduces ability model projection across all phases, with cross-task belief correlation falling from $0.69$ to $0.31$, $p<0.001$ and remaining significantly lower at interim and final elicitation phases. Effects on difficulty projection are smaller and not statistically distinguishable from zero at any phase. Taken together, these results suggest that removing anthropomorphic framing primarily disrupts the single-index inferential structure underlying ability model projection, while the tendency to anchor on human difficulty appears more robust to framing interventions.
	When perceptions of human difficulty are accurate, lower reliance on the ability model leads to higher shares of optimal adoption.

	\sss{Delegation behavior}
	Finally, Appendix Table~\ref{tab:delegation-signals} shows that \emph{Black Box} treatment 
	reduces delegation to the AI during training, consistent with \emph{Black Box} framing inducing lower initial beliefs. Conditional on delegating, however, observed AI performance is similar across framings, suggesting the treatment mainly shifts delegation propensity rather than the feedback received.
	
	%training phase behavior: BB reveal fewer easy problems with AI, about same for hard (14 vs. 10, and 22.5 vs. 21). Share of mistakes is the same: 38 for ai eays, 34 for ai hard, 

	\sss{Takeaway}
	Taken together, the results are consistent with the \emph{Black Box} framing weakening Human Projection. While we do not claim the effect is driven entirely by HP (framings typically shift several things at once), all patterns we would expect under reduced HP do arise: lower all-or-nothing adoption, weaker single-index inference (the crucial force behind Theorem~\ref{thm:BN-E}), and partial adoption that tracks participants' own difficulty perceptions. When perceptions of human difficulty are accurate, weaker HP translates into higher shares of optimal adoption.

	%------------------------------------------------%
	%                     PARENTDATA                 %
	%------------------------------------------------%
	
	\section{Consequences of HP for Engagement with AI Chatbots}
	\label{sec:parentdata}

	We examine a potential implication of HP for real-world \emph{engagement} 
	with an AI chatbot providing parenting advice, hosted on 
	ParentData.org---a website providing information related to pregnancy and child-rearing. The chatbot works by trying to match the question asked with one of about 8,000 pre-written (human-vetted) 
	answers.\footnote{When a question does not match any of the possible answers, the chatbot responds by saying ``It looks like we don't have an answer in the ParentData archives.''} It can fail by misinterpreting a question and returning an answer that, while useful for a \emph{different} question, contains no useful information for the question  posed.
	
	Here the relevant human feature is the \emph{reasonableness} of a wrong answer---its similarity to what a useful human response would have been. Conditional on a human misunderstanding a question, the reasonableness of their mistake matters: a particularly unreasonable misunderstanding is a stronger signal of low 
	competence, which further reduces our willingness to consult them again. Under HP, users apply the same logic to the chatbot, treating human 
	reasonableness of its errors as informative of overall performance---even when it is not. Consider, for example, a real conversation from the experiment: a user asks ``Which is the best car seat brand?'' One 
	failure answers about \emph{where} to install a car seat (front vs.\ back seat); another discusses the best baby food brand. The former is typically judged more reasonable---it shares the relevant context---but both reflect the same underlying system behavior: returning the highest-scoring pre-written match to the user's prompt. Absent projection, conditional on observing a failure, human reasonableness need not affect beliefs about AI performance.\footnote{No alternative, lower-score answer should be interpreted as ``more reasonable'' than the top-score answer returned by the system.}

	%In Appendix~\ref{app:HP_reason} we formalize this logic. Mirroring Assumption~\ref{assn:human-model}, we define reasonableness as the similarity of a wrong answer to the nearest useful answer, and impose amonotone likelihood ratio condition under which higher-ability agents produce more reasonable answers in expectation. This yields an analog of Prediction~\ref{pred:cross}, which we test in the following experiment:
	In Appendix~\ref{app:HP_reason} we formalize this logic, yielding an analog of Prediction~\ref{pred:cross}:
	\begin{prediction}\label{pred:parentdata}
		When a user observes an AI failure, if the answer is less reasonable 
		from a human perspective: (i) beliefs in AI performance decrease more; 
		(ii) trust in AI decreases more; (iii) likelihood of subsequent engagement 
		is lower.
	\end{prediction}

	%---------------- EMPIRICS -----------------%
	\subsection{Engagement Experiment}
	\label{sec:parentdata-empirics}

	We test Prediction \ref{pred:parentdata} in a framed field experiment \citep{harrison2004field} with a sample designed to mirror ParentData's  user base. We recruit online participants who match the platform's core user demographics and show them a sequence of real parent-chatbot conversations in an ecological survey interface, varying between participants  the \emph{reasonableness} of the chatbot's misunderstandings. We then elicit beliefs in performance, trust, and measure engagement both through a revealed-preference link choice and actual subsequent usage.

	%----------- PROCESS FOR PAIRS ------------%
	\subsubsection{Design and Procedures}
	\label{ref:parentdata-empirics-pairs}
	
	To isolate the role of human reasonableness, we constructed matched \emph{pairs} of real ParentData conversations in which the chatbot misunderstands the user's question. Using a pre-registered procedure, we formed candidate pairs of same-intent conversations and, in two separate rating studies with the target population (current and expectant parents), elicited perceived reasonableness and answer usefulness. We retained five pairs for which the two sides are same-intent, equally unhelpful, and sharply different in reasonableness. Appendix~\ref{appendix:parentdata} provides the full procedure, instructions, and selected conversations with their ratings.
	
	The experiment is between-subjects. After instructions, participants view five real human-chatbot conversations. The first three are ``successes'' (median usefulness 4--5) and are identical across treatments; the last two are ``failures'' (median usefulness 1--2, held fixed within each pair), drawn from the matched pairs above. Treatment assignment determines which side is shown: participants in \emph{Reasonable} see the more reasonable misunderstanding, those in \emph{Unreasonable} see the less reasonable one. Each conversation page initially displays the user query with a \textit{Generate} button; clicking it reveals the AI answer via a typewriter animation mimicking the actual chatbot. After each conversation, we elicit beliefs about AI performance and trust.\footnote{For beliefs: ``What is the \% chance the chatbot answers a random parenting question correctly?''; 0--100 scale. For trust: ``How much do you trust the chatbot?''; 1--7 scale.} Beliefs are incentivized by paying for accuracy against an external benchmark: we draw 100 conversations at random and elicit usefulness ratings from a distinct sample of potential users.
	
	Immediately after the last conversation, participants choose to receive a link either to the chatbot or to a ``large list (500+) of good-quality parenting articles.'' As an actual engagement measure, we track subsequent chatbot usage by matching experiment IP addresses to those appearing in ParentData chatbot logs.\footnote{For confidentiality, we do not observe the content of post-experiment questions. The link to the chatbot is \url{https://parentdata.org/ask-a-question/}, and the one to parenting articles is \url{https://parentdata.org/articles/}. The LLM powering the chatbot was trained on a corpus that includes the full list of parenting articles.}
	
	The procedures, target sample sizes, and main hypotheses were pre-registered. All participants were recruited on Prolific in July--August 2024, targeting current or expectant parents (85\% female; U.S.-based; age 18--45; in a relationship; screened in-survey for parental status). As pre-registered, participants who indicated familiarity with ParentData.org were dropped. Excluding those failing attention and comprehension checks, the final sample is $n=654$ (315 \emph{Reasonable}, 339 \emph{Unreasonable}). The survey lasted about 5 minutes, with mean earnings = \$0.87.

	% ---------------- RESULTS EXP ------------- %
	\subsubsection{Results}
	\label{ref:parentdata-empirics-results}

\begin{figure}[t]
		\centering
		\caption{Effect of Reasonableness of Chatbot Failures}
		\label{fig:parent-main}
		\begin{subfigure}{.49\textwidth}
			\centering
			\caption{Belief Measures}
			\includegraphics[width=\textwidth]{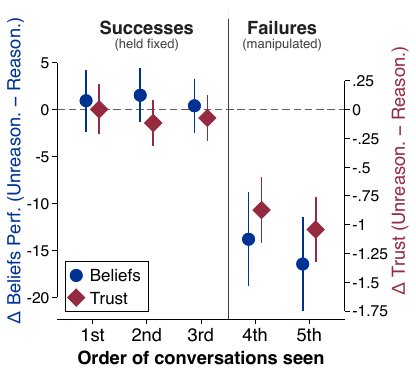}
		\end{subfigure}
		\begin{subfigure}{.49\textwidth}
			\centering
			\caption{Engagement Measures}
			\includegraphics[width=\textwidth]{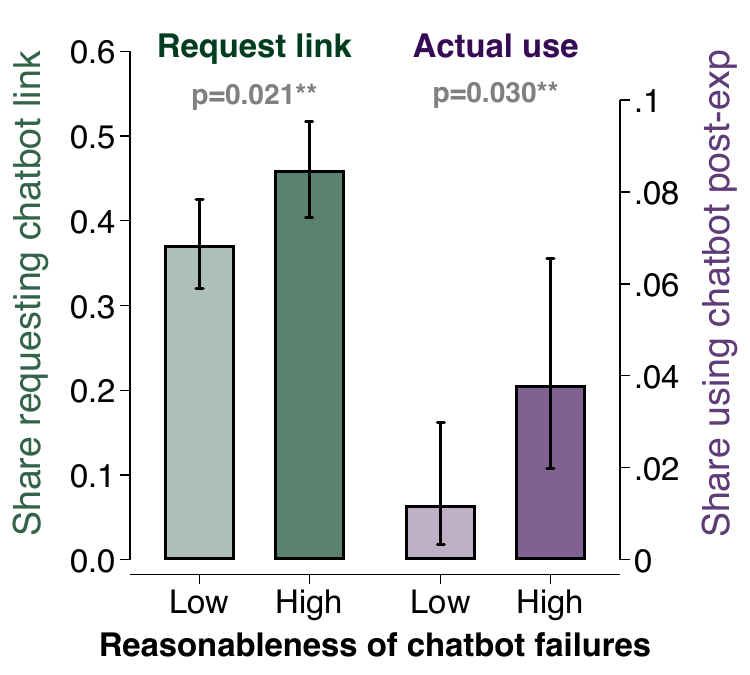}
		\end{subfigure}
		\floatfoot{\textit{Notes:} Panel (a) plots the difference between average beliefs (0--100\% scale) and trust (1--7 scale) in the chatbot across treatments (\emph{Unreasonable} minus \emph{Reasonable}). The first three points are elicited after successful conversations (median usefulness of 4 or 5 out of 5); the last two after failures (median usefulness of 1 or 2). Error bars: 95\% CIs. Panel (b) plots, for each treatment, the share of participants who requested a link to the chatbot (green bars) and the share whose IP address appears on the website at least once in the three weeks following the experiment (purple bars). $p$-values are from a two-sided test of proportions. Sample sizes are $n=315$ for \emph{Reasonable} and $n=339$ for \emph{Unreasonable}.}
	\end{figure}

	We first examine beliefs about performance and trust.
	Consistent with Prediction \ref{pred:parentdata}, Panel A in Figure \ref{fig:parent-main} shows that beliefs and trust decline after failures in both treatments, but the drop is around twice as large under \emph{Unreasonable} for both beliefs in performance and trust.
	%beliefs fall by about 18 percentage points more (33 pp vs.\ 15 pp) and trust by 1 point more (1.7 points vs.\ 0.7).
	Table \ref{tab:parent-main} shows that treatment effects are robust to adding demographic controls and fixed effects for the matched failure pairs shown in the 4th and 5th conversations.
	
	\begin{table}[t]
		\caption{Effect of Reasonableness -- OLS Estimates}
		\label{tab:parent-main}
		\centering
		\resizebox{\textwidth}{!}{
			\begin{threeparttable}
				\begin{tabular}{lcccccccc} \toprule
& \multicolumn{4}{c}{\textbf{Belief measures}} & \multicolumn{4}{c}{\textbf{Engagement measures}} \\
\cmidrule(lr){2-5} \cmidrule(lr){6-9}
& \multicolumn{2}{c}{\textbf{Belief in perf.}} & \multicolumn{2}{c}{\textbf{Trust}} & \multicolumn{2}{c}{\textbf{Request link}} & \multicolumn{2}{c}{\textbf{Actual use}} \\
\cmidrule(lr){2-3}\cmidrule(lr){4-5}\cmidrule(lr){6-7}\cmidrule(lr){8-9}
                    &\multicolumn{1}{c}{(1)}   &\multicolumn{1}{c}{(2)}   &\multicolumn{1}{c}{(3)}   &\multicolumn{1}{c}{(4)}   &\multicolumn{1}{c}{(5)}   &\multicolumn{1}{c}{(6)}   &\multicolumn{1}{c}{(7)}   &\multicolumn{1}{c}{(8)}   \\
\hline
Unreasonable        &     -15.208***&     -15.881***&      -0.964***&      -1.000***&      -0.092** &      -0.096** &      -0.026** &      -0.028** \\
                    &     (2.237)   &     (2.202)   &     (0.128)   &     (0.126)   &     (0.038)   &     (0.038)   &     (0.012)   &     (0.013)   \\
\midrule
Controls            &  \checkmark   &  \checkmark   &  \checkmark   &  \checkmark   &  \checkmark   &  \checkmark   &  \checkmark   &  \checkmark   \\
Pair FE             &               &  \checkmark   &               &  \checkmark   &               &  \checkmark   &               &  \checkmark   \\
R-squared           &       0.089   &       0.131   &       0.102   &       0.135   &       0.025   &       0.056   &       0.017   &       0.025   \\
Observations        &         654   &         654   &         654   &         654   &         654   &         654   &         654   &         654   \\
\bottomrule \end{tabular}

				\smallskip\footnotesize
				\begin{tablenotes}
					\item \textit{Notes:} This table reports OLS estimates of the effect of assignment to the \emph{Unreasonable} (low-reasonableness) failure condition. For beliefs and trust, the dependent variables are averages of the post-failure outcomes after the 4th and 5th conversations. Engagement outcomes are indicators for requesting the chatbot link and subsequent chatbot usage within three weeks. Controls include age, gender, education, race indicator, and prior familiarity with AI. Columns marked ``Pair FE'' include fixed effects for the matched failure pairs shown in the 4th and 5th conversations. Robust standard errors in parentheses. Significance levels: * $p<0.1$, ** $p<0.05$, *** $p<0.01$
				\end{tablenotes}
		\end{threeparttable}}
	\end{table}
	Engagement moves in the same direction. Figure~\ref{fig:parent-main}, Panel B, shows that low reasonableness decreases both the share of participants requesting the chatbot link by 8.9 percentage points ($p=0.021$) and the probability of subsequent usage (asking at least one question within three weeks) by 2.6 percentage points ($p=0.030$).\footnote{Even conditional on requesting the chatbot link, follow-through is significantly lower under \emph{Unreasonable} than under \emph{Reasonable}.} %(2.3\% vs.\ 6.5\%; $p=0.048$).} 
%Excluding participants who report prior familiarity with the website yields qualitatively similar patterns, with larger level gaps in engagement (link choice: 49\% vs.\ 39\%, $p=0.005$; usage: 3.4\% vs.\ 0.9\%, $p=0.010$).

Therefore, consistent with HP, the human 
reasonableness of chatbot errors strongly shapes inferences and 
engagement, even though---as a separate rating exercise confirms---the 
chatbot is generally useful (mean rating 3.5/5) and failures are 
rare (14\% of conversations have median usefulness $\leq 2$). The design does not rule out alternative accounts, but the pattern holds across beliefs, trust, and revealed engagement behavior, and follows the logic of the HP framework developed in earlier sections.

%------------------------------------------------%
%                     DISCUSSION                 %
%------------------------------------------------%

\section{Conclusion}
\label{sec:conclusion}

When evaluating an entity that communicates and behaves like a human, it is only natural to rely on familiar human performance models.\footnote{In principle, the underlying logic  extends beyond AI: people import inferential structures learned in familiar contexts to unfamiliar targets that appear similar enough to invite their use in other  settings. Grammar mistakes by a non-native speaker may be read as signs of poor education when ``basic'' is judged from a native speaker's perspective; visitors may read local behavior through the cultural frame they brought with them, drawing inferences that the local context need not warrant.} This paper formalizes and documents this tendency---\emph{Human Projection}---showing that people evaluate AI by relying on human-relevant features such as task difficulty and error reasonableness. Beliefs about AI exhibit clear anthropomorphic structure, in both priors and updating. HP-driven beliefs then shape adoption: in the lab, de-anthropomorphizing AI reduces all-or-nothing adoption when AI is better only on some tasks; in the field, less humanly reasonable chatbot failures reduce subsequent engagement with an otherwise useful tool.

Our results carry three practical implications. First, they provide a rationale for attending to features that shape user generalizations from observed performance---task difficulty and error reasonableness---during post-training procedures such as Reinforcement Learning with Human Feedback (RLHF). Second, they point to an inherent tradeoff in anthropomorphic design: anthropomorphic cues can increase trust and engagement \citep{chugunova2022we}, but may simultaneously amplify HP-driven distortions in beliefs about capability across tasks. Third, HP may distort the path of technological progress itself. While our analysis focuses on choices between AI and humans, the same logic applies to choices \emph{between} models---a setting increasingly relevant as AI capabilities grow. 
If users generalize from performance through the HP lens, demand---and subsequent development---may skew toward models that \emph{appear} capable under projection rather than those that \emph{are}. The same generalization risk applies to benchmarks: because they are constructed around human-relevant domains, extrapolating from benchmark scores should be done with caution, motivating richer approaches to capability evaluation \citep{zhou2025lost}.

How will HP evolve in the future? Jagged performance appears to be a stable feature of current models, unresolved by capability alone 
(Section~\ref{sec:math}). Whether \emph{beliefs} will adjust is less clear. 
Experience with AI could sharpen users' intuitions about where 
these systems are and are not reliable. But AI systems are also becoming 
increasingly anthropomorphic by design, making the human--machine boundary harder to perceive. These effects are also unlikely to be uniform, raising concerns about equity in who adopts AI and who benefits from its gains.  
The stakes extend well beyond performance evaluation: 
younger users turn to AI not only for school or work, but increasingly for emotional support and companionship,\footnote{A large share of adolescents already engage with AI  companions for emotionally sensitive conversations; see Common Sense Media, 
	\emph{Talk, Trust, and Trade-Offs} (2025). \url{https://www.commonsensemedia.org/sites/default/files/research/report/talk-trust-and-trade-offs_2025_web.pdf}.} where anthropomorphic cues can carry  risks of miscalibrated trust and dependency \citep{defreitas2025emotional}. 
We hope that continued research will improve our understanding of how experience and design jointly shape the trajectory of HP---and 
human perceptions of technology more broadly.

%------------------------------------------------%
%                    BIBLIOGRAPHY                %
%------------------------------------------------%

% References: single-spaced only here
\begingroup
\renewcommand{\baselinestretch}{1}\selectfont
\bibliographystyle{aer}
\bibliography{bib_ai_draft_latest}
\endgroup

%\end{document}
%------------------------------------------------%

%------------------------------------------------%
%                     APPENDIX                   %
%------------------------------------------------%
\clearpage

\appendix
% Number appendix tables/figures by appendix section: A.1, A.2, B.1, ...
\counterwithin{table}{section}
\counterwithin{figure}{section}

\renewcommand{\thetable}{\thesection.\arabic{table}}
\renewcommand{\thefigure}{\thesection.\arabic{figure}}

\begin{center}
	\LARGE{ONLINE APPENDIX \\ \emph{Human Learning about AI}}
\end{center}

This appendix presents formal details and proofs of the framework and experimental predictions (Section \ref{app:theory}); details on the construction of the domain of mathematical tasks and procedures for measuring human and AI performance (Section \ref{appendix:math}); separate sections for additional results and details about the beliefs experiment (Section \ref{appendix:beliefs}), adoption experiment (Section \ref{appendix:adoption}), and engagement experiment (Section \ref{appendix:parentdata}).
Further experimental instructions and materials are available at our \href{https://www.dropbox.com/scl/fi/a8gua4s23yur7dmcj946w/ai_draft_exp_materials.pdf?rlkey=cplykye5bo1vnekoidd45hpwj&dl=0}{Experimental Materials Appendix}.

%------------- PROOFS ---------------------%
\section{Theoretical Appendix}\label{app:theory}

\subsection{Microfoundation for Difficulty Projection}
Assume the DM does not perfectly observe $\delta^A(t)$ but receives a noisy signal instead,
\[
z_t = \delta^A(t) + \varepsilon,\qquad \varepsilon \sim \mathcal{N}\big(0,\sigma_s^2\big).
\]
Human difficulty is a natural anchor when the DM assesses AI difficulty, so we assume a normal prior centered around $\delta^H(t)$, i.e., $\delta_0^{A}(t) \sim \mathcal{N}\big(\delta^{H}(t),\sigma_{H}^2\big).$
By conjugacy, the posterior mean of AI difficulty is
\[
\tilde{\delta}^{A}(t)
= \lambda\,\delta^{H}(t) + (1-\lambda)\,z_t,
\qquad\text{where}\quad
\lambda = \frac{\sigma_{s}^2}{\sigma_{s}^2+\sigma_{H}^2}.
\]
Taking expectations with respect to the signal noise yields
\[
\mathbb{E}[\tilde{\delta}^{A}(t)\mid \delta^{H}(t),\delta^{A}(t)]
= \lambda\,\delta^{H}(t) + (1-\lambda)\,\delta^{A}(t).
\]
This is a particular functional form that satisfies main-text Assumption~2 under mild restrictions.\footnote{This holds, for example, if true AI difficulty has the same ordering as human difficulty, or if the anchoring weight \(\lambda\) is sufficiently large relative to any ordering reversals in \(\delta^A\).} Note that $\lambda$ is increasing in the precision of the prior, $1/\sigma_{H}^2$, and decreasing in the precision of the signal, $1/\sigma_{s}^2$.

\subsection{Proofs}
\begin{proof}[Proof of Proposition 1] Fix any prior $F^g$. By part 2 of main-text Assumption 1, $p^g(\theta,\delta)$ is strictly decreasing in $\delta$ for all $\theta$.
	
	For $g=H$: $\delta^H(t)>\delta^H(t')$ implies $p^H(\theta,\delta^H(t))< p^H(\theta,\delta^H(t'))$ for all $\theta$. Taking expectations over $\theta_i\sim F^H$ yields $\mathbb{E}_{F^H}[p^H(\theta_i,\delta^H(t))]< \mathbb{E}_{F^H}[p^H(\theta_i,\delta^H(t'))]$.
	
	For $g=A$: by Difficulty Projection (main-text Assumption 2), $\delta^H(t)>\delta^H(t')$ implies $\delta^A(t)>\delta^A(t')$. Applying the same argument as above with $F^A$ and $p^A$ yields the result. \end{proof}

\begin{proof}[Proof of Proposition 2]

	Let $\Theta=[\underline\theta,\overline\theta]$ with prior density $f$ and let $t',t''$ where $\delta^H(t')>\delta^H(t'')$.  We prove part (1) for a generic success rate function $p(\theta,\delta)$;\footnote{For \(g=H\), the ordering \(\delta^H(t')>\delta^H(t'')\) is immediate. For \(g=A\), difficulty projection implies \(\delta^A(t')>\delta^A(t'')\). The argument below applies to either group after replacing \(p,\delta,F\) by \(p^g,\delta^g,F^g\).} part (2) is analogous, replacing the success rate $p(\theta,\delta)$ with the failure rate $1-p(\theta,\delta)$.
	
	After observing $(\tilde t,1)$, the posterior CDF is
	\[
	F(\theta\mid (\tilde t,1))
	= \frac{\int_{\underline\theta}^{\theta} f(s)\,p(s,\delta(\tilde t))\,ds}
	{\int_{\underline\theta}^{\overline\theta} f(u)\,p(u,\delta(\tilde t))\,du}
	= \frac{1}{1+R(\theta,\delta(\tilde t))},
	\]
	where
	\begin{equation}\label{ratio}
		R(\theta,\delta)
		= \frac{\int_{\theta}^{\overline\theta} f(u)\,p(u,\delta)\,du}
		{\int_{\underline\theta}^{\theta} f(s)\,p(s,\delta)\,ds}
		= \frac{\int_{\theta}^{\overline\theta} f(u)\,\frac{p(u,\delta)}{p(\theta,\delta)}\,du}
		{\int_{\underline\theta}^{\theta} f(s)\,\frac{p(s,\delta)}{p(\theta,\delta)}\,ds}.
	\end{equation}
	By MLRP (main-text Assumption 1 part 3), for each $u>\theta$, the likelihood ratio $p(u,\delta)/p(\theta,\delta)$ is increasing in $\delta$, while for each $s<\theta$, the likelihood ratio $p(s,\delta)/p(\theta,\delta)$ is decreasing in $\delta$. Therefore, the integrand at each point of integration in the numerator of \eqref{ratio} is increasing in $\delta$, while the integrand at each point of integration in the denominator is decreasing in $\delta$. Hence the numerator is increasing in $\delta$, the denominator is decreasing in $\delta$, and  $R(\theta,\delta)$ is increasing in $\delta$. Thus $F(\theta\mid (\tilde t,1))$ is  decreasing in the difficulty of task $\tilde t$, for all $\theta$. 
	Thus \(F(\cdot\mid (t',1))\) first-order stochastically dominates
	\(F(\cdot\mid (t'',1))\), in the sense that
	\[
	F(\theta\mid (t',1))\leq F(\theta\mid (t'',1)),
	\quad\forall\theta.
	\]
	Since $p(\theta,\delta(t))$ is increasing in $\theta$ and $\pi$ is an expected value of $p(\theta,\delta(t))$, this gives $\pi(t\mid (t'',1))\leq\pi(t\mid (t',1))$.
\end{proof}

% --------- BN-E MATH ------------ %
\subsection{Adoption Framework}

\subsubsection{Proof of Main Theorem}
\begin{proof}
	We first derive the KL minimizer given a delegation profile, then the optimal delegation given perceived ability, and finally combine these to characterize equilibria and prove statements (i)--(iii).
	Fix a domain of $n$ tasks $\mathcal T$, AI performance $\mathbf{q}^A\in(0,1)^n$, and a human ability level $\theta^H$ inducing $q_j^H=p(\theta^H,\delta_j)\in(0,1)$. We assume throughout that \(\Theta\) is compact, \(\theta^H\) lies in its interior, and \(p(\theta,\delta)\) is continuously differentiable in \(\theta\).
	
	\medskip\noindent
	\textbf{KL minimization.}
	Let $x$ be a delegation profile with $x_j=A$ for at least one task (we treat the case of no adoption separately below). Expanding the KL divergence,
	\begin{align*}
		K(x,\theta)
		&= \sum_{j:\,x_j=A}\!\left[q_j^A\ln\frac{q_j^A}{p(\theta,\delta_j)}+(1-q_j^A)\ln\frac{1-q_j^A}{1-p(\theta,\delta_j)}\right]\\
		&=-\sum_{j:\,x_j=A}\!\left[q_j^A\ln p(\theta,\delta_j)+(1-q_j^A)\ln\!\big(1-p(\theta,\delta_j)\big)\right]+C,
	\end{align*}
	where $C$ is constant in $\theta$. Because $p(\theta,\delta_j)$ and $1-p(\theta,\delta_j)$ are log-concave in $\theta$ by assumption, the negative log terms are each convex, so $K(x,\cdot)$ is strictly convex and admits a unique minimizer $\theta^*(x)$.
	Differentiating with respect to $\theta$:
	\[
	\frac{\partial K(x,\theta)}{\partial \theta}
	= \sum_{j:\,x_j=A} h_j(\theta)\,\big(p(\theta,\delta_j)-q_j^A\big),
	\qquad
	h_j(\theta)=
	\frac{\partial_\theta\, p(\theta,\delta_j)}{p(\theta,\delta_j)\big(1-p(\theta,\delta_j)\big)}.
	\]
	Main-text Assumption~1 ensures $p(\theta,\delta_j)\in(0,1)$ and $\partial_\theta\, p>0$, so $h_j(\theta)>0$ for every $j$.
	Evaluating at $\theta=\theta^H$ and noting $p(\theta^H,\delta_j)=q_j^H$:
	\[
	\frac{\partial K(x,\theta^H)}{\partial \theta}
	= \sum_{j:\,x_j=A} h_j(\theta^H)\big(q_j^H-q_j^A\big).
	\]
	By strict convexity, $\theta^*(x)>\theta^H$ if and only if this derivative is negative, i.e.,
	\begin{equation}\label{eq:kl-direction}
		\sum_{j:\,x_j=A} h_j(\theta^H)\,q_j^A
		\;\gtrless\;
		\sum_{j:\,x_j=A} h_j(\theta^H)\,q_j^H
		\quad\Longleftrightarrow\quad
		\theta^*(x)\;\gtrless\;\theta^H.
	\end{equation}
	Note that by definition of $\mathcal{H}$, $h_j(\theta^H)=w_j$; this will be used in Parts~2 and~3.
	
	\medskip\noindent
	\textbf{Optimal delegation given $\theta$.}
	Given belief $\theta$, the subjective success probability of AI on task $j$ is $p(\theta,\delta_j)$, while the human's is $q_j^H=p(\theta^H,\delta_j)$. Since $p$ is strictly increasing in its first argument (main-text Assumption~1),
	\[
	\theta>\theta^H \;\Longrightarrow\; p(\theta,\delta_j)>q_j^H
	\quad\text{for all } j,
	\qquad
	\theta<\theta^H \;\Longrightarrow\; p(\theta,\delta_j)<q_j^H
	\quad\text{for all } j.
	\]
	Because $\Pi(y)$ is separable across tasks, so is $\mathbb{E}_{Q_\theta}\Pi(y)$, and therefore the unique best response is
	\[
	x_j=
	\begin{cases}
		A & \text{for all } j \quad\text{if } \theta>\theta^H,\\
		H & \text{for all } j \quad\text{if } \theta\leq\theta^H,
	\end{cases}
	\]
	where we apply the tie-breaking rule (tasks go to the human when $\theta=\theta^H$).
	\medskip\noindent

	\sss{Part~1: Partial adoption is never a BN-E}
	Fix any delegation profile $x$ with partial adoption, i.e., $x_j=A$ for some $j$ and $x_{j'}=H$ for some $j'$. Since at least one task is delegated, $\theta^*(x)$ is uniquely defined. If $\theta^*(x)>\theta^H$, the best response assigns every task to AI; if $\theta^*(x)\leq\theta^H$, the best response assigns every task to the human. In either case $x$ is not a best response to $\theta^*(x)$, so no partial-adoption profile can be part of a Berk--Nash equilibrium.
	\medskip\noindent

	\sss{Part~2: Full adoption above $\mathcal{H}$}
	Let $x^{\mathrm{full}}$ denote full adoption ($x_j=A$ for all $j$). Applying~\eqref{eq:kl-direction} with $h_j(\theta^H)=w_j$:
	\[
	\sum_{j=1}^n w_j\,q_j^A \;>\; \sum_{j=1}^n w_j\,q_j^H
	\quad\Longleftrightarrow\quad
	\theta^*(x^{\mathrm{full}})>\theta^H.
	\]
	When $\mathbf{q}^A$ lies strictly above $\mathcal{H}$ the left-hand inequality holds, so $\theta^*(x^{\mathrm{full}})>\theta^H$ and full adoption is the unique best response; hence, $(x^{\mathrm{full}},\theta^*(x^{\mathrm{full}}))$ is a BN-E.
	Conversely, if \(\mathbf q^A\) does not lie strictly above \(\mathcal H\), then
	\[
	\sum_{j=1}^n w_j q_j^A \leq \sum_{j=1}^n w_j q_j^H \quad\Longleftrightarrow\quad \theta^*(x^{\mathrm{full}})\leq \theta^H
	\]
	Given the tie-breaking rule favoring the human, full adoption is then not a best response; hence, no BN-E with full adoption exists.
	
	\medskip\noindent
	\sss{Part~3: No adoption is always an equilibrium; below $\mathcal{H}$, it is the only one}
	Let $x^{0}$ denote no adoption ($x_j=H$ for all $j$). Since $\{j:x_j^0=A\}=\emptyset$,
	\[
	K(x^0,\theta)=0
	\qquad\text{for every } \theta \text{ and } \mathbf{q}^A.
	\]
	Intuitively, no delegation to AI generates no data on AI performance, so any belief can be sustained. Take any $\theta^0\leq\theta^H$. The KL divergence is identically zero, so $\theta^0$ trivially minimizes it. And because $\theta^0\leq\theta^H$, we have $p(\theta^0,\delta_j)\leq q_j^H$ for all $j$, so $x^0$ maximizes expected profit under $Q_{\theta^0}$. Thus $(x^0,\theta^0)$ is a BN-E.
	Finally, we have shown that full adoption is supported only when $\mathbf{q}^A$ lies above $\mathcal{H}$, and partial adoption is never supported. Therefore, when $\mathbf{q}^A$ lies below $\mathcal{H}$, all equilibria involve no adoption.
\end{proof}

\subsection{Extension: Dynamics of Adoption}
\label{app:adoption-model-dyn}
We extend the static adoption model to characterize the adoption path under Human Projection as AI technology improves over time, comparing it to the optimal benchmark.

\sss{Setup}
Consider a continuous-time model over $\tau \in [0, \overline{\tau}]$. Human performance is fixed at $\boldsymbol{q}^H \in (0,1)^n$ throughout. AI technology progresses over time: in each period $\tau$, the set of feasible AI success-rate vectors is $\mathcal{P}(\tau) \subseteq [0,1]^n$, with $\mathcal{P}(0) = {\{\boldsymbol 0\}}$ and $\mathcal{P}(\overline{\tau}) = [0,1]^n$, representing progress from no capability to universal capability. Throughout, vector inequalities are understood coordinatewise. The \emph{technological frontier} $\mathcal{P}^*(\tau)$ is the set of undominated points in $\mathcal{P}(\tau)$:

\[\mathcal{P}^*(\tau) = \{\boldsymbol{q} \in \mathcal{P}(\tau) : \nexists\; \boldsymbol{q}' \in \mathcal{P}(\tau) \text{ with } \boldsymbol{q}' \geq \boldsymbol{q},\; \boldsymbol{q}' \neq \boldsymbol{q}\}.\]

We impose the following structure on the shape of technological progress:
\begin{assumption} \label{ass:frontier}
	The correspondence $\mathcal{P} : [0,\overline{\tau}] \rightrightarrows [0,1]^n$ satisfies:
	\begin{enumerate}
		\item (Convexity and compactness) $\mathcal{P}(\tau)$ is convex and compact for every $\tau$.
		\item(Expanding feasibility) $\tau' > \tau \implies \mathcal{P}(\tau) \subset \mathcal{P}(\tau')$.
		\item (Dominance) If $\boldsymbol{q} \in \mathcal{P}(\tau)$ and $\boldsymbol{q}' \leq \boldsymbol{q}$, then $\boldsymbol{q}' \in \mathcal{P}(\tau)$.
		\item (Continuity) $\mathcal{P}$ is upper and lower hemicontinuous.
		\item (Non-degenerate frontier) $|\mathcal{P}^*(\tau)| > 1$ for every $\tau\in(0,\overline\tau)$.
	\end{enumerate}    
\end{assumption}

Part 1 requires the set of feasible technologies in each period to be convex and compact. Parts 2 and 3 impose two monotonicity conditions: part 2 requires technological progress to be persistent, so feasible technologies are not ``forgotten,'' while part 3 requires feasibility to be downward closed, so that if a technology is feasible, then any technology it dominates is feasible as well. Part 4 imposes continuity of technological progress, ruling out jumps. Part 5 requires a non-degenerate frontier: for every $\tau > 0$, there remains a tradeoff at the (non-degenerate) frontier, so no single technology dominates all others across all tasks.

First, we characterize how the set of available technologies compare with human performance as technology progresses:

\begin{theorem}[Progression Regimes]
	\label{thm:path}
	Let $\boldsymbol q^H\in(0,1)^n$, and fix a technological progress mapping $\mathcal P$ satisfying Assumption \ref{ass:frontier}. There exist $0<\tau_1<\tau_2<\overline\tau$ such that:
	\begin{enumerate}
		\item For $\tau\in[0,\tau_1]$, every feasible AI vector satisfies $\boldsymbol q\le \boldsymbol q^H$.
		\item For $\tau\in(\tau_1,\tau_2)$, there exists a feasible AI vector that exceeds $\boldsymbol q^H$ on some task, but no feasible AI vector satisfies $\boldsymbol q\ge \boldsymbol q^H$.
		\item For $\tau\in[\tau_2,\overline\tau]$, there exists a feasible AI vector satisfying $\boldsymbol q\ge \boldsymbol q^H$.
	\end{enumerate}
\end{theorem}

This allows us to characterize the optimal adoption path absent projection, then compare it to the equilibrium path under Human Projection. 

\begin{corollary}[Optimal path]
	\label{cor:optimal-path}
	There exists $\tau_3\in[\tau_2,\overline\tau)$ such that
	\[
	\begin{array}{ll}
		\text{for } \tau\in[0,\tau_1], & \text{no adoption is optimal},\\[2pt]
		\text{for } \tau\in(\tau_1,\tau_2), & \text{partial adoption is optimal},\\[2pt]
		\text{for } \tau\in[\tau_2,\tau_3), & \text{the optimum may be partial or full},\\[2pt]
		\text{for } \tau\in[\tau_3,\overline\tau], & \text{full adoption is optimal}.
	\end{array}
	\]
\end{corollary}

\noindent The optimal path follows an intuitive progression: as AI improves, it first surpasses human performance on some tasks (triggering partial adoption), then eventually full adoption becomes feasible. The optimum may still alternate between partial and full adoption for a while, but eventually full adoption is optimal forever.

Under Human Projection, however, this pattern is disrupted:
\begin{theorem}[Adoption Path Under Human Projection]
	\label{thm:BN-E-path}
	Under Human Projection there exists $\tilde{\tau} \in (\tau_1, \tau_2]$ such that:
	\begin{enumerate}
		\vspace{-.4cm}\item If $\tau \leq \tilde{\tau}$, then for any $\mathbf{q}^A \in \mathcal{P}^*(\tau)$, all BN-E imply no adoption.
		\vspace{-.4cm}\item If $\tau > \tilde{\tau}$, then there exists $\mathbf{q}^A \in \mathcal{P}^*(\tau)$ such that full adoption is a BN-E.
	\end{enumerate}
\end{theorem}

\noindent This theorem shows two distortions generated by HP, relative to the optimal path: (i) \emph{delayed adoption} for $\tau \in (\tau_1, \tilde{\tau})$, some adoption is optimal yet all BN-E imply no adoption; and (ii) potential \emph{over-adoption} for $\tau \in (\tilde{\tau}, \tau_2)$, where full adoption is a BN-E even though partial adoption is optimal. Thus, conditional on adoption occurring, Human Projection leads decision-makers to assign too many tasks to AI prematurely. 
\subsubsection{Proofs}
\begin{proof}[Proof of Theorem~\ref{thm:path}]
	Let
	\[
	D=\{\boldsymbol q\in[0,1]^n:\boldsymbol q\le \boldsymbol q^H\},
	\;\;
	S=\{\tau:\mathcal P(\tau)\subseteq D\},
	\;\;
	T=\Bigl\{\tau:\exists\,\boldsymbol q\in\mathcal P(\tau)\text{ with }\boldsymbol q\ge \boldsymbol q^H\Bigr\},
	\]
	and define
	\[
	\tau_1=\sup S,
	\qquad
	\tau_2=\inf T.
	\]
	Since $\mathcal P(0)=\{\boldsymbol 0\}\subseteq D$ and $\boldsymbol 1\in\mathcal P(\overline\tau)$, both $S$ and $T$ are nonempty. By expanding feasibility, $S$ is downward closed and $T$ is upward closed. All openness statements below are relative to $[0,1]^n$.
	
	\medskip\noindent
	\textbf{Step 1:} $0<\tau_1$ and $0<\tau_2<\overline\tau$.
	
	Choose $\varepsilon\in(0,\min_i q_i^H)$ and set $U=[0,\varepsilon)^n$. Then $U$ is an open neighborhood of $\boldsymbol 0$ in $[0,1]^n$, $U\subseteq D$, and no point of $U$ satisfies $\boldsymbol q\ge \boldsymbol q^H$. By upper hemicontinuity at $0$, $\mathcal P(\tau)\subseteq U$ for all $\tau$ sufficiently close to $0$. Hence such $\tau$ belong to $S$ and do not belong to $T$, so $\tau_1>0$ and $\tau_2>0$.
	
	Next choose $\eta\in(0,\min_i(1-q_i^H))$ and set $V=(1-\eta,1]^n$. Then $V$ is an open neighborhood of $\boldsymbol 1$ in $[0,1]^n$, and every $\boldsymbol q\in V$ satisfies $\boldsymbol q\ge \boldsymbol q^H$. Since $\boldsymbol 1\in\mathcal P(\overline\tau)$, lower hemicontinuity at $\overline\tau$ implies that $\mathcal P(\tau)\cap V\neq\varnothing$ for all $\tau<\overline\tau$ sufficiently close to $\overline\tau$. Hence such $\tau$ belong to $T$, so $\tau_2<\overline\tau$.
	
	\medskip\noindent
	\textbf{Step 2:} The endpoint properties are
	\[
	\mathcal P(\tau_1)\subseteq D
	\qquad\text{and}\qquad
	\exists\,\boldsymbol q^*\in\mathcal P(\tau_2)\text{ with }\boldsymbol q^*\ge \boldsymbol q^H.
	\]
	
	For the first claim, suppose $\mathcal P(\tau_1)\not\subseteq D$. Then some $\boldsymbol q\in\mathcal P(\tau_1)$ has $q_i>q_i^H$ for some $i$. The set
	\[
	O_i=\{\boldsymbol x\in[0,1]^n:x_i>q_i^H\}
	\]
	is open and meets $\mathcal P(\tau_1)$. By lower hemicontinuity at $\tau_1$, there exists $\tau<\tau_1$ with $\mathcal P(\tau)\cap O_i\neq\varnothing$. Since $\tau_1=\sup S$, there is some $s\in S$ with $s>\tau$. Expanding feasibility gives $\mathcal P(\tau)\subseteq\mathcal P(s)\subseteq D$, contradicting $\mathcal P(\tau)\cap O_i\neq\varnothing$.
	
	For the second claim, suppose no point of $\mathcal P(\tau_2)$ weakly dominates $\boldsymbol q^H$. Then
	\[
	\mathcal P(\tau_2)\subseteq \bigcup_{i=1}^n \{\boldsymbol x\in[0,1]^n:x_i<q_i^H\},
	\]
	and the right-hand side is open in $[0,1]^n$. By upper hemicontinuity at $\tau_2$, the same inclusion holds for all $\tau>\tau_2$ sufficiently close to $\tau_2$. For such $\tau$ there is no $\boldsymbol q\in\mathcal P(\tau)$ with $\boldsymbol q\ge \boldsymbol q^H$, contradicting $\tau_2=\inf T$. Thus some $\boldsymbol q^*\in\mathcal P(\tau_2)$ satisfies $\boldsymbol q^*\ge \boldsymbol q^H$.
	
	\medskip\noindent
	\textbf{Step 3:} $\tau_1<\tau_2$.
	
	If $\tau_2\le \tau_1$, then by expanding feasibility and Step~2,
	\[
	\mathcal P(\tau_2)\subseteq \mathcal P(\tau_1)\subseteq D.
	\]
	But Step~2 also gives $\boldsymbol q^*\in\mathcal P(\tau_2)$ with $\boldsymbol q^*\ge \boldsymbol q^H$, so in fact $\boldsymbol q^*=\boldsymbol q^H$. Therefore every $\boldsymbol q\in\mathcal P(\tau_2)$ satisfies $\boldsymbol q\le \boldsymbol q^H$, and hence $\boldsymbol q^H$ dominates every feasible point at time $\tau_2$. It follows that
	\[
	\mathcal P^*(\tau_2)=\{\boldsymbol q^H\},
	\]
	contradicting Assumption~\ref{ass:frontier}, since $\tau_2>0$ by Step~1. Thus $\tau_1<\tau_2$.
	
	\medskip\noindent
	\textbf{Step 4:} Verification of the three claims.
	
	If $\tau\le \tau_1$, then expanding feasibility and Step~2 give
	\[
	\mathcal P(\tau)\subseteq \mathcal P(\tau_1)\subseteq D,
	\]
	so every feasible AI vector satisfies $\boldsymbol q\le \boldsymbol q^H$.
	
	If $\tau\in(\tau_1,\tau_2)$, then $\tau\notin S$, so $\mathcal P(\tau)\not\subseteq D$; equivalently, some feasible AI vector exceeds $\boldsymbol q^H$ on at least one task. Also $\tau\notin T$, so no feasible AI vector satisfies $\boldsymbol q\ge \boldsymbol q^H$.
	
	Finally, by Step~2 there is $\boldsymbol q^*\in\mathcal P(\tau_2)$ with $\boldsymbol q^*\ge \boldsymbol q^H$. Expanding feasibility implies $\boldsymbol q^*\in\mathcal P(\tau)$ for every $\tau\ge \tau_2$.
\end{proof}

\begin{proof}[Proof of Corollary~\ref{cor:optimal-path}]
	By Theorem~\ref{thm:path}, for $\tau\le\tau_1$ every feasible $\boldsymbol q$ satisfies $\boldsymbol q\le \boldsymbol q^H$, so replacing any human task by AI weakly lowers profit. Hence no adoption is optimal on $[0,\tau_1]$.
	
	For \(\tau\in(\tau_1,\tau_2)\), Theorem~\ref{thm:path} gives some feasible \(\mathbf q\) with \(q_i>q_i^H\) for at least one task \(i\), so adopting AI on all and only the tasks for which \(q_j>q_j^H\) yields a strict gain over no adoption. Moreover, since no feasible \(\mathbf q\) satisfies \(\mathbf q\geq \mathbf q^H\), every feasible AI vector has at least one task \(k\) with \(q_k<q_k^H\). Full adoption with that vector is strictly dominated by assigning task \(k\) to the human while keeping the same AI vector on the remaining tasks. Hence partial adoption is optimal on \((\tau_1,\tau_2)\).
	
	It remains to prove eventual full adoption. At $\overline\tau$ the vector $\boldsymbol 1$ is feasible, so full adoption attains profit $\sum_j R_j$. Any partial-adoption policy leaves at least one task $k$ with the human, and therefore earns at most
	\[
	\sum_{j=1}^n R_j - R_k(1-q_k^H)
	\le
	\sum_{j=1}^n R_j - \delta,
	\qquad
	\delta=\min_j R_j(1-q_j^H)>0.
	\]
	Thus at $\overline\tau$ full adoption strictly dominates every partial-adoption policy by a positive margin. Choose $\varepsilon>0$ small enough that any $\boldsymbol q$ with $q_j>1-\varepsilon$ for all $j$ still yields full-adoption profit strictly above $\sum_j R_j-\delta$. By lower hemicontinuity at $\overline\tau$ and the fact that $\boldsymbol 1\in\mathcal P(\overline\tau)$, there exists $\tau_3<\overline\tau$ and $\boldsymbol q\in\mathcal P(\tau_3)$ with $q_j>1-\varepsilon$ for all $j$. By expanding feasibility, $\boldsymbol q\in\mathcal P(\tau)$ for all $\tau\ge\tau_3$, so full adoption is optimal on $[\tau_3,\overline\tau]$.
	
	Finally, for $\tau\in[\tau_2,\tau_3)$, full adoption is feasible by Theorem~\ref{thm:path}, but need not yet dominate all partial-adoption choices. Hence the optimum on that interval may be either partial or full.
\end{proof}

\begin{proof}[Proof of Theorem~\ref{thm:BN-E-path}]
	Define the linear score function
	\[
	\sigma(\mathbf{q}) = \sum_{j=1}^n w_j(q_j - q_j^H),
	\qquad w_j = \frac{\partial p(\theta^H,\delta_j)/\partial\theta}{q_j^H(1-q_j^H)}>0,
	\]
	so that $\mathbf{q}^A$ lies strictly above $\mathcal{H}$ iff $\sigma(\mathbf{q}^A)>0$.
	By the main Theorem, full adoption is a BN-E iff $\sigma(\mathbf{q}^A)>0$;
	partial adoption is never a BN-E; no adoption is always a BN-E.
	
	Define $\sigma^*(\tau)=\max_{\mathbf{q}\in\mathcal{P}(\tau)}\sigma(\mathbf{q})$.
	Since $\sigma$ is continuous and linear and $\mathcal{P}(\tau)$ is compact, the maximum
	exists and is attained at a frontier point (since all $w_j>0$).
	By expanding feasibility, $\sigma^*$ is non-decreasing;
	by the Maximum Theorem, $\sigma^*$ is continuous.
	
	We establish values at the boundaries. From Theorem~\ref{thm:path}, $\mathcal{P}(\tau_1)\subseteq D
	=\{\mathbf{q}:\mathbf{q}\le\mathbf{q}^H\}$.
	If $\mathbf{q}^H\in\mathcal{P}(\tau_1)$, then $\mathbf{q}^H$ dominates every
	element of $\mathcal{P}(\tau_1)\subseteq D$, which implies
	$\mathcal{P}^*(\tau_1)=\{\mathbf{q}^H\}$---contradicting
	Assumption~\ref{ass:frontier} part 5 since $\tau_1>0$. Therefore, we must have $\mathbf{q}^H\notin\mathcal{P}(\tau_1)$,
	and every $\mathbf{q}\in\mathcal{P}(\tau_1)$ satisfies
	$\mathbf{q}\le\mathbf{q}^H$, $\mathbf{q}\neq\mathbf{q}^H$,
	hence $\sigma(\mathbf{q})<0$. By compactness, $\sigma^*(\tau_1)<0$.
	Meanwhile, Theorem~\ref{thm:path} (Step~2) gives
	$\mathbf{q}^*\in\mathcal{P}(\tau_2)$ with $\mathbf{q}^*\ge\mathbf{q}^H$,
	so $\sigma^*(\tau_2)\ge\sigma(\mathbf{q}^*)\ge 0$.
	
	Since $\sigma^*$ is continuous and non-decreasing with $\sigma^*(\tau_1)<0\le\sigma^*(\tau_2)$, define
	\[
	\tilde\tau = \sup\bigl\{\tau\in[0,\overline\tau]:\sigma^*(\tau)\le 0\bigr\}.
	\]
	Then $\tilde\tau\in(\tau_1,\tau_2]$, with $\sigma^*(\tilde\tau)=0$ by continuity and $\sigma^*(\tau)>0$ for all $\tau>\tilde\tau$ by monotonicity.
	For $\tau\leq\tilde\tau$: $\sigma^*(\tau)\le 0$, so $\sigma(\mathbf{q}^A)\le 0$ for every $\mathbf{q}^A\in\mathcal{P}^*(\tau)$. The main Theorem then implies full adoption is not a BN-E and partial adoption is never a BN-E, while no adoption is always a BN-E. Hence every BN-E features no adoption. For $\tau>\tilde\tau$: $\sigma^*(\tau)>0$, so some $\mathbf{q}^A\in\mathcal{P}^*(\tau)$ achieves $\sigma(\mathbf{q}^A)>0$, and full adoption is a BN-E by main-text Theorem~1.
	
	Relative to the optimal path (Theorem~\ref{thm:path}), two distortions emerge. For $\tau\in(\tau_1,\tilde\tau)$, partial adoption is optimal yet every BN-E features no adoption---\emph{delayed adoption}. For (nonempty) $\tau\in(\tilde\tau,\tau_2)$, partial adoption is optimal yet there exists a technology such that full adoption is a BN-E---\emph{over-adoption}, conditional on adopting.
\end{proof}

\iffalse	
%adoption path figures
\begin{figure}[H]
	\centering
	\caption{Adoption Path under HP vs.\ Optimal Adoption Path}
	\label{fig:BN-E-path}
	
	\begin{tikzpicture}[y=0.5cm]
		% Draw the lines
		\draw[thick,->] (0,-2) -- (15,-2);
		\draw[thick,->] (0,-6) -- (15,-6);
		%\draw[thick,->] (0,-7) -- (15,-7);
		% Add ticks for the segments
		\foreach \x/\xtext in {0/\tau_0,  5.5/\tilde{\tau}, 15/\overline{\tau}}{
			\draw (\x,-1.8) -- (\x,-2.2);
			\node[below] at (\x, -2.2) {$\xtext$};
		}
		\foreach \x/\xtext in {0/\tau_0, 3/\tau_1, 7/\tau'_2, 12/\tau_2, 15/\overline{\tau}}{
			\draw (\x,-5.8) -- (\x,-6.2);
			\node[below] at (\x, -6.2) {$\xtext$};
		}
		
		% Add labels above the line for Berk Nash
		\node[above] at (2.75, -2) {No Adoption};
		\node[above] at (10.25, -2) {Full Adoption*};
		\node[above=, align=center] at (7.5, 0) {\textbf{Berk-Nash Equilibrium}};
		% Add labels above the line for Full info
		\node[above] at (1.5, -6) {No Adoption};
		\node[above] at (5, -6) {Partial Adoption};
		\node[above] at (9.5, -6) {Partial/Full Adoption};
		\node[above] at (13.5, -6) {Full Adoption};
		\node[above, align=center] at (7.5, -4) {\textbf{Optimal Adoption}};
	\end{tikzpicture}
	\vspace{-0.5cm}\floatfoot{\textit{Notes:}  This timeline represents the different stages of adoption under HP relative to the optimal action. ``Full Adoption*'' represents the stage where full adoption is a Berk-Nash equilibrium, but recall that by Theorem \ref{thm:BN-E}, no-adoption is always a Berk-Nash equilibrium.}	
\end{figure}
\fi

\subsection{Framework in Field Context} \label{app:HP_reason}
\sss{Setup}
Assume a similar setup to that of the basic framework, where a DM is trying to predict the performance of an agent over a domain of tasks.\footnote{The agent belongs to a population which can be ordered by ability $\theta$ as before. We omit agent subscripts for clarity.} 
Let $\mathcal{Z}$ and $\mathcal{W}$ denote the (finite) sets of \emph{questions} and \emph{answers}: for each question $z_i\in \mathcal{Z}$, assume there exists a subset of useful answers $U_i\subset \mathcal{W}$. 
Given a question--answer pair $(z_i,w_j)$, the DM's utility is $\overline{u}>0$ if $w_j\in U_i$, and $0$ otherwise.
Assume a \emph{similarity} function $S:\mathcal{W}\times\mathcal{W}\to[0,1]$ 
with $S(w,w')=1$ if and only if $w=w'$.\footnote{Equivalently, one may interpret elements of $\mathcal{W}$ as equivalence classes of substantively identical answers. Faithfulness of $S$ then ensures that useless answers have reasonableness strictly below $1$.} The function $S$ captures the human similarity of answers, depending on semantic overlap, contextual proximity, shared meaning, etc.

Define the \emph{reasonableness} of answer $w_j$ to question $z_i$ by
\[
r_{ij} \;=\; \max_{w\in U_i} S(w_j,w)\in[0,1].
\]
In words, an answer's reasonableness is its similarity to the most similar useful answer---the most ``forgiving'' measure of proximity to any useful answer. Useful answers have $r_{ij}=1$ by construction; faithfulness of $S$ implies $r_{ij}<1$ for every useless answer $w_j\notin U_i$. The framework thus focuses on inference from \emph{failures}: cases where the agent provides an answer that is not useful.

Let $\mathcal{R}\subset[0,1]$ denote the common support of observed reasonableness scores, with $1\in\mathcal{R}$. Let $R$ denote the random variable that captures the reasonableness of the agent's answer when asked a question. We now specify how ability determines the distribution of $R$. Let
\[
\tilde p(\theta,r)\;=\;\Pr(R=r\mid \theta), \qquad r\in\mathcal{R}.
\]

\begin{assumption}[Universal reasonableness distribution]\label{assn:answer-symmetry}
	For every ability level $\theta$, the distribution $\tilde p(\theta,\cdot)$ over $\mathcal{R}$ is the same for all questions $z_i\in\mathcal{Z}$.
\end{assumption}

This question-invariance allows the DM to pool observations from different questions to learn about $\theta$.\footnote{This assumption abstracts from question-specific difficulty: conditional on ability, all questions induce the same distribution of reasonableness scores and hence the same probability of a useful answer.} 

The probability of a useful answer is then $\tilde p(\theta,1)$ for every question, and the probability of a useless answer is $1-\tilde p(\theta,1)$.

\begin{assumption}[MLRP]\label{assn:field-mlrp}
	For every $\theta$ and every $r\in\mathcal{R}$, $\tilde p(\theta,r)>0$. Moreover, for any $\theta'>\theta$ and $r'>r$,
	\[
	\frac{\tilde p(\theta',r')}{\tilde p(\theta',r)}
	\;>\;
	\frac{\tilde p(\theta,r')}{\tilde p(\theta,r)}.
	\]
\end{assumption}
In words, higher-ability agents are relatively more likely to give more reasonable answers.

Now suppose the DM has observed the agent answer a question in a previous instance, yielding a reasonableness score $r$, and wishes to predict performance on a (possibly different) question $z_i$. Let $F_{\mid r}$ denote the DM's posterior over $\theta$ after this observation.\footnote{More generally, data may consist of scores $D=(r^{(1)},\dots,r^{(T)})$ from $T$ past episodes, potentially involving different questions. Because $\tilde p(\theta,\cdot)$ is question-invariant (Assumption~\ref{assn:answer-symmetry}), the posterior $F_{\mid D}$ depends on which scores were observed but not on which questions generated them.} The subjective probability of a useful answer to $z_i$ is
\[
\pi(z_i\mid r)
\;=\;
\mathbb{E}_{F_{\mid r}}\!\big[\tilde p(\theta,1)\big].
\]

We obtain the following result, which is the formal statement of main-text Prediction~4.\footnote{For the strict inequality below, assume the DM's prior over $\theta$ is nondegenerate. Without this, the conclusion holds weakly.}
\begin{proposition}\label{prop:parentdata}
	For any $r'>r$ and any $z_i\in \mathcal{Z}$, $\; \pi(z_i\mid r') > \pi(z_i\mid r).$
\end{proposition}
Observing a less reasonable failure (lower $r$) induces stronger negative inference about $\theta$, which lowers the perceived chance of a useful answer to any question. Further assuming that trust and willingness to delegate are increasing in the perceived likelihood of a useful answer, we obtain the remaining parts of main-text Prediction~4.

\begin{proof}[Proof of Proposition~\ref{prop:parentdata}]
	We first show that $\tilde p(\theta,1)$ is strictly increasing in $\theta$, then apply the MLRP to rank posteriors.
	
	\medskip\noindent\emph{Step 1.}
	Consider the odds ratio
	\[
	\frac{1-\tilde p(\theta,1)}{\tilde p(\theta,1)}
	= \sum_{r\in\mathcal{R}\setminus\{1\}}
	\frac{\tilde p(\theta,r)}{\tilde p(\theta,1)}.
	\]
	Every summand has $r<1$. Applying Assumption~\ref{assn:field-mlrp} with $(r',r)=(1,r)$, for $\theta'>\theta$,
	\[
	\frac{\tilde p(\theta',r)}{\tilde p(\theta',1)}
	\;<\;
	\frac{\tilde p(\theta,r)}{\tilde p(\theta,1)}.
	\]
	Each summand is strictly decreasing in $\theta$, so the odds ratio is strictly decreasing in $\theta$, and therefore $\tilde p(\theta,1)$ is strictly increasing in $\theta$.
	
	\medskip\noindent\emph{Step 2.}
	Since $\tilde p(\theta,r)$ is the likelihood of observing score $r$ given $\theta$, Assumption~\ref{assn:field-mlrp} and the standard MLRP--FOSD argument (as in main-text Proposition~2) give
	\[
	r'>r \;\Longrightarrow\; F_{\mid r'} \succ_{\mathrm{FOSD}} F_{\mid r}.
	\]
	Since $\tilde p(\theta,1)$ is increasing in $\theta$ by Step~1, taking expectations yields
	\[
	\pi(z_i\mid r')
	= \mathbb{E}_{F_{\mid r'}}\!\big[\tilde p(\theta,1)\big]
	>
	\mathbb{E}_{F_{\mid r}}\!\big[\tilde p(\theta,1)\big]
	= \pi(z_i\mid r),
	\]
	which completes the proof.
\end{proof}

%----------------- APP: MATH TASKS -----------------%

\section{Performance in Mathematics}\label{appendix:math}

\subsection{Task Dataset}

We construct a benchmark of standardized mathematics problems from released items in the \textit{Trends in International Mathematics and Science Study} (TIMSS). All released items were accessed through the \href{https://timssandpirls.bc.edu/timss-landing.html}{TIMSS portal}, and items for the 2015 and 2019 test waves were obtained through direct request to the IEA. 
We manually re-transcribe all selected items into a uniform format, while trying as much as possible to preserve wording, tables, and symbols used. We restrict attention to multiple-choice questions with a unique objectively correct answer (4 or 5 options, labeled A--E), which yields a binary performance measure. We excluded items with visual components that could not be processed in textual form by ChatGPT at the time.
Our final benchmark contains 414 items spanning 4\textsuperscript{th} grade (29\%), 8\textsuperscript{th} grade (58\%), and high-school (13\%), and covers the main topic categories in the TIMSS frameworks. 
To allow for better comparison between human and AI performance, we experimented beforehand to make sure ChatGPT was able to correctly process inputs and was not tricked by minor formatting issues. We prompted ChatGPT with one task at a time, following the order used in the released items documents. Full logs of conversations are available upon request.

\subsection{Human Performance}

\begin{figure}[t]
	\centering
	\caption{Item Difficulty and Average Human Performance by Difficulty}
	\label{fig:diff-grade-and-perf}
	\begin{subfigure}[t]{0.49\textwidth}
		\caption{Histogram of Item Difficulty}
		\centering
		\includegraphics[width=\textwidth]{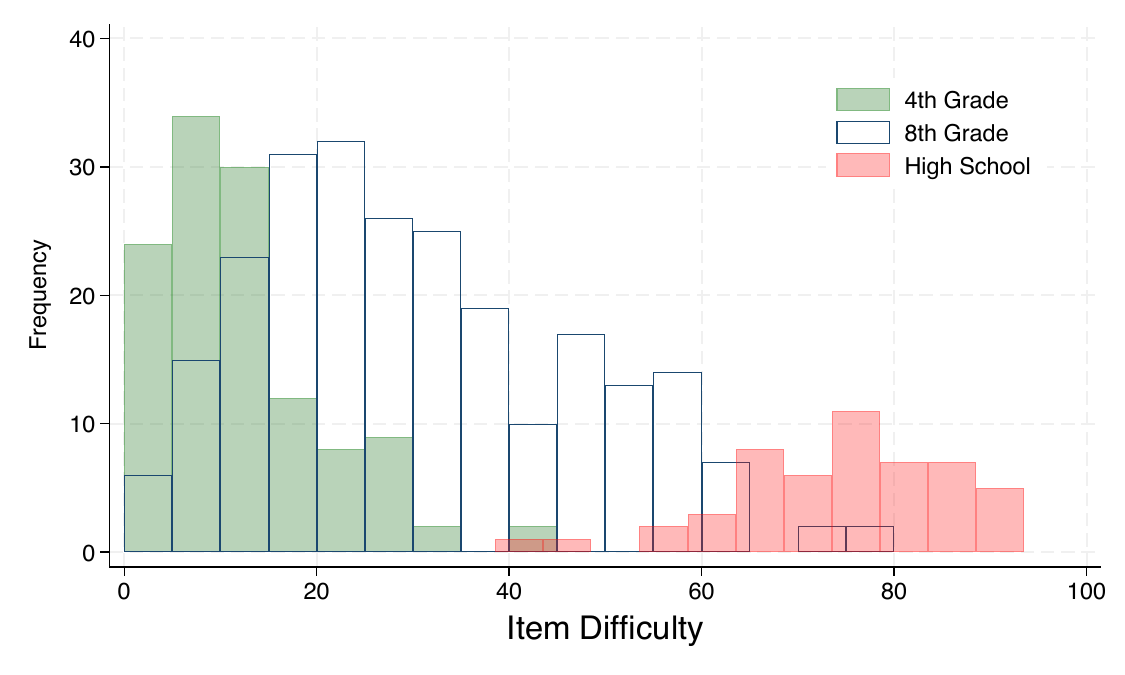}
	\end{subfigure}\hfill
	\begin{subfigure}[t]{0.49\textwidth}
		\caption{Average Human Perf. by Difficulty Decile}
		\centering
		\includegraphics[width=\textwidth]{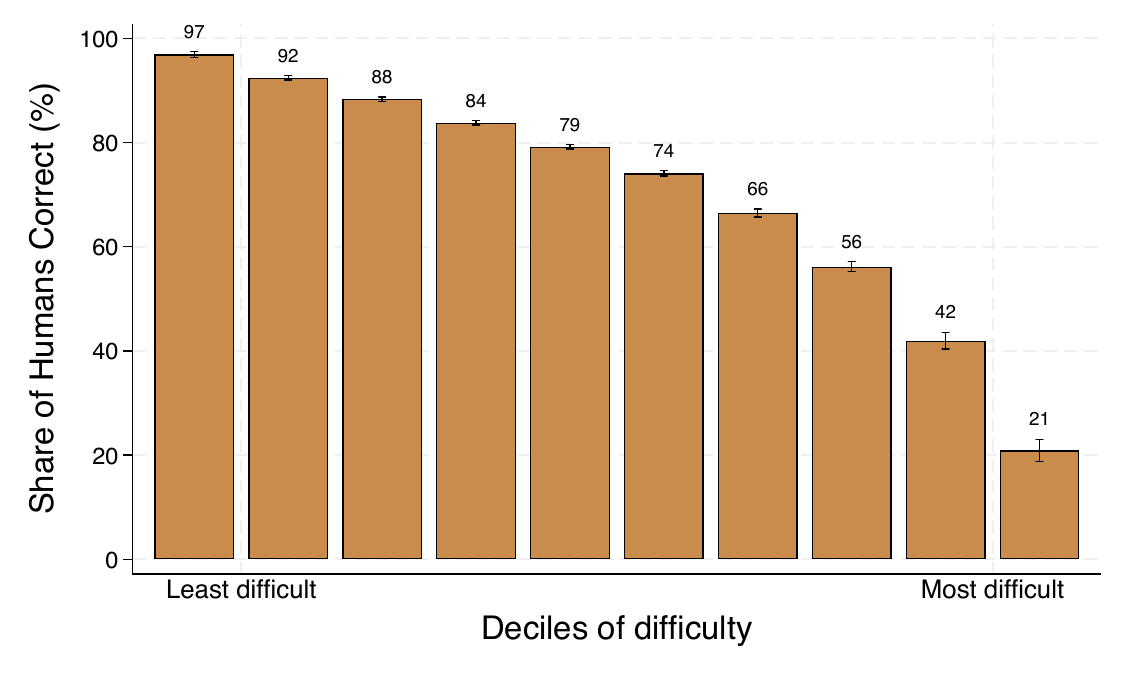}
	\end{subfigure}
	
	\vspace{-0.4cm}
	\floatfoot{\textit{Notes:} \textbf{Left panel:} Overlaid histograms of item difficulty by grade level (4th grade, 8th grade, high-school). The y-axis reports the number of questions and the x-axis reports human difficulty. \textbf{Right panel:} Average share of participants answering items correctly within each decile of human difficulty. Deciles are constructed from the item difficulty measure; each decile contains around 41 items.}
\end{figure}

\sss{Descriptive Results}
For human performance data, the number of answers per item ranges from 24 to 191, with an average of 41.5 and a median of 34. High-school problems tend to have more answers per question, while 8\textsuperscript{th}-grade problems have fewer, because (i) the  released problems are not evenly split by grade level (see Figure \ref{fig:diff-grade-and-perf}); (ii) the beliefs experiment sampled H.S. items more heavily to ensure each participant sees the whole gradient of difficulty.

%these diff numbers are checked (diff9)
The average task difficulty in the sample is 30.3. There are 7 items with a difficulty level of 0 (meaning every participant who attempted it answered correctly) and the highest difficulty is 92. 
%A total of 29 items have a difficulty level of more than 75, which is more than random guessing would imply. This is consistent with the TIMSS item guidelines, which encourage the use of plausible incorrect response options (called ``distractors'') when designing multiple-choice items.\footnote{See ``Plausibility of Distractors'' section in \href{https://timssandpirls.bc.edu/timss2019/pdf/T19-item-writing-guidelines.pdf}{guidelines} for the TIMSS 2019 edition.} 
Panel A of Figure \ref{fig:diff-grade-and-perf} shows the origin of questions for each difficulty level, while panel B plots the average share of correct answers for each difficulty decile: the 10\% easiest items have a success rate of around 97\%, while the 10\% most difficult ones have around 21\%.

\begin{figure}[t]
	\centering
	\caption{ChatGPT 3.5 Performance by Human Difficulty Deciles (API)}
	\label{fig:perf-gpt-api}
	\includegraphics[width=.95\textwidth]{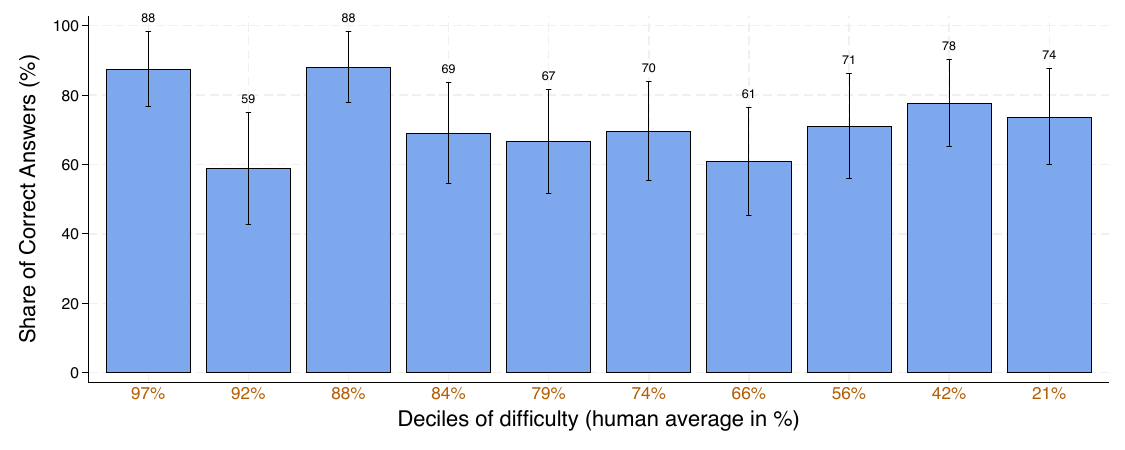}
	\vspace{-0.5cm}\floatfoot{\textit{Notes:} The figure plots GPT-3.5 Turbo's  share of correct answers (collected in May 2025 via the API) within each decile of human difficulty. Each decile contains between 40 and 42 items. Numbers under each bar report the average human success rate in the corresponding decile. Human difficulty is computed from an average of 41.5 answers per item.}
\end{figure}

\subsection{Human-likeness of AI Performance}

\sss{Approach}
We combine item-level human response distributions from the TIMSS benchmark with item-level correctness and answer choices for several LLMs of various AI labs (GPT-3.5 Turbo, GPT-4o, Claude 3.7, Mistral Large, Llama 3.1, o3-mini; all accessed through their respective APIs). For each of 414 items, we observe (i) the share of human test-takers answering correctly, which we treat as the item's empirical ``easiness'' for humans, and (ii) the full distribution of human answer choices (modal option and its share), alongside the model's chosen option and whether it matches the correct key. We compare model outputs across multiple prompt variants: text prompts (standard prompt vs.\ one suppressing chain-of-thought by asking models to only output the final answer) and image prompts (a screenshot of the problem with a standardized layout). 

\sss{Measures} We then compute several complementary notions of ``human-likeness,'' all at the item level. First, we compare how model success co-varies with human success across items using two notions of correlation: the Spearman rank correlation between model correctness and average human correctness, capturing whether models tend to succeed on the same items humans find easier, and  a tetrachoric correlation between model correctness and a binary ``easy-for-humans'' indicator (human accuracy at least $50\%$). 
Then, we summarize the ``shape'' of model performance as a function of human performance by regressing model correctness on  human correctness at the item level and reporting the slope $\beta$, standard error, and $R^2$. 
Finally, we compute two measures of agreement between human and AI answers.
The first ``agreement rate'' measures how much of the mass of the human distribution of answers lies in the model's chosen option, meaning it is high when the model tends to pick options that many humans pick; we compute it both overall (among all items) and conditional on the model picking the wrong option.
The second is  the probability the model chooses the most common human answer, i.e., the human mode. 
We compute it both overall and conditional on model error, the probability it selects the most common human \emph{incorrect} option, to see whether humans and AIs tend to fail in the same way.

\begin{table}[t]
	\caption{Human-likeness of GenAI Performance}
	\label{tab:humanlike}
	\centering
	\resizebox{\textwidth}{!}{
		\begin{threeparttable}
			\begin{tabular}{lccccccccc}
\toprule
 & & \multicolumn{2}{c}{\textbf{Correlations}} & \multicolumn{2}{c}{\textbf{OLS}} & \multicolumn{2}{c}{\textbf{Agreement Rate}} & \multicolumn{2}{c}{\textbf{AI Picks Human Mode}}\\
\cmidrule(lr){3-4} \cmidrule(lr){5-6} \cmidrule(lr){7-8} \cmidrule(lr){9-10}
Model & Mean Perf. & Spearman & Tetrachoric & \(\beta\) (s.e.) & R-squared & Overall & AI wrong & Overall & AI  wrong \\
\hline
\addlinespace[0.25em]
\multicolumn{10}{l}{\textit{Text prompt (standard)}} \\
\addlinespace[0.25em]
GPT-3.5 Turbo & 0.725 & 0.036 & -0.061 & 0.033 (0.095) & 0.000 & 0.568 & 0.126 & 0.718 & 0.380 \\
GPT-4o & 0.978 & 0.060 & 0.181 & 0.040 (0.031) & 0.004 & 0.675 & 0.149 & 0.886 & 0.286 \\
Claude 3.7 & 0.990 & 0.015 & -1.000 & 0.000 (0.021) & 0.000 & 0.679 & 0.301 & 0.894 & 0.250 \\
Mistral Large & 0.928 & 0.018 & 0.125 & 0.014 (0.055) & 0.000 & 0.673 & 0.137 & 0.885 & 0.429 \\
Llama 3.1 & 0.935 & 0.133 & 0.224 & 0.154 (0.052) & 0.021 & 0.677 & 0.192 & 0.887 & 0.545 \\
o3-mini & 0.993 & -0.075 & -1.000 & -0.025 (0.018) & 0.005 & 0.678 & 0.041 & 0.896 & 0.500 \\
\midrule
\multicolumn{10}{l}{\textit{Text prompt (no chain-of-thought)}} \\
\addlinespace[0.25em]
GPT-3.5 Turbo & 0.432 & 0.089 & 0.181 & 0.228 (0.104) & 0.011 & 0.376 & 0.112 & 0.428 & 0.335 \\
GPT-4o & 0.650 & 0.141 & 0.241 & 0.281 (0.100) & 0.019 & 0.549 & 0.257 & 0.688 & 0.285 \\
Claude 3.7 & 0.915 & 0.107 & 0.209 & 0.136 (0.059) & 0.013 & 0.647 & 0.172 & 0.845 & 0.257 \\
Mistral Large & 0.763 & 0.243 & 0.351 & 0.452 (0.087) & 0.061 & 0.584 & 0.157 & 0.728 & 0.448 \\
Llama 3.1 & 0.693 & 0.255 & 0.348 & 0.508 (0.094) & 0.066 & 0.589 & 0.135 & 0.736 & 0.386 \\
o3-mini & 0.995 & -0.043 & -1.000 & -0.013 (0.015) & 0.002 & 0.679 & 0.031 & 0.896 & 0.000 \\
\midrule
\multicolumn{10}{l}{\textit{Image prompt}} \\
\addlinespace[0.25em]
GPT-4o & 0.630 & 0.208 & 0.352 & 0.452 (0.100) & 0.047 & 0.618 & 0.117 & 0.777 & 0.308 \\
Claude 3.7 & 0.696 & 0.180 & 0.289 & 0.357 (0.096) & 0.032 & 0.621 & 0.128 & 0.782 & 0.353 \\
Mistral Large & 0.618 & 0.233 & 0.371 & 0.503 (0.100) & 0.058 & 0.604 & 0.137 & 0.759 & 0.333 \\
Llama 3.1 & 0.585 & 0.207 & 0.227 & 0.406 (0.102) & 0.037 & 0.590 & 0.156 & 0.720 & 0.458 \\
o3-mini & 0.693 & 0.245 & 0.348 & 0.486 (0.095) & 0.060 & 0.623 & 0.117 & 0.787 & 0.278 \\
\bottomrule
\end{tabular}

			\smallskip\footnotesize
			\begin{tablenotes}
				\item \textit{Notes:} Each row reports item-level ``human-likeness'' statistics comparing a model to humans on the same multiple-choice items ($n=414$). Average human performance is 0.697 overall, to be compared with each model's mean performance. \emph{Correlations}  columns report Spearman's rank correlation between human and model correctness, and the tetrachoric correlation between model correctness and a human-defined ``easy'' indicator equal to one if human accuracy is at least $50\%$. \emph{OLS} columns report the slope (s.e.) and $R^2$ from regressing model correctness on human accuracy across items. \emph{Agreement Rate} columns report the average human mass on the model's chosen option (Overall), and conditional on model error (AI wrong). \emph{AI Picks Human Mode} columns report the probability the model selects the human modal option (Overall), and conditional on model error, the probability it selects the most common human \emph{incorrect} option (AI wrong). Higher values generally indicate closer alignment with human difficulty and answer-choice patterns.
			\end{tablenotes}
	\end{threeparttable}}
\end{table}

\sss{Results}
Results are reported in Table \ref{tab:humanlike}. Under the standard text prompt, the most advanced models achieve extremely high mean accuracy (roughly $0.93$ to $0.99$), yet item-level alignment with human difficulty is close to zero (Llama 3.1 being the highest): Spearman correlations are tiny (often near $0$ and sometimes negative), OLS slopes are near $0$, and $R^2$ is essentially null. Thus, \emph{which} items humans find easy versus hard barely predicts \emph{which} items the models answer correctly---strong evidence that ``difficulty'' is not a shared ordering even when the model nearly saturates the benchmark.
The tetrachoric boundary value ($\rho=-1$) for Claude~3.7 and o3-mini under the standard prompt reflects a near-separation, small-error regime and is substantively consistent with the raw counts: all observed errors ($4/4$ and $3/3$) occur on human-easy items, with none on human-hard items.

Overall answer-choice overlap is fairly high, but weakens markedly on mistakes: the agreement rate is large unconditionally yet much lower conditional on model error.
A similar divergence is seen for agreement with the human mode. 
Although models often match the human modal answer overall, their mistakes are only partially human-like: conditional on error, agreement with the most common human wrong option is far from one and varies substantially across models ($\approx 0.25$ to $0.55$ under the standard prompt), implying heterogeneous divergence in error modes.
Finally, we note that prompting choices have large consequences for performance. 
Suppressing chain-of-thought substantially reduces accuracy for most models (except for o3-mini, consistent with this manipulation constraining visible output rather than internal reasoning). It tends to increase alignment with humans among difficulty measures (higher Spearman and steeper $\beta$, with larger $R^2$) but decrease it among answer-choice alignment measures (reduction in overall agreement rates). 
Image prompting further reduces average accuracy and, relative to the standard text prompt, increases correlation-based difficulty alignment while modestly reducing overall choice-based overlap.

% --------------- APP: BELIEFS EXP ----------------- %
\section{Beliefs Experiment}\label{appendix:beliefs}

\subsection{Additional Results}\label{appendix:beliefs-results}

\begin{figure}[h]
	\centering
	\caption{Average Beliefs about AI vs. Optimistic Beliefs about Humans}
	\label{fig:bin-humanlike}
	\includegraphics[width=.5\textwidth]{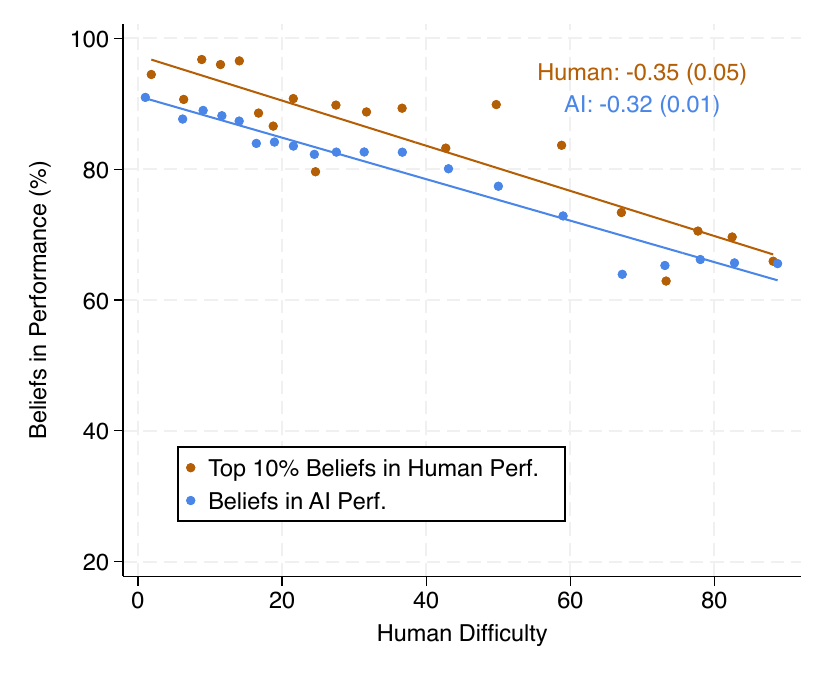}
	\vspace{-0.5cm}\floatfoot{\textit{Notes:}  This figure compares all beliefs about ChatGPT with a subset of the most optimistic participants in the \textit{Human} treatment. We order participants by their average reported belief and keep the top 10\%, then plot all of their beliefs. Belief-level observations are $n=254$ for humans, and $n=10340$ for AI. The average difficulty of questions seen by a participant is added as control in the regression. We report coefficient estimates and their standard errors on the top right.}
\end{figure}

\sss{Separating performance levels and gradient}
The binned scatterplot figure in the main text shows that beliefs about AI have both higher level  (80\% vs.\ 64\%) and a smaller slope with respect to human difficulty (-0.313 vs.\ -0.668).  In Figure \ref{fig:bin-humanlike} we plot the top 10\% most optimistic participants in \emph{Human}, based on their mean reported belief, against the full sample in \emph{AI}. As we control for the mean difficulty of tasks seen, this is a way to proxy for prior belief in agent ability, which determines performance level. Comparing these two groups, we find beliefs about performance to be highly similar, both in slope and levels. In other words, beliefs about AI performance closely resemble highly optimistic beliefs about humans. While not causal, this finding is  consistent with the smaller slope in \emph{AI} being driven by a more optimistic prior over AI ability, rather than incomplete projection.

\sss{Updating sensitivity by projection strength}
We estimate a random-slope mixed model (by Maximum Likelihood) within the AI arm, regressing beliefs on standardized task difficulty with participant-level random intercepts and slopes. This yields Empirical Bayes (BLUP) posterior means of individual slopes, which combine each participant's data with the estimated population distribution and shrink noisier estimates toward the arm mean; we define projection strength as the negative of this slope, so higher values correspond to greater belief sensitivity to difficulty.
We then split the AI-arm sample at the median into low- and high-projection participants (results are qualitatively similar using OLS slopes). Consistent with main-text Prediction~2, higher-projection participants exhibit a stronger easy-hard asymmetry in belief updating, with the differential effect most pronounced following successes.

\begin{figure}[h]
	\centering
	\caption{Belief Updating by Participant's Projection Strength}
	\label{fig:belief-updating-proj}
	\includegraphics[width=\textwidth]{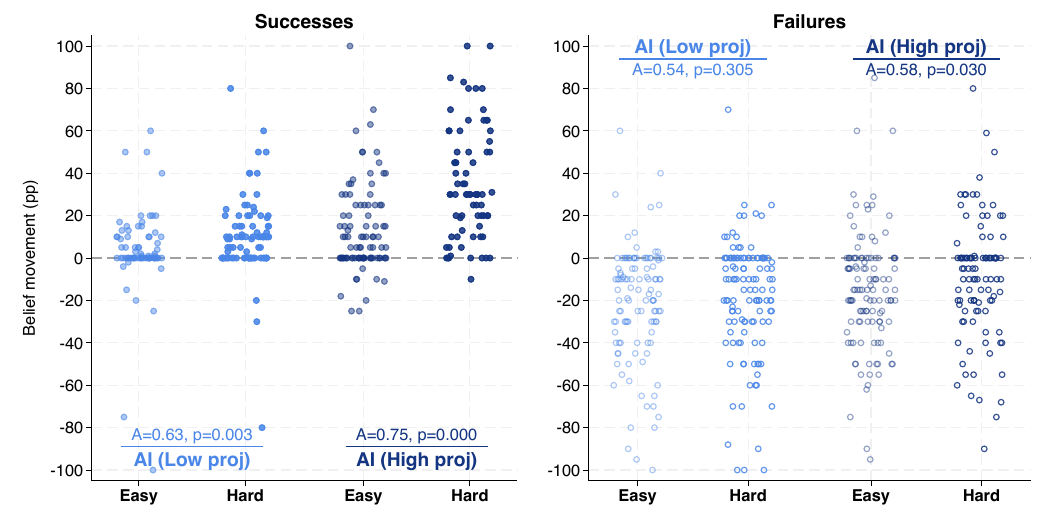}
	\vspace{-0.9cm}\floatfoot{\textit{Notes:} The figure presents jittered dot plots of raw belief movement, defined as the difference between posterior and prior belief (in pp). The AI-arm sample is split at the median of participant-level projection strength, estimated via empirical Bayes. Sample sizes from left to right are $n=73$, $n=87$, $n=94$, $n=74$, $n=112$, $n=120$, $n=118$, $n=106$. Colored horizontal brackets span the easy--hard pair within each group. Reported $A = \Pr(\Delta_H > \Delta_E) + 0.5\Pr(\Delta_H = \Delta_E)$ is the common-language effect size from a Mann--Whitney $U$-test ($A = 0.5$ under the null). $p$-values are from two-sided Wilcoxon rank-sum tests
		within each group and signal.}
\end{figure}

\begin{figure}[h]
	\caption{Belief--Difficulty Gradient by Participants' Own Performance}
	\label{fig:hetero_slope}
	\centering
	\includegraphics[width=.75\textwidth]{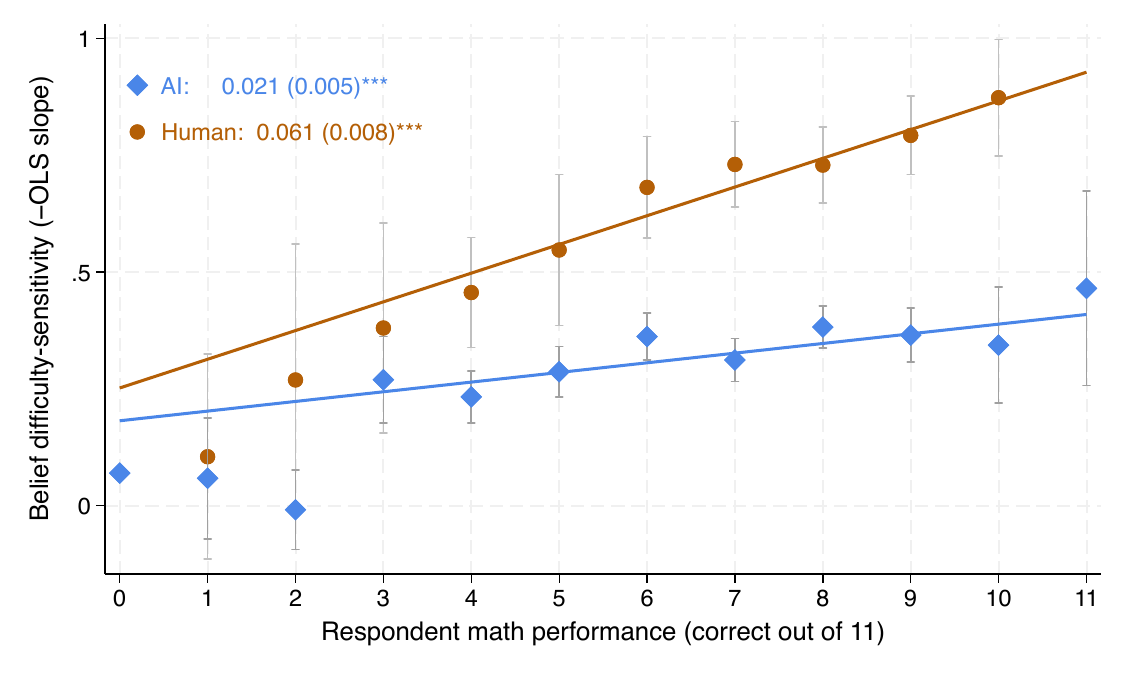}
	\vspace{-0.5cm}
	\floatfoot{\textit{Notes:} Belief difficulty-sensitivity is the participant-level statistic $-\hat\beta_i$, where $\hat\beta_i$ is the OLS slope from regressing the participant's stated success beliefs on human item difficulty. Larger values indicate a steeper decline in beliefs with difficulty.
		Each point is the mean difficulty-sensitivity among participants with a given math score, by treatment arm.
		Lines plot fitted values from OLS regressions of difficulty-sensitivity on own performance (interacted with the arm indicator), controlling for effort proxies (log time spent, cheating indicators, attention-failure indicator); the slope labels correspond to these regressions.
		Vertical bars are 95\% confidence intervals around the binned means.}
\end{figure}

\sss{Heterogeneity in Belief Difficulty-Sensitivity}
Figure~\ref{fig:hetero_slope} documents heterogeneity in belief 
difficulty-sensitivity across participants. Participants who perform better 
on the math items exhibit steeper belief-difficulty gradients in both arms: 
the slope of difficulty-sensitivity on own test performance is $0.061$ 
($0.008$, $p<0.001$) in the \emph{Human} arm and $0.021$ ($0.005$, 
$p<0.001$) in \emph{AI}, with a cross-arm attenuation of $-0.041$ ($0.009$, 
$p<0.001$). The strong relationship observed in both arms suggests higher scorers perceive item difficulty more sharply and track it regardless of who they predict. The cross-arm attenuation suggests higher scorers may project human difficulty onto AI less strongly, although we view this finding as descriptive. 

Neither self-reported AI familiarity nor any 
demographic characteristic we examine (math background, education, income, 
gender, or age) is significantly correlated with belief-difficulty sensitivity.

%predicting power of belifs on perf
\sss{Belief accuracy}
Regressing actual performance on elicited beliefs (with participant fixed 
effects), estimated coefficients are positive and significant for both 
humans and AI (0.58 and 0.06 respectively), but predictive power 
differs sharply: $R^2 = 0.391$ for humans versus $0.002$ for AI. 
Participants rely on human difficulty to predict largely uncorrelated AI 
performance, making their beliefs only weakly predictive of actual AI performance.

% priors
\begin{figure}[h]
	\centering
	\caption{Histograms of Prior Beliefs in Performance}
	\label{fig:histogram-beliefs}
	\includegraphics[width=\textwidth]{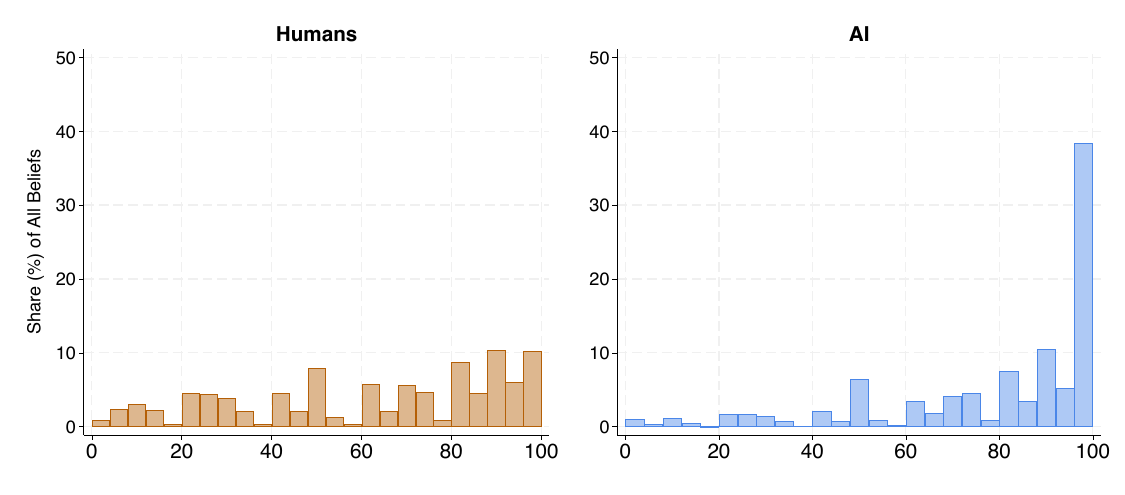}
	\vspace{-1.2cm}\floatfoot{\textit{Notes:} The figure plots histograms of prior beliefs in agent performance. On the x-axis are beliefs in \%.  On the y-axis is the share of all beliefs contained in each bin. $n=2541$ for humans, and $n=10340$ for AI.}
\end{figure}

\begin{table}[h]
	\caption{Beliefs Experiment - Balance Table}
	\label{balance-prior}
	\centering
	\resizebox{.6\textwidth}{!}{
		\begin{threeparttable}
			\begin{tabular}{lcccc} \toprule
 & \multicolumn{2}{c}{\textbf{Humans}} & \multicolumn{2}{c}{\textbf{AI}} \\
\cmidrule(lr){2-3}\cmidrule(lr){4-5}
\textbf{Variable} & \textit{Mean} & \textit{S.E.} & \textit{Mean} & \textit{S.E.} \\ \hline
\multicolumn{5}{l}{\textbf{Gender}} \\
\hspace{2mm}Female & 46.2 &  3.2 & 49.4 &  1.6 \\
\hspace{2mm}Male   & 51.7   &  3.2   & 49.3   &  1.6   \\
\addlinespace[0.35em]\midrule
\multicolumn{5}{l}{\textbf{Race}} \\
\hspace{2mm}White & 73.3 &  2.9 & 78.3 &  1.3 \\
\hspace{2mm}Black & 13.8 &  2.2 & 13.4 &  1.1 \\
\hspace{2mm}Asian & 12.9 &  2.2 &  8.2 &  0.9 \\
\hspace{2mm}Other &  2.9 &  1.1 &  3.4 &  0.6 \\
\addlinespace[0.35em]\midrule
\multicolumn{5}{l}{\textbf{Highest Education}} \\
\hspace{2mm}HS or more         & 98.8       &  0.7       & 99.6       &  0.2       \\
\hspace{2mm}College or more    & 62.1 &  3.1 & 58.6 &  1.6 \\
\hspace{2mm}Master's or more   & 20.8   &  2.6   & 16.4   &  1.2   \\
\addlinespace[0.35em]\midrule
\multicolumn{5}{l}{\textbf{Highest Math Class}} \\
\hspace{2mm}High school or more    & 96.7  &  1.2  & 97.1  &  0.5  \\
\hspace{2mm}College or more        & 49.2 &  3.2 & 46.4 &  1.6 \\
\hspace{2mm}Graduate level or more &  1.2 &  0.7 &  2.8 &  0.5 \\
\addlinespace[0.35em]\midrule
\multicolumn{5}{l}{\textbf{AI Familiarity}} \\
\hspace{2mm}Never heard of AI                & -- & -- &  2.1      &  0.5      \\
\hspace{2mm}Heard of AI, no further familiarity                 & -- & -- & 34.9       &  1.5       \\
\hspace{2mm}Some understanding, never used   & -- & -- & 38.3  &  1.6  \\
\hspace{2mm}Interacted with AI               & -- & -- & 24.7    &  1.4    \\
\addlinespace[0.35em]\hline
Mean Age & 41.8 &  0.9 & 43.6 &  0.4 \\
\textbf{Observations} & \multicolumn{2}{c}{240} & \multicolumn{2}{c}{957} \\
\bottomrule \end{tabular}

			\smallskip\footnotesize
			\begin{tablenotes}
				\item \textit{Notes:} This table provides demographics of participants in the beliefs experiment, run in December 2023. The AI familiarity question was only asked in the \emph{AI} treatment. The sample includes all recruited participants, including those who fail comprehension checks for priors and belief updating. The race question allows participants to select several answers, meaning the sum exceeds 100.
			\end{tablenotes}
	\end{threeparttable}}
\end{table}	

\clearpage

%========================
% APPENDIX (place in Appendix \ref{appendix:beliefs}, after the human-likeness table)
%========================
\subsection{Counterfactual HP Distortion under Fixed 2023 Beliefs}
\label{appendix:distortion}

\begin{table}[h]
	\caption{Counterfactual HP-distortion under fixed 2023-experiment beliefs}
	\label{tab:distortion}
	\centering
	\resizebox{\textwidth}{!}{
		\begin{threeparttable}
			\begin{tabular}{lccccc}
\toprule
 & \multicolumn{2}{c}{\textbf{Performance}} & \multicolumn{3}{c}{\textbf{Distortion measures}} \\
\cmidrule(lr){2-3} \cmidrule(lr){4-6}
Model & Mean Perf. & \# Errors & Level wedge & Slope gap & Avg. belief at errors \\
\midrule
GPT-3.5 Turbo & 0.725 &       114 & -0.091 & -0.280 &  0.813 \\
GPT-4o & 0.978 &         9 &  0.163 & -0.273 &  0.784 \\
Claude 3.7 & 0.990 &         4 &  0.175 & -0.313 &  0.815 \\
Mistral Large & 0.928 &        30 &  0.112 & -0.299 &  0.812 \\
Llama 3.1 & 0.935 &        27 &  0.119 & -0.160 &  0.775 \\
o3-mini & 0.993 &         3 &  0.177 & -0.338 &  0.873 \\
\bottomrule
\end{tabular}

			\begin{tablenotes}[flushleft]
				\footnotesize
				\item \textit{Notes:} Counterfactual diagnostics computed by holding fixed the belief rule $\hat{p}_i = \hat{a} + \hat{b}\cdot d_i$ estimated from AI-arm priors in the beliefs experiment (Appendix~\ref{appendix:beliefs}), where $d_i = 1 - \bar{h}_i$ is item $i$'s human difficulty  and $\hat{p}_i$ is the belief-predicted AI accuracy for item $i$. Let $\bar{\mu}$ and $\bar{p}_m$ denote item averages of $\hat{p}_i$ and realized model accuracy $p_{mi}$, respectively; $\bar{\mu} = 0.815$.
				\textit{Mean Perf.}\ and \textit{\# Errors} summarize realized performance (out of 414 items).
				\textit{Level wedge}: $\bar{p}_m - \bar{\mu}$, average realized minus belief-predicted accuracy.
				\textit{Slope gap}: $\hat{b} - \hat{\beta}_m$, where $\hat{\beta}_m$ is the OLS slope of $p_{mi}$ on $d_i$; negative values indicate that beliefs are steeper in difficulty than realized performance.
				\textit{Avg. belief at errors}: $\frac{1}{E_m}\sum_{i:\,p_{mi}=0}\hat{p}_i$, the mean
				belief-predicted accuracy on items the model answers incorrectly, where $E_m$ is the total number of errors.
				%values above $\bar{\mu} = 0.815$ indicate that errors are over-represented on items beliefs rate as relatively easy.
			\end{tablenotes}
	\end{threeparttable}}
\end{table}

Table~\ref{tab:distortion} applies the belief rule estimated from the belief experiment  (mapping human item difficulty to expected AI accuracy) to each model's realized item-level performance on the same benchmark, holding beliefs fixed. Results from this exercise should be viewed as a qualitative proof of concept (rather than a precise quantification) since beliefs plausibly update across models and over time.

Findings across three diagnostics (defined in table notes) point in a similar direction. First, the \textit{level wedge} flips sign from GPT-3.5 Turbo ($-0.091$) to more recent models ($+0.112$ to $+0.177$): beliefs from our 2023 experimental participants would now systematically underestimate frontier accuracy on average. Second, the \textit{slope gap} is large and negative for every model ($-0.160$ to $-0.338$), with no monotone reduction as capability improves: realized accuracy is flatter with respect to human difficulty than beliefs predict across all systems tested, and the gap does not narrow among the highest-performing models. 
Third, the average predicted accuracy of errors---the mean belief-predicted success rate on items a model actually fails---is at or above the item average of $\bar{\mu}=0.815$ for Claude 3.7 and o3-mini, the highest-performing models in our benchmark. This suggests that, even near benchmark saturation, the remaining errors need not concentrate on items people perceive as difficult. Together, the difficulty-ordering distortion---the core HP misspecification---persists across all models we test, with no sign of convergence toward a more human-like error pattern.

%------------- ADOPTION  ---------------------%
\section{Adoption Experiment}
\label{appendix:adoption}
% ============================================================
\subsection{Misperceptions of Human Difficulty and Adoption}
\label{app:misperceptions}

\begin{table}[t]
	\caption{Misperceptions about Human Difficulty: Beliefs Proxy Validity}
	\label{tab:misperceptions}
	\centering
	\resizebox{\textwidth}{!}{
		\begin{threeparttable}
			\begin{tabular}{lcccc}
\toprule
Proxy: $\Delta^H\equiv p^H(hard)-p^H(easy)$ & Estimate & s.e. & $p$-value & $N$ \\
\midrule
\multicolumn{1}{l}{\textbf{Panel A: Misperceptions proxy ($\Delta^H\geq0$)}} \\
Baseline wave mean (\%)      & 55.9  & (2.8) &  & 306 \\
Replication wave mean (\%)   & 17.2  & (3.5) &  & 116 \\
Replication -- Baseline (pp) & -38.6  & (4.5) &  0.000    & 422 \\
\addlinespace[0.35em]
\multicolumn{1}{l}{\textbf{Panel B: Treatment does not affect the proxy}} \\
\emph{Black Box} effect in baseline (pp)      & -1.0 & (5.7) &  0.866   & 422 \\
\emph{Black Box} effect in replication (pp)   & 2.9 & (7.0) &  0.685   & 422 \\
Diff.\ across waves: $Black Box\times Rep$ (pp) & 3.8 & (9.1) &  0.674 & 422 \\
\bottomrule
\end{tabular}

			\begin{tablenotes}[flushleft]
				\footnotesize
				\item \textit{Notes:} The misperceptions proxy equals one if the participant reports initial beliefs in human performance on the hard pool at least as high as on the easy pool (ties counted as misperceptions; they are rare and mostly in baseline: $N=46$ overall: $42$ baseline, $4$ replication).
				Panel~A reports wave means and the replication--baseline difference from a pooled regression of the proxy on indicators for Black Box and Replication.
				Panel~B reports within-wave \emph{Black Box} effects and the across-wave difference from a pooled regression of the proxy on indicators for Black Box, Replication and their interaction.
				Robust standard errors are in parentheses.
			\end{tablenotes}
	\end{threeparttable}}
\end{table}

\sss{Proxy for misperceptions} Table~\ref{tab:misperceptions} documents the large share of misperceptions in the baseline sample (55.9\%), proxied by $\Delta^H\geq 0$: the difference between the initial beliefs about \emph{human} performance on easy vs hard tasks, \emph{before} the delegation phase, and thus before receiving endogenous signals of performance. The replication wave, which made human task difficulty more salient within the same design, strongly reduces these misperceptions (to 17.2\%) but they remain non-trivial. Importantly, the treatment does not affect the share of misperceptions either in baseline or replication, lending credibility to the proxy; results are robust to classifying ties as correct perceptions, and to using interim or final beliefs (instead of initial beliefs) as proxy.

\begin{table}[t]
	\caption{Belief-Conditional Treatment Effects on Adoption -- Initial Beliefs Proxy}
	\label{tab:adoption-regs-pr}
	\centering
	\resizebox{\textwidth}{!}{
		\begin{threeparttable}
			{\setlength{\tabcolsep}{3.2pt}
\renewcommand{\arraystretch}{1.1}
\begin{tabular}{l @{\hspace{10pt}} *{4}{c} *{2}{c}  p{6pt} r}
\toprule
 & \multicolumn{4}{c}{\textbf{\emph{Black Box} effect for participants with:}} & & & & \\
\cmidrule(lr){2-5}\addlinespace[0.15em]
 & \multicolumn{2}{c}{Incorrect perception} & \multicolumn{2}{c}{Correct perception} & \multicolumn{2}{c}{\textbf{Difference}} & & \\
\cmidrule(lr){2-3}\cmidrule(lr){4-5}\cmidrule(lr){6-7}
Outcome & Coefficient & s.e. & Coefficient & s.e. & Coeff. & s.e. & & N \\
\midrule
\textbf{All-or-nothing adoption:} & -16.9**\phantom{*} & (6.8) & -15.6*** & (5.7) & \phantom{-}1.3\phantom{***} & (8.9) & & 422 \\ \addlinespace[0.1em]
\hspace{1em}--- Full Adoption & -18.7*** & (6.7) & -17.2*** & (5.3) & \phantom{-}1.6\phantom{***} & (8.6) & & 422 \\
\hspace{1em}--- No Adoption & \phantom{-}1.8\phantom{***} & (2.0) & \phantom{-}1.6\phantom{***} & (2.7) & -0.3\phantom{***} & (3.4) & & 422 \\ \addlinespace[0.5em]
\textbf{Partial adoption:} & \phantom{-}16.9**\phantom{*} & (6.8) & \phantom{-}15.6*** & (5.7) & -1.3\phantom{***} & (8.9) & & 422 \\ \addlinespace[0.1em]
\hspace{1em}--- Only Hard & -5.9\phantom{***} & (7.2) & \phantom{-}12.7**\phantom{*} & (6.2) & \phantom{-}18.5*\phantom{**} & (9.5) & & 422 \\
\hspace{1em}--- Only Easy & \phantom{-}22.8*** & (5.6) & \phantom{-}2.9\phantom{***} & (4.5) & -19.8*** & (7.2) & & 422 \\
\bottomrule
\end{tabular}}

			\smallskip\footnotesize
			\begin{tablenotes}
				\item \textit{Notes:} Each row reports coefficients (in pp) from pooled linear regressions of the form
				$y=\alpha+\beta\,BB+\gamma\,Correct+\delta(BB\times Correct)+\rho\,Rep+X'\kappa+\varepsilon$,
				estimated separately for each adoption outcome $y$.
				$BB$, $Rep$, $Correct$,  are indicators for Black Box framing, replication wave, and correct perception of human difficulty ($=1$ if the participant reports initial beliefs in human performance lower on human-hard tasks than on human-easy tasks) respectively.
				Demographic controls $X$ include age, gender, bachelor's degree indicator, white indicator, and self-reported AI familiarity.
				``Incorrect perception'' column reports $\beta$, ``correct'' reports $\beta+\delta$, and ``difference'' reports $\delta$, i.e., the change in the BB effect when moving from incorrect to correct perceptions.
				Robust standard errors in parentheses, with significance: *** $p<0.01$, ** $p<0.05$, * $p<0.10$.
			\end{tablenotes}
	\end{threeparttable}}
\end{table}

\sss{Belief-conditional partial adoption}
Using this proxy, Table~\ref{tab:adoption-regs-pr} presents the main evidence on the role of misperceptions.
Each row reports estimates of the \emph{Black Box} treatment effect from pooled linear specifications of the form:
\begin{equation*}
	y=\alpha+\beta\,BB+\gamma\,Correct+\delta(BB\times Correct)+\rho\,Rep+X'\kappa+\varepsilon,
\end{equation*}
estimated separately for each adoption outcome $y$.
$BB$, $Rep$, and $Correct$ are respectively indicators for Black Box, replication wave, and the proxy (based on initial beliefs) for correct perceptions of human difficulty.
First, the overall reduction in all-or-nothing adoption is stable: \emph{Black Box} reduces Full Adoption and increases partial adoption by similar magnitudes for correct and incorrect perceivers (interactions near zero). Second, misperceptions strongly moderate the \emph{direction} of partial adoption. Among participants with incorrect initial beliefs about human difficulty, \emph{Black Box} reallocates mass primarily into the normatively wrong margin (``Only Easy'' increases substantially), while ``Only Hard'' does not rise. Among participants with correct priors, the reallocation flips: ``Only Hard'' increases, while ``Only Easy'' is essentially unchanged. The interaction coefficients quantify this reversal directly, and the pattern is stable with and without demographic controls. Our conservative interpretation is that \emph{Black Box} reliably reduces all-or-nothing adoption, consistent with our hypothesis, but that whether resulting partial adoption is normatively correct hinges on correct perceptions of the human task difficulty.
As a robustness check, Table~\ref{tab:adoption-regs-pt} reports highly similar results using the final reported beliefs (elicited right after the final adoption decision) as misperception proxy, instead of the initial beliefs.

\begin{table}[t]
	\caption{Belief-Conditional Treatment Effects on Adoption -- Final Beliefs Proxy}
	\label{tab:adoption-regs-pt}
	\centering
	\resizebox{\textwidth}{!}{
		\begin{threeparttable}
			{\setlength{\tabcolsep}{3.2pt}
\renewcommand{\arraystretch}{1.1}
\begin{tabular}{l @{\hspace{10pt}} *{4}{c} *{2}{c}  p{6pt} r}
\toprule
 & \multicolumn{4}{c}{\textbf{\emph{Black Box} effect for participants with:}} & & & & \\
\cmidrule(lr){2-5}\addlinespace[0.15em]
 & \multicolumn{2}{c}{Incorrect perception} & \multicolumn{2}{c}{Correct perception} & \multicolumn{2}{c}{\textbf{Difference}} & & \\
\cmidrule(lr){2-3}\cmidrule(lr){4-5}\cmidrule(lr){6-7}
Outcome & Coefficient & s.e. & Coefficient & s.e. & Coeff. & s.e. & & N \\
\midrule
\textbf{All-or-nothing adoption:} & -19.3**\phantom{*} & (7.8) & -14.0*** & (5.1) & \phantom{-}5.3\phantom{***} & (9.3) & & 422 \\ \addlinespace[0.1em]
\hspace{1em}--- Full Adoption & -22.0*** & (7.6) & -15.2*** & (4.9) & \phantom{-}6.8\phantom{***} & (9.1) & & 422 \\
\hspace{1em}--- No Adoption & \phantom{-}2.6\phantom{***} & (3.1) & \phantom{-}1.1\phantom{***} & (2.1) & -1.5\phantom{***} & (3.7) & & 422 \\ \addlinespace[0.5em]
\textbf{Partial adoption:} & \phantom{-}19.3**\phantom{*} & (7.8) & \phantom{-}14.0*** & (5.1) & -5.3\phantom{***} & (9.3) & & 422 \\ \addlinespace[0.1em]
\hspace{1em}--- Only Hard & -11.2\phantom{***} & (7.0) & \phantom{-}12.8**\phantom{*} & (5.6) & \phantom{-}24.0*** & (8.9) & & 422 \\
\hspace{1em}--- Only Easy & \phantom{-}30.6*** & (7.0) & \phantom{-}1.2\phantom{***} & (3.5) & -29.3*** & (7.8) & & 422 \\
\bottomrule
\end{tabular}}

			\smallskip\footnotesize
			\begin{tablenotes}
				\item \textit{Notes:} Each row reports coefficients (in pp) from pooled linear regressions of the form
				$y=\alpha+\beta\,BB+\gamma\,Correct+\delta(BB\times Correct)+\rho\,Rep+X'\kappa+\varepsilon$,
				estimated separately for each adoption outcome $y$.
				$BB$, $Rep$, $Correct$,  are indicators for Black Box framing, replication wave, and correct perception of human difficulty ($=1$ if the participant reports final beliefs in human performance lower on human-hard tasks than on human-easy tasks) respectively.
				Demographic controls $X$ include age, gender, bachelor's degree indicator, white indicator, and self-reported AI familiarity.
				``Incorrect perception'' column reports $\beta$, ``correct'' reports $\beta+\delta$, and ``difference'' reports $\delta$, i.e., the change in the BB effect when moving from incorrect to correct perceptions.
				Robust standard errors in parentheses, with significance: *** $p<0.01$, ** $p<0.05$, * $p<0.10$.
			\end{tablenotes}
	\end{threeparttable}}
\end{table}

% ============================================================
\subsection{Additional Results}
\label{app:additional}

\sss{Delegation behavior and realized AI signals}
Table~\ref{tab:delegation-signals} shows that \emph{Black Box} reduces delegation to the AI by 9.7 pp overall ($p<0.001$), with a larger decline on the easy pool (14.8 pp) than on the hard pool (4.5 pp)---so the easy--hard delegation gap widens by around 10 pp under \emph{Black Box} ($p<0.001$). Conditional on delegating to the AI, realized AI failure rates are statistically similar across framings, both overall and by pool (Panel~B). %Thus, the framing primarily shifts the propensity to delegate rather than generating large differences in realized AI performance conditional on delegation.

\begin{table}[H]
	\caption{Adoption Experiment -- Delegation Behavior}
	\label{tab:delegation-signals}
	\centering
	\resizebox{.75\textwidth}{!}{
		\begin{threeparttable}
			\begin{tabular}{lcccc}
\toprule
 & \multicolumn{2}{c}{\textbf{Treatment}} &  &  \\
\cmidrule(lr){2-3} 
 & \textit{Anthropomorphic} & \textit{Black Box} & \textbf{Diff.} & $p$ \\
\midrule
\multicolumn{1}{l}{\textbf{Panel A: Delegation to AI (\%)}} \\
AI delegation (overall) &  61.4 &  51.7 & \textbf{ -9.7} & <0.001 \\
AI delegation (easy pool) &  47.4 &  32.5 & \textbf{-14.8} & <0.001 \\
AI delegation (hard pool) &  75.3 &  70.8 & \textbf{ -4.5} & 0.009 \\
Easy--hard delegation gap & -27.9 & -38.3 & \textbf{-10.4} & <0.001 \\
\addlinespace[0.25em]
\multicolumn{1}{l}{\textbf{Panel B: AI failures among delegations (\%)}} \\
AI failure rate (overall) &  34.1 &  34.6 & \textbf{  0.5} & 0.280 \\
AI failure rate (easy pool) &  35.6 &  38.3 & \textbf{  2.6} & 0.139 \\
AI failure rate (hard pool) &  33.7 &  34.4 & \textbf{  0.7} & 0.211 \\
\bottomrule
\end{tabular}

			\smallskip
			\begin{tablenotes}[flushleft]
				\footnotesize
				\item \textit{Notes:} The table reports mean outcomes by treatment arm and their difference. Panel~A reports the share of tasks delegated to the AI (in pp), overall and separately for the easy vs.\ hard task pools; the ``easy--hard delegation gap'' is the easy-pool delegation rate minus the hard-pool delegation rate. Panel~B reports realized AI failure rates (in \%) conditional on delegating to the AI. $p$-values are from two-sided $t$-tests by treatment arm.
			\end{tablenotes}
	\end{threeparttable}}
\end{table}

\begin{table}[H]
	\caption{Adoption Experiment -- Balance Table}
	\label{tab:balance-adoption}
	\centering
	\resizebox{.75\textwidth}{!}{
		\begin{threeparttable}
			\begin{tabular}{lcccc p{6pt} cccc} \toprule
 & \multicolumn{4}{c}{\textbf{Baseline Sample}} & & \multicolumn{4}{c}{\textbf{Replication Sample}} \\
\cmidrule(lr){2-5}\cmidrule(lr){7-10}
 & \multicolumn{2}{c}{\textit{Anthropomorphic}} & \multicolumn{2}{c}{\textit{Black Box}} & & \multicolumn{2}{c}{\textit{Anthropomorphic}} & \multicolumn{2}{c}{\textit{Black Box}} \\
\cmidrule(lr){2-3}\cmidrule(lr){4-5}\cmidrule(lr){7-8}\cmidrule(lr){9-10}
 & Value & s.e. & Value & s.e. & & Value & s.e. & Value & s.e. \\
\midrule
\multicolumn{1}{l}{\textbf{Demographics}} & & & & & & & & & \\
Age (mean) & 41.5 & ( 1.0) & 37.8 & ( 1.0) & & 39.2 & ( 1.8) & 37.7 & ( 1.7) \\
Female (\%) & 52.3 & ( 4.1) & 49.0 & ( 4.0) & & 49.1 & ( 6.6) & 54.2 & ( 6.5) \\
White (\%) & 58.4 & ( 4.0) & 59.9 & ( 3.9) & & 73.7 & ( 5.8) & 64.4 & ( 6.2) \\
Black (\%) & 34.9 & ( 3.9) & 31.8 & ( 3.7) & & 22.8 & ( 5.6) & 32.2 & ( 6.1) \\
\addlinespace[0.2em]
\multicolumn{1}{l}{\textbf{Socioeconomic and education}} & & & & & & & & & \\
Income (mean category) &  2.6 & ( 0.1) &  2.5 & ( 0.1) & &  2.2 & ( 0.2) &  2.5 & ( 0.2) \\
HS or more (\%) & 100.0 & ( 0.0) & 99.4 & ( 0.6) & & 100.0 & ( 0.0) & 98.3 & ( 1.7) \\
Bachelor or more (\%) & 73.8 & ( 3.6) & 76.4 & ( 3.4) & & 61.4 & ( 6.4) & 74.6 & ( 5.7) \\
Graduate or more (\%) & 25.7 & ( 3.6) & 21.8 & ( 3.3) & & 15.8 & ( 4.8) & 23.7 & ( 5.5) \\
\addlinespace[0.2em]
\multicolumn{1}{l}{\textbf{Math background}} & & & & & & & & & \\
HS math or more (\%) & 99.3 & ( 0.7) & 98.7 & ( 0.9) & & 100.0 & ( 0.0) & 98.3 & ( 1.7) \\
College math or more (\%) & 52.3 & ( 4.1) & 63.1 & ( 3.9) & & 47.4 & ( 6.6) & 64.4 & ( 6.2) \\
Graduate math or more (\%) &  4.0 & ( 1.6) &  5.1 & ( 1.8) & &  3.5 & ( 2.4) &  3.4 & ( 2.4) \\
\addlinespace[0.2em]
\multicolumn{1}{l}{\textbf{AI familiarity}} & & & & & & & & & \\
Never heard of AI (\%) &  0.7 & ( 0.7) &  1.3 & ( 0.9) & &  0.0 & ( 0.0) &  0.0 & ( 0.0) \\
Heard of AI, no further familiarity (\%) & 22.1 & ( 3.4) & 26.1 & ( 3.5) & & 19.3 & ( 5.2) & 22.0 & ( 5.4) \\
Some understanding, never used (\%) & 42.3 & ( 4.0) & 41.4 & ( 3.9) & & 40.4 & ( 6.5) & 32.2 & ( 6.1) \\
Interacted with AI (\%) & 34.9 & ( 3.9) & 31.2 & ( 3.7) & & 40.4 & ( 6.5) & 45.8 & ( 6.5) \\
\midrule
\textbf{Observations} & \multicolumn{2}{c}{149} & \multicolumn{2}{c}{157} & & \multicolumn{2}{c}{ 57} & \multicolumn{2}{c}{ 59} \\
\bottomrule \end{tabular}

			\smallskip\footnotesize
			\begin{tablenotes}[flushleft]
				\footnotesize
				\item \textit{Notes:} The table reports baseline covariate balance by treatment arm, separately for the Baseline sample and the Replication sample. ``Value'' is the cell mean; ``s.e.'' is the standard error of the mean (for binary indicators, computed from the Bernoulli standard error). Shares are shown in percent; age is in years; income is the categorical index used in the survey.
			\end{tablenotes}
	\end{threeparttable}}
\end{table}

%------------ PARENTDATA ---------------------%
\section{Engagement Experiment}
\label{appendix:parentdata}

\subsection{Institutional Context and Interaction Data}\label{appendix:parentdata-context}

This appendix provides institutional background and descriptive details for the ParentData setting that are not required for the main text exposition.\footnote{The full list of conversations used in the experiment (five pairs plus the baseline successes), along with their usefulness and reasonableness ratings, is available in the separate \href{https://www.dropbox.com/scl/fi/a8gua4s23yur7dmcj946w/ai_draft_exp_materials.pdf?rlkey=cplykye5bo1vnekoidd45hpwj&dl=0}{Experimental Materials Appendix}.}

\sss{ParentData.org}
ParentData.org is a self-described ``data-driven guide through pregnancy, parenthood, and beyond.'' Created in 2020 by American economist Emily Oster---and launched as its own website in 2023---its goal is to translate scientific evidence into rigorous answers to questions asked by current or expectant parents. The platform provides multiple formats (articles, newsletters, podcasts) with several tiers of paid subscription. It also offers an AI chatbot service that is free to use and openly accessible.

\sss{User base}
The pool of users is almost exclusively composed of people who are trying to conceive, currently expecting, or parents of young children. The website does not collect user demographics. A  survey from July 2024 and discussions with the ParentData team describe the prototypical user as (i) a woman in her 30s living in the U.S.; (ii) either expecting or a mother of young children; (iii) educated; and (iv) with higher income. We recruit all experimental participants on the basis of these demographic features.

\sss{The AI chatbot}
The website hosts Dewey, an ``AI librarian,'' which is an LLM-based AI chatbot.\footnote{This service is provided by the Dewey Labs company, of which ParentData.org is a client. Dewey has since the time of the experiment been renamed ``Petey.''} Dewey was built by first summarizing parenting materials (books and articles) into a set of questions and answers, which were then verified and vetted by humans from the ParentData team. Dewey is then available to answer users' questions by matching each incoming query to the archive of pre-made questions using a confidence score, and displaying the highest-score answer if it exceeds a confidence threshold. The matching relies on a measure of textual similarity which is related but not identical to what humans understand by ``similarity.''\footnote{The matching process is a custom approach to vector embeddings, whose details were not disclosed. We confirm that the AI score correlates with various textual similarity measures. Below the confidence threshold, the chatbot either suggests related questions or displays the message: ``I'm sorry. It looks like we don't have an answer in the ParentData archives. However, your question has caught our attention and will be shared with Emily and the team. It could be a great topic for a future newsletter!''}

\sss{Interaction data snapshot}
Our interaction data consist of approximately 30{,}000 conversations, where a ``conversation'' is defined as one user query and one AI answer, occurring between December 31\textsuperscript{st} 2023 and April 30\textsuperscript{th} 2024. The Experimental Materials Appendix  presents descriptive statistics for this dataset, including the most frequent questions asked to the chatbot.

\subsection{Selecting Matched Conversation Pairs}\label{appendix:parentdata-pairs}

This appendix documents the pre-registered procedure used to construct matched \emph{pairs} of real ParentData conversations that differ in the \emph{human reasonableness} of the chatbot's misunderstanding while holding constant the user's intent and the (low) usefulness of the answer. Throughout, a ``conversation'' refers to one user query and one chatbot answer.

\sss{Step 1: Initial coding and candidate pairs}
We began from the corpus of real ParentData conversations. We manually coded each conversation for:
(i) \emph{intent} (a short label capturing what the user is asking), and
(ii) a binary indicator for whether the chatbot \emph{misunderstood} the question (a proxy for the answer being unhelpful/useless for that specific query).
We then created an initial list of candidate \emph{same-intent} pairs in which \emph{both} answers were coded as misunderstandings. This produced approximately 40 candidate pairs.

\sss{Step 2: Reasonableness ratings}
We recruited participants from the population of interest (parents of young children or currently expecting) to rate the \emph{reasonableness} of each misunderstood answer. The survey presented a single conversation at a time in a display mimicking the ParentData website. To avoid within-subject contrast effects, participants did not see both sides of a pair: we split each pair into two pools and randomly drew conversations from one pool per participant. Instructions emphasized that the object is not whether the answer is helpful, but whether the \emph{specific misunderstanding} is a plausible human error.
\emph{Reasonableness prompt (verbatim).} ``The AI gave answers that were deemed unhelpful. What do you think is the percent (\%) chance that a \textbf{reasonable human} would misunderstand the questions in the way the AI did? Choose a \% between 0 and 100.''

\sss{Step 3: Usefulness ratings}
In a separate study with the same target population and a nearly identical display, we elicited perceived \emph{usefulness} of each answer for the specific question on a 1--5 scale. We included a small number of genuinely useful conversations in the rating pool both (i) to obtain usefulness ratings for the ``success'' conversations used in the main experiment and (ii) to avoid an unnatural rating environment in which all answers are useless. Instructions stressed that usefulness is about answering the specific query, not about sounding generally sensible or containing good generic parenting advice.
\emph{Usefulness prompt (verbatim).}  ``Assess the answer's \textbf{usefulness}: does its content answer \emph{that specific question?} \textbf{Read carefully:} some answers may \emph{appear} useful at a glance, even though they are not! Indicate your answer on the [1--5] scale.''

\sss{Final selection rule and implemented pairs}
We retained five final pairs satisfying three criteria:
(i) the two conversations within a pair share the same intent (from Step 1),
(ii) both sides are comparably unhelpful (same \emph{median} usefulness, equal to 1 or 2), and
(iii) the pair exhibits a large difference in \emph{median} reasonableness.
As pre-registered, medians are the primary matching criterion to reduce sensitivity to outliers and scale use. As a robustness check, we check that (a) average usefulness differences within a pair are not statistically significant at the 90\% level, while (b) average reasonableness differences are statistically significant. We additionally set aside three ``success'' conversations (median usefulness 4 or 5) used as common baseline content in the engagement experiment.

%\subsection{Additional Experimental Results}
%\label{appendix:parentdata-results}

\begin{table}[H]
	\caption{Demographics of Participants in Engagement Experiment}
	\label{balance-pd}
	\centering
	\resizebox{.65\textwidth}{!}{
		\begin{threeparttable}
			\begin{tabular}{lcccc} \toprule
 & \multicolumn{2}{c}{\textit{Unreasonable}} & \multicolumn{2}{c}{\textit{Reasonable}} \\
\cmidrule(lr){2-3}\cmidrule(lr){4-5}
 & Value & std. err. & Value & std. err. \\
\midrule
\multicolumn{5}{l}{\textbf{Gender}} \\
Female & 85.5 & ( 1.9) & 83.8 & ( 2.1) \\
Male & 13.3 & ( 1.8) & 15.9 & ( 2.1) \\
\addlinespace[0.2em]
\multicolumn{5}{l}{\textbf{Race}} \\
White & 73.5 & ( 2.4) & 76.8 & ( 2.4) \\
Black & 17.1 & ( 2.0) & 17.5 & ( 2.1) \\
\addlinespace[0.2em]
\multicolumn{5}{l}{\textbf{Education}} \\
HS or more & 99.7 & ( 0.3) & 99.4 & ( 0.4) \\
Bachelor or more & 56.9 & ( 2.7) & 58.4 & ( 2.8) \\
\addlinespace[0.2em]
\multicolumn{5}{l}{\textbf{Parenting status}} \\
Trying to conceive & 23.6 & ( 2.3) & 20.3 & ( 2.3) \\
Currently expecting &  7.1 & ( 1.4) &  7.0 & ( 1.4) \\
Have children below 18 & 80.2 & ( 2.2) & 79.0 & ( 2.3) \\
\addlinespace[0.2em]
\multicolumn{5}{l}{\textbf{AI familiarity}} \\
\hspace{2mm}Never heard of AI &  1.5 & ( 0.7) &  2.2 & ( 0.8) \\
\hspace{2mm}Heard of AI, no further familiarity & 20.9 & ( 2.2) & 20.0 & ( 2.3) \\
\hspace{2mm}Some understanding, never used & 39.5 & ( 2.7) & 40.6 & ( 2.8) \\
\hspace{2mm}Interacted with AI & 38.1 & ( 2.6) & 37.1 & ( 2.7) \\
\midrule
Age (years) & 34.5 & ( 0.3) & 34.1 & ( 0.3) \\
\midrule
\textbf{Observations} & \multicolumn{2}{c}{339} & \multicolumn{2}{c}{315} \\
\bottomrule \end{tabular}

			\smallskip\footnotesize
			\begin{tablenotes}
				\item \textit{Notes:} This table provides demographics of participants (current or expecting parents) in the engagement experiment. ``Value'' reports cell mean and ``std. err.'' reports the standard error of the mean.
			\end{tablenotes}
	\end{threeparttable}}
\end{table}

\end{document}